\documentclass[graybox]{svmult}

\usepackage{physics,amssymb,hyperref}

\usepackage{type1cm}        
\usepackage{makeidx}         
\usepackage{graphicx}        
\usepackage{multicol}        
\usepackage[bottom]{footmisc}

\usepackage{newtxtext}       %
\usepackage[varvw]{newtxmath}       

\newcommand\ion[2]{\text{#1\,\textsc{\lowercase{#2}}}} 
\newcommand\lya{Ly$\alpha$} 
\newcommand\mlya{Ly\alpha} 
\newcommand\ha{H$\alpha$} 
\newcommand\mha{H\alpha} 
\newcommand\heii{He\,\textsc{\lowercase{II}}} 
\newcommand\heiii{He\,\textsc{\lowercase{III}}} 
\newcommand\mean[1]{\langle #1 \rangle} 
\newcommand\sbunit{erg~s$^{-1}$~cm$^{-2}$~arcsec$^{-2}$}

\makeindex             

\begin{document}

\title*{The multiphase circumgalactic medium and its relation to galaxies: an observational perspective}
\titlerunning{Observing the multiphase CGM}
\author{Michele Fumagalli}
\authorrunning{M. Fumagalli}
\institute{Michele Fumagalli \at Universit\`a degli Studi di Milano-Bicocca, Piazza della Scienza 3, 20126 Milano, Italy.
\email{michele.fumagalli@unimib.it}}

%
%
\maketitle

\abstract{The circumgalactic medium (CGM) is a vital element in galaxies, as it mediates the baryon cycle essential for regulating galaxy activity. It is also highly complex due to the intricate distributions of temperature, density, metallicity, and ionization that make the CGM a multiphase medium. Therefore, learning about the CGM requires combining various observational techniques. This contribution starts by reviewing how absorption spectroscopy, together with modeling of the ionization conditions, yields critical insights into the underlying physical properties of the CGM. Next, the rapidly growing application of imaging and integral field spectroscopy for studying the halo gas in emission, using hydrogen and metal lines as tracers, is examined. Finally, the essential role of the CGM in galaxy evolution is highlighted by considering current studies that directly link galaxies to their halo gas. The novel dimension of how the environment affects the CGM and alters the evolution of galaxies is also investigated.}

\section{Introduction}
\label{sec:absorption}

The circumgalactic medium (CGM) is the gas that extends beyond the gas and stellar disc of a galaxy and up to scales comparable to the size of the virial overdensity of halos. This gas is multiphase. The fundamental physical processes that occur in the CGM (mainly inflows, outflows, and environmental interactions) are indeed characterized by, or responsible for, order-of-magnitude differences in the gas's physical density, temperature, and metallicity, thus making the CGM a highly dynamic and complex medium to study. 

Observations of the CGM in absorption using background sightlines as gas probes reveal multiple absorption lines of hydrogen and metal ions in various ionization states, ranging from neutral species (e.g., \ion{H}{I} or \ion{C}{I}) that can be found only in cold and dense gas with temperature $T\lesssim 10^4$\,K and hydrogen density up to $n_H \approx 10-100$\,cm$^{-3}$ \cite{Neeleman2015}, or highly ionized elements (e.g., \ion{Ne}{VIII} or \ion{Mg}{X}) which are indicative of much hotter ($T\gtrsim 10^{5.5}$\,K) and underdense ($n_H \lesssim 10^{-4}$\,cm$^{-3}$) plasma \cite{Oppenheimer2012}. Remarkably, such a diversity of physical states can be observed in individual sightlines, unambiguously revealing the presence of a highly multiphase medium even on small spatial scales \cite{Tripp2011}.

\begin{figure}
\sidecaption
\centering
\includegraphics[scale=.2]{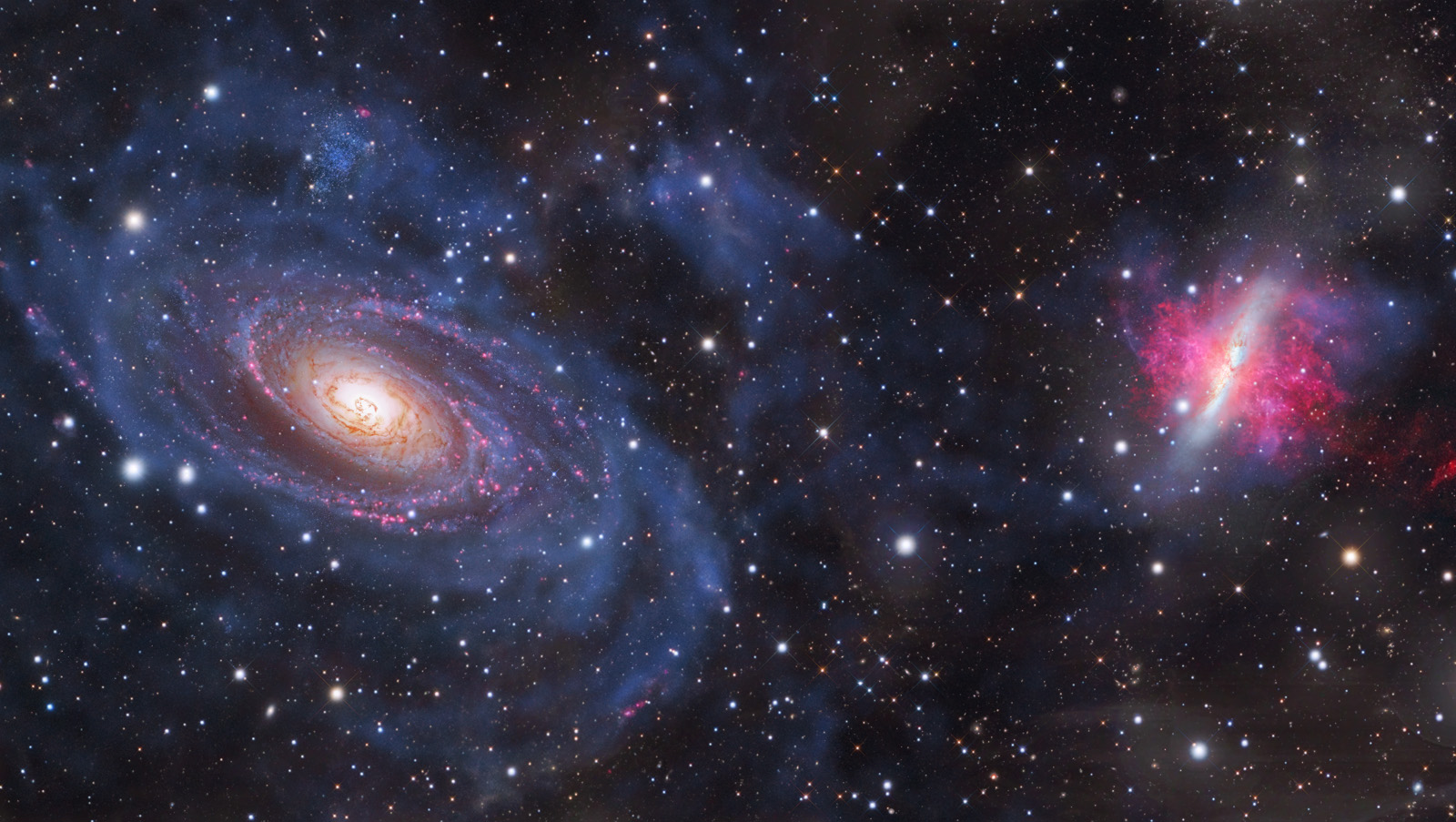}
\caption{Composite image of the M81 group, with the M81 galaxy to the left and the M82 galaxy to the right. The multiphase CGM, containing neutral (in blue) and ionized (in red) gas,
is visible in multi-wavelength data. The image has been obtained by combining observations from the Hubble Space Telescope, the Subaru Telescope, the Spitzer Telescope, and the Very Large Array. {\it \href{http://www.robgendlerastropics.com/M81-82-HST-Subaru-H1.html}{Credit}: R. Gendler, R. Croman, R. Colombari; Acknowledgement: R. Jay GaBany; VLA Data: E. de Block (ASTRON)}.}
\label{fig:m81}       
\end{figure}

\begin{figure}[b]
\centering
\includegraphics[scale=.3]{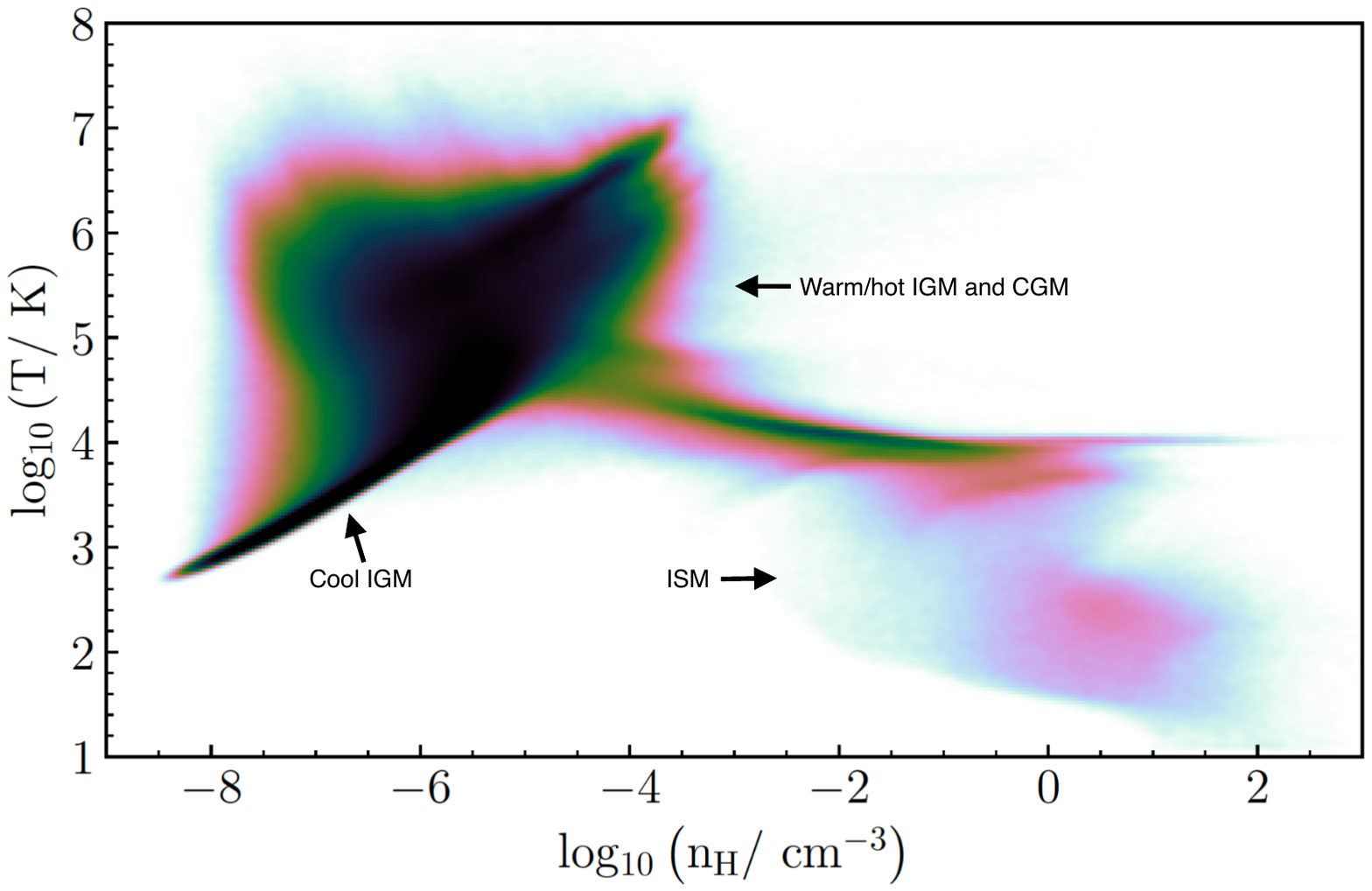}
\caption{Density-temperature diagram for a hydrodynamical cosmological simulation at $z\approx 0$. The high-density cold medium at $T\lesssim 10^{4}$\,K and $n_H \gtrsim 1$\,cm$^{-3}$ is related to the interstellar medium, or ISM; the thin locus of gas stretching below $n_H \lesssim 10^{-5}$\,cm$^{-3}$ with temperature proportional to the density is associated to the IGM; finally, the gas above $T\approx 10^{4}-10^{7}$\,K and between $n_H \lesssim 10^{-7}-10^{-2}$\,cm$^{-3}$ is connected in part to the CGM and in part to the warm-ionized IGM. The color coding reflects the relative number of resolution elements that occupy each region, where dark colors are for the largest fractions. {\it Credit: Dr. Alejandro Benitez-Llambay}.}
\label{fig:simrhoT}       
\end{figure}

The multiphase nature of the CGM is also manifest in emission, particularly in the nearby Universe where imaging and spectroscopy at high spatial resolution are available. The M81 group is a particularly iconic example. An atomic gas phase extending across the entire group is visible via \ion{H}{I} observations \cite{deBlok2018} (blue in Fig.~\ref{fig:m81}), molecular gas beyond the disc radius of M82 is visible in CO($1-0$) \cite{Krieger2021}, and a warm/hot phase from the M82 outflow (red in Fig.~\ref{fig:m81}) appears in optical and X-ray observations \cite{Lehnert1999}.   
Finally, cosmological simulations that capture the hydrodynamics and thermodynamics of the CGM predict a wide variety of densities and temperatures for the halo gas, ranging between $T\approx 10^{4}-10^{7}$\,K and $n_H \lesssim 10^{-7}-10^{-2}$\,cm$^{-3}$ (see Figure~\ref{fig:simrhoT}) 

The first part of this chapter provides an overview of the current techniques employed to gain observational insight into the physics of the multiphase CGM through absorption spectroscopy of background sources (Section~\ref{sec:absorption}) and via imaging and spectroscopy of UV emission lines (Section~\ref{sec:Hemission} and Section~\ref{sec:Zemission}). It also provides a summary of some representative studies in the field. The second part of this contribution (Section~\ref{sec:galaxies}) reviews current efforts in linking the multiphase CGM to the associated galaxy populations and the surrounding intergalactic medium (IGM). A discussion in Section~\ref{sec:environment} on the role of the wider-scale galaxy environment on the CGM -- an emerging and fast-developing field of research -- concludes this chapter.

\section{Observing the multiphase CGM in absorption}
\label{sec:absorption}

The traditional and still widely adopted technique for observing the multiphase CGM is through high-resolution spectroscopy of bright background sources (typically quasars), which are used to measure the properties of the low-density CGM in absorption along their line of sight. By targeting the rest-frame UV transitions of the principal elements (e.g., hydrogen, carbon, silicon, oxygen, iron), observers measure the column density of a given set of ions $N_{j,i}$ where the index $j$ refers to the ionization state and the index $i$ refers to the element. As gas at different densities and temperatures gives rise to distinct ions (see fig.\,6 in \cite{Tumlinson2017}), absorption spectroscopy constrains the physical properties of the CGM only through a technique that maps the observed ions into the underlying values of temperature, density, and metallicity.

This Section introduces an overview of how a model of the gas ionization state can be computed and used to learn about the multiphase CGM (Section~\ref{subsec:ioncor}), provides example applications to the study of absorption line systems (Section~\ref{subsec:als}), and analyzes the main shortcomings and limitations of this approach (Section~\ref{subsec:ionclimit}).

\subsection{Modeling ionization corrections}\label{subsec:ioncor}

In practice, through an assumed model for the ionization state of the gas, we can derive ionization fractions $X_{j,i}(T,n_H)$ (see fig. 2 of \cite{Fumagalli2016}) that map the column density of an observed ion to the parent element as a function of density and temperature:
\begin{equation}
N_i = X^{-1}_{j,i}(T,n_H) N_{j,i}\,.
\end{equation}
The ionization fractions are often referred to as ionization corrections because they can be applied to an observed ion to correct the observation and derive the original element abundance. Further assuming an abundance pattern $A_i$ for each element relative to hydrogen, we can derive a set of equations 
\begin{equation}
N_{j,i} = X_{j,i}(T,n_H) A_i N_H\,.
\end{equation}
Comparing an observed set of ions, $\{N_{j,i} \}_o$, with a model set, $\{N_{j,i} \}_m$, one can constrain the temperature and density of the gas, and its metallicity in solar units 
\begin{equation}
[M/H] = \log (N_M/N_H) - \log (N_M/N_H)_\odot\,.
\end{equation}
To fully specify $X_{j,i}$, it is necessary to understand the detailed ionization state of the gas by solving the balance between photoionization, collisional ionization, and recombination. Solving the gas thermal state is also required by computing the heating and cooling rates.

\subsubsection{The gas ionization state}

The first step in computing the ionization corrections is solving equations for the recombination and ionization rate due to radiation and collisions. This is a complex calculation to tackle even numerically because of the need to specify the geometry and composition of the material and the radiation field. 
To make the problem more tractable, equilibrium conditions are often assumed by solving for the steady-state conditions in which ionizations balance recombinations (see Section~\ref{subsec:nonequilibrium} for a discussion on non-equilibrium effects). 

A cloud of gas illuminated by radiation is subject to photoionization (bound-free transitions), recombination (free-bound transitions), and radiative de/excitations (bound-bound transitions). For a general reaction in which $A + B \rightarrow C$, the C production rate is set by the products of the number density of reagents, the cross-section of the interaction, and the reaction speed. Considering the photoionization of hydrogen as a simple case\footnote{The formalism that follows can be generalized to any element, by tracking the ionization and recombination with the suitable cross-sections and coefficients derived from atomic physics.}, for which $H^0 + h\nu \rightarrow H^{+} + e^{-}$, the rate of production of ions is set by
\begin{equation}
   \dv{n_{H^+}}{t} =\sigma_\nu n_{H^0} n_\gamma c
\end{equation}
where $\sigma_\nu$ is the frequency-dependent hydrogen photoionization cross-section (see fig.~13.1 in \cite{Drain2011}), $n_\gamma$ is the photon number density, and the speed of light $c$ gives the velocity of the reaction. 
By transforming the photon number density in a mean intensity $J(\nu)$ via the energy density $u(\nu)$
\begin{equation}
n_\gamma = \frac{u(\nu)}{h\nu} d\nu = \frac{4\pi J(\nu)}{c h\nu} d\nu\,,
\end{equation}
the photoionization rate equation becomes 
\begin{equation}\label{eq:ionization}
   \dv{n_{H^+}}{t} =n_{H^0} \int_{\nu_i}^\infty \sigma_\nu \frac{4\pi J(\nu)}{h\nu}d\nu \equiv \Gamma_{H^0}n_{H^0}\,,
\end{equation}
where $\nu_i$ is the threshold frequency at which ionization occurs and $ \Gamma_{H^0}$ is the photoionization rate. From this equation, it becomes apparent that a spectrum for the radiation must be assumed in the calculation.

Considering the opposite reaction for recombination, $H^{+} + e^{-}\rightarrow H^0 + h\nu$, the rate of production of neutral atoms (hereafter, the neutrals) is set by 
\begin{equation}\label{eq:recombination}
    \dv{n_{H^0}}{t} = \alpha(T)n_{H^+}n_{e^-}\,,
\end{equation}
with $\alpha(T)$ the temperature-dependent recombination coefficient (see fig.~14.1 in \cite{Drain2011}).
When recombining onto the ground state, ionizing photons are produced, which could either escape the cloud or be absorbed again, causing a new ionization. Two cases can be considered for the recombination coefficient: case A and case B. Under case A, the cloud is optically thin, and ionizing photons can escape. Therefore, recombinations can occur onto every $n,l$ state, including the ground:
\begin{equation}
    \alpha_A (T) = \sum_{n=1}^{\infty}\sum_{l=0}^{n-1} \alpha_{n,l}(T)\,.
\end{equation}
Conversely, under case B, the cloud is optically thick, and the ionizing photons produced by recombinations on the ground state cannot escape the cloud as they are immediately reabsorbed. Thus, the effective recombination coefficient is the sum of all allowed recombinations except the one to the ground state
\begin{equation}
    \alpha_B (T) = \alpha_A(T) - \alpha_{1s}(T)\,.
\end{equation}

By imposing equilibrium, there is no net rate of production of neutrals and ions, and Equations~\ref{eq:ionization} and \ref{eq:recombination} can be equated, leading to 
\begin{equation}
        \Gamma_{H^0}n_{H^0} = \alpha(T)n_{H^+}n_{e^-}\,.
\end{equation}
The solution of this equation is non-trivial, as it requires knowledge of the photoionization cross-section and the recombination coefficient from quantum mechanics. Moreover, the solution requires knowledge of the local radiation field, which, under most circumstances, is not specified for the small element of gas under consideration and must be computed by solving the radiative transfer equation, which accounts for the change of intensity $I_\nu$ due to absorption and emission along a ray
\begin{equation}\label{eq:RTeq}
    dI_\nu = (-\alpha_\nu I_\nu + j_\nu) ds\,,
\end{equation}
where $\alpha_\nu$ is the absorption coefficient, $j_\nu$ is the emission coefficient, and $ds$ is the path along the ray. As intensity is present on both sides, the solution of this equation is non-trivial unless some simplifying assumptions are invoked. For instance, most applications assume simple geometry, like a plane-parallel gas slab. 

In addition to photoionization, collisions between electrons and atoms are another ionization channel, $H^0 + e^- \rightarrow H^+ + e^- + e^-$. The rate of production of ions due to collisions is 
\begin{equation}
    \dv{n_{H^+}}{t} = \Gamma_{e}(T) n_{e^-}n_{H^0}\,,
\end{equation}
with $\Gamma_e$ the collisional ionization coefficient.
Including collisions, the equilibrium condition becomes
\begin{equation}
        \Gamma_{H^0}n_{H^0} + \Gamma_{e}(T) n_{e^-}n_{H^0}= \alpha(T)n_{H^+}n_{e^-}\,.
\end{equation}
From the above equation, it becomes apparent that the gas temperature enters the calculation of the recombination and collisional ionization coefficients $\alpha$ and $\Gamma_e$. 
The thermal state of the gas must also be explicitly calculated, as described in the next Section.

\begin{figure}[b]
\centering
\includegraphics[scale=0.35]{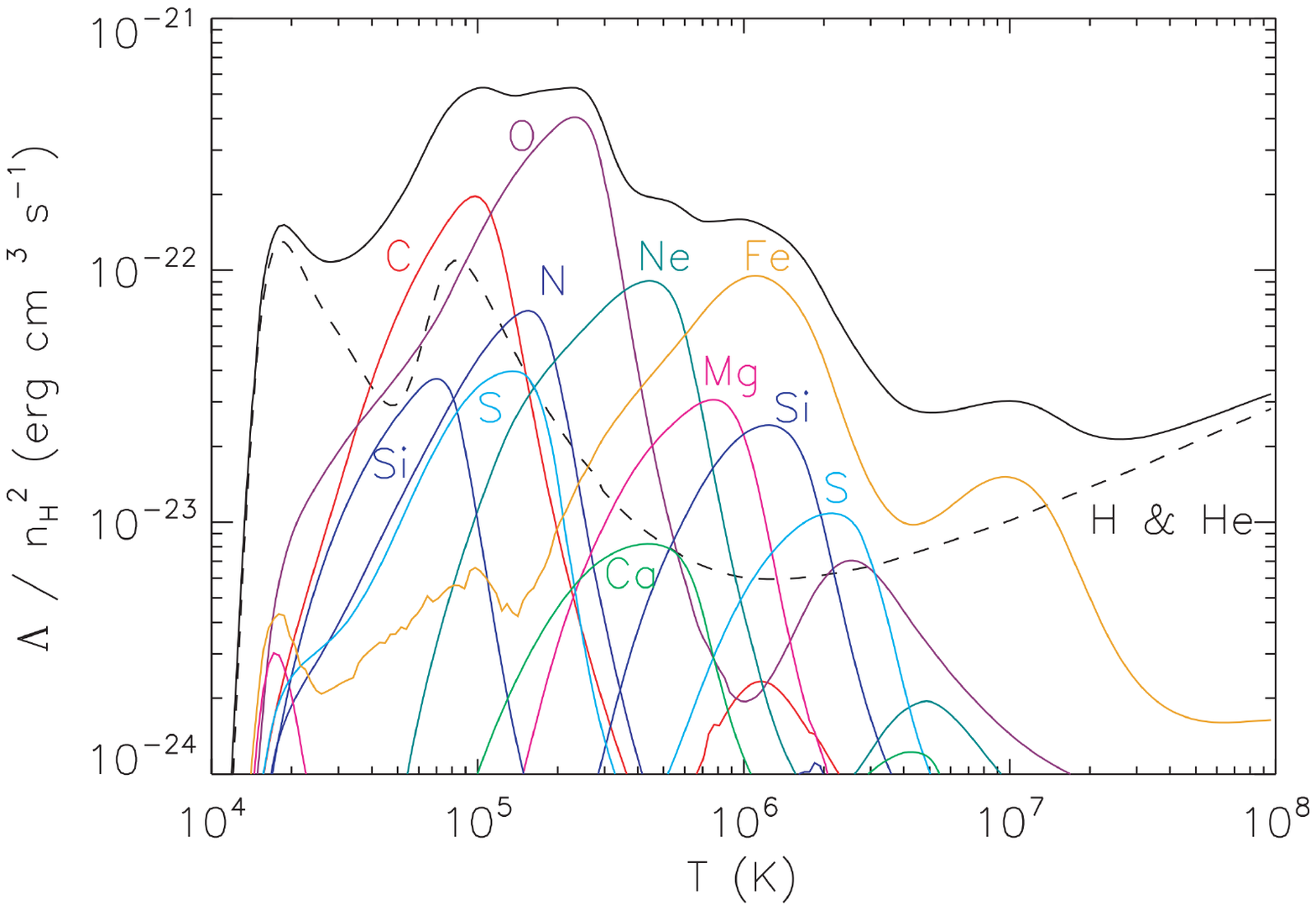}
\caption{Normalized cooling rates as a function of temperature for a collisionally-excited gas cloud. The contribution of each element to the cooling rate is highlighted with different colors, as labeled. The dashed line represents the cooling rate at primordial composition, including only hydrogen and helium. The solid black line shows the total cooling rate in gas at solar composition. {\it Credit: Wiersma R.P.C. et al., 2009, MNRAS, 393, 99. Reproduced with permission.}} 
\label{fig:cooling}       
\end{figure}

\subsubsection{The gas thermal state}

Under equilibrium conditions, the gas's thermal state can be computed by balancing the photoheating and cooling rates, typically expressed as energy per unit time and unit volume.
Considering again for simplicity the case of hydrogen, photons with energy above the ionization potential $h\nu>h\nu_i = 13.6$\,eV produce electrons with net kinetic energy $E_k = h\nu -h\nu_i$. Through collisions, these electrons contribute to the heating of the gas at a rate
\begin{equation}
    H = n_{H^0} \int_{\nu_i}^\infty  \frac{4\pi J(\nu)}{h\nu} \sigma_\nu (h\nu -h\nu_i) d\nu\,.
\end{equation}
At first glance, this equation seems to suggest that the heating rate depends on the intensity of the radiation field. However, the number density of neutrals is inversely proportional to $J(\nu)$, canceling the dependence on the intensity. The heating rate is thus primarily dependent only on the shape of the radiation field, where a harder field (i.e., with more photons emitted at high frequencies) leads to more photoheating.

A primary channel for cooling is radiation from collisionally ionized or excited gas: a collisionally excited or ionized atom that radiatively decays contributes to the cooling of the gas through the loss of photons. Considering, as a simple example, a two-level system with ground state 0 and excited state 1, the rate with which state 1 is populated is given by 
\begin{equation}
    \dv{n_1}{t} = n_e n_0 k_{01} - n_e n_1 k_{10} - n_1 A_{10}\,,
\end{equation}
where $k$ is the collisional (de)excitation coefficient, and $A$ is the radiative transition probability. 
In a steady state, 
\begin{equation}
  \frac{n_1}{n_0} = \frac{n_e k_{01}}{n_e k_{10} + A_{10}}\,.
\end{equation}
Two limits can be identified. At high densities $n_ek_{10} \gtrsim A_{10}$, and de-excitations occur primarily through collisions. Radiative recombinations become the primary de-excitation mechanism at lower densities -- like those of the CGM -- because $n_ek_{10} \lesssim A_{10}$. Thus, free electrons can transfer kinetic energy to an atom by exciting it, and the subsequent recombinations generate photons that escape from the cloud, leading to a net cooling $\Lambda \propto n_1 n_e k_B T$. Here, $k_B$ is the Boltzmann constant. 
Generalizing to the atomic structure of a mixture of elements, one can compute numerically the detailed cooling function and its variation with the temperature, finding similar results to what is shown in Figure~\ref{fig:cooling} (see \cite{Wiersma2009}). 
The cooling function is proportional to the gas metallicity because is related to the number of recombinations occurring in metals.

\begin{figure}[b]
\includegraphics[scale=.4]{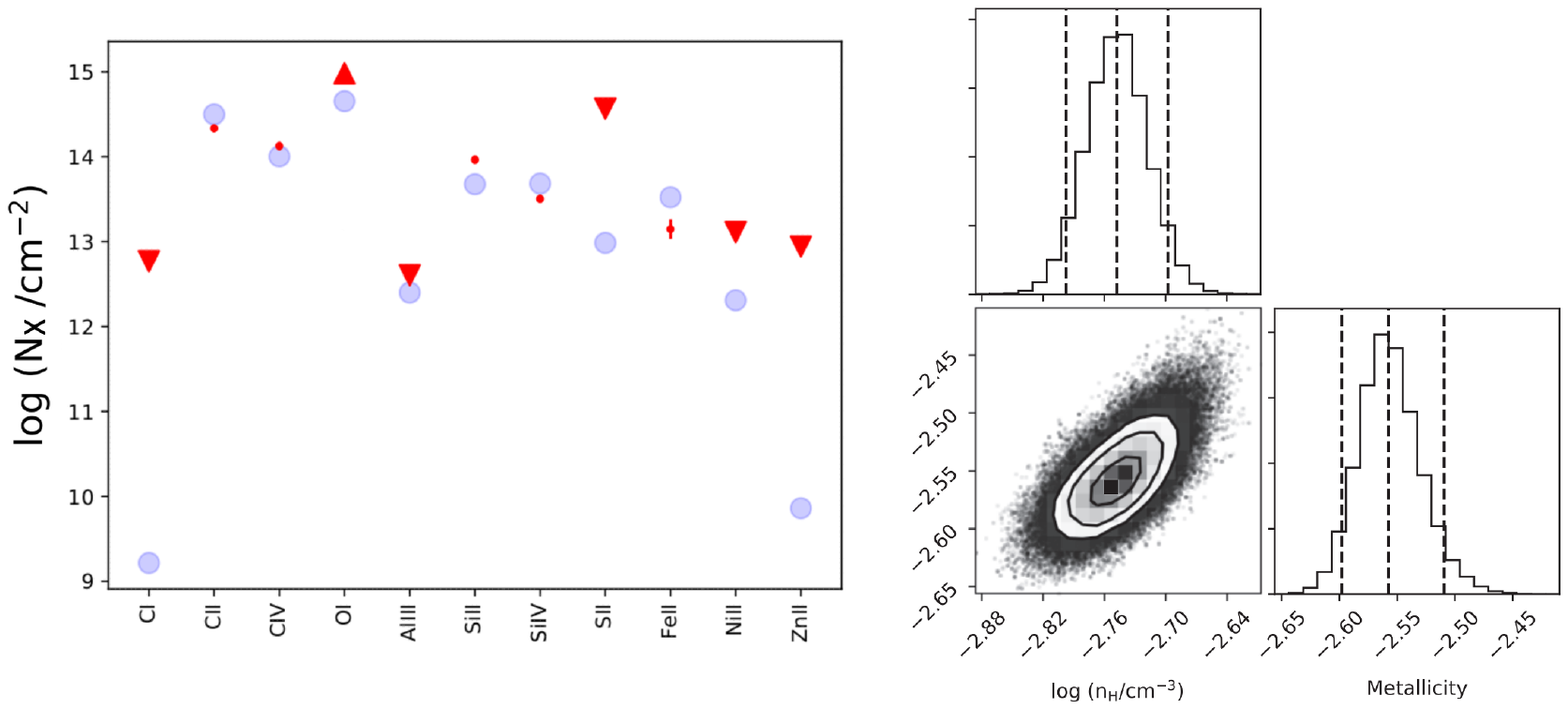}
\caption{Example of application of the ionization correction technique on a $z\approx 3$ LLS in the sample of Lofthouse et al. \cite{Lofthouse2023}. Left: Observational column densities of multiple ions (red points, with error bars), including upper and lower limits (downward and upward triangles). The predicted model column densities for the most likely values of parameters are also shown as circles. Neutral hydrogen is included in the analysis but not shown. Right: Corner plot showing the posterior probability distribution for the free parameters of the ionization model (density and metallicity in this example). When column densities across a significant range of ionization stages are well contained in data, this technique yields precise estimates of the principal underlying physical parameters of the model. {\it Adapted from: Lofthouse E.K. et al., 2023, MNRAS, 518, 305. Reproduced with permission.}}
\label{fig:ioncorr}       
\end{figure}

\subsection{Application to absorption line systems}\label{subsec:als}

Applied to a gas of known chemical composition illuminated by a specified radiation field, the formalism outlined in the previous sections allows the determination of the detailed ionization conditions of each element. This non-trivial task requires the solution of a large set of coupled equations, and it is the realm of numerical radiative transfer codes (e.g., {\sc Cloudy} \cite{cloudy}). With the ionization model in hand (i.e., once the ionization fractions $X_{j,i}$ are known), the problem of inferring the underlying physical conditions of the CGM can be tackled by comparing the observed column densities of ions in absorption line systems with a large grid of radiative transfer calculations.  
This approach has been widely applied in the literature over the past decade and has been used at a large scale since the introduction of Bayesian analysis methods \cite{Fumagalli2016,Crighton2015} that allow deriving the posterior probability distribution functions of model parameters (e.g., density, temperature, metallicity, radiation field) given a set of observed ion column densities (Fig.~\ref{fig:ioncorr}).

Some example applications are reviewed next, but more can be found in the literature. From the analysis of $\approx 150$ Lyman limit systems (LLSs, i.e., optically-thick gas clouds) at $z\approx 2-4$ \cite{Fumagalli2016}, which trace the densest part of cold filaments across and within the CGM of galaxies \cite{Fumagalli2011,Faucher2011,Lofthouse2023}, constraints on the density and metallicity of this cool phase were found to be $n_H\approx 10^{-3}-10^{-2}$\,cm$^{-3}$ and $Z/Z_\odot = 10^{-3}-10^{-1}$ with a peak at $Z/Z_\odot = 10^{-2}$ (see also \cite{Lehner2022}). Applying photoionization models to very metal-poor optically-thick absorbers at $z~\approx 3-4$,  Saccardi et al. \cite{Saccardi2023} unveiled possible signatures of first stars, raising the intriguing possibility that optically-thick and diffuse absorbers can remain relatively unpolluted due to their inability of forming subsequent generations of stars. At $z<1$, Zahedy et al. \cite{Zahedy2021} performed a detailed ionization modeling of individual \ion{H}{I} and metal components, inferring typical sizes of order of $\approx 150$\,pc. At similar redshifts, Lehner et al. \cite{Lehner2013} analyzed optically thick absorbers, inferring a bimodal metallicity distribution, with a metal-rich branch possibly linked to outflows and a metal-poor branch consistent with fresh accretion. These examples demonstrate the power of this technique to infer the chemical properties, sizes, and astrophysical nature of the gas clouds that give rise to absorption line systems.

\subsubsection{A universal self-shielding relation}

\begin{figure}[b]
\centering
\includegraphics[scale=.4]{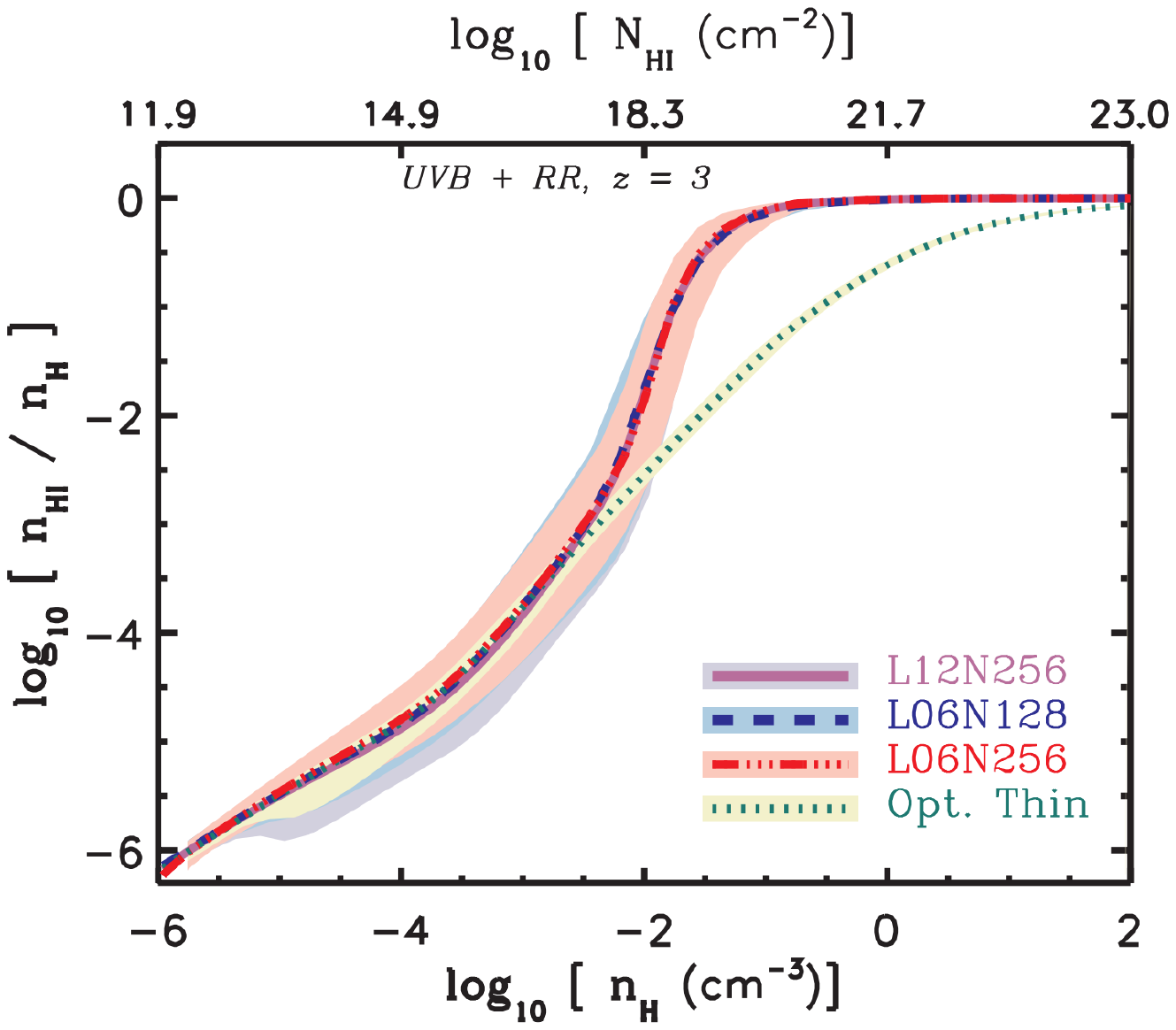}
\caption{Fraction of neutral hydrogen as a function of total hydrogen volume density, computed for a cosmological simulation at $z\approx 3$ that incorporates radiative transfer effects from the UV background and recombination radiation. Purple, red, and blue lines are for different box sizes and resolutions, highlighting that no difference is found. The green dotted line marks the expected solution for the optically thin limit. The onset of self-shielding is evident for densities $n_H\gtrsim 10^{-2}~\rm cm^{-3}$, where the result of the radiative transfer calculation starts diverging from what is expected in the optically thin limit. A sharp transition around this density is a general prediction of different simulations and radiative transfer codes. {\it Credits: Rahmati A. et al., 2013, MNRAS, 430, 2427. Reproduced with permission.}}
\label{fig:selfshielding}       
\end{figure}

The formalism outlined in this section, when applied to absorption line systems, results in a universal relation that predicts the onset of self-shielding. The self-shielding threshold is set at the boundary where the gas cloud becomes sufficiently optically thick to ionizing radiation and retains a dominant fraction of gas in the neutral phase. 
The post-processing of cosmological simulations with radiative transfer codes reveals a sharp increase in the neutral fraction of gas structures around a volume density of $n_H = 10^{-2}-10^{-1}$\,cm$^{-3}$ (Fig.~\ref{fig:selfshielding}) \cite{Fumagalli2011,Faucher2010,Rahmati2013}. 
Numerically, this threshold can be predicted by computing the gas photoionization rate as a function of density. Following the derivation by Rahmati et al. \cite{Rahmati2013},
\begin{equation}
   \frac{\Gamma_{H^{0}}}{\Gamma_{uvb}}=0.98 \left[1+\left(\frac{n_H}{n_{H,ssh}} \right)^{1.64} \right]^{-2.28} + 0.02 \left[1+\frac{n_H}{n_{H,ssh}} \right]^{-0.84}\,,
\end{equation}
where 
\begin{equation}
    n_{H,ssh} \approx 6.73 \times 10^{-3} \mathrm{cm^{-3}} \left( \frac{\sigma_\nu}{2.49\times 10^{-18} \mathrm{cm^{2}}} \right)^{-2/3} T_4^{0.17}\Gamma^{2/3}_{-12}\left(\frac{f_g}{0.17}\right)^{-1/3}
\end{equation}
is the density threshold at which the gas optical depth is $\tau = 1$ for typical values of the gas temperature ($T_4$ in units of 10$^4$\,K), photoionization rate ($\Gamma_{-12}$ in units of 10$^{-12}$\,s$^{-1}$), baryon fractions $f_g$, and photoionization cross-section $\sigma_\nu$ (for further details, see \cite{Rahmati2013}). $\Gamma_{uvb}$ is the photoionization rate from the (local) radiation background at the face of the cloud before attenuation.

As the optical depth depends on the column density and not directly on the volume density, a mapping between volume and column density needs to be specified for a universal scaling relation to hold. 
This mapping stems from the following argument \cite{schaye2001}.
Gravity-confined clouds (i.e., in hydrostatic equilibrium) intersected along a line of sight have a typical size of the order of the local Jeans length. For clouds with universal baryon fraction that are photoionized, the total column density is a function of the volume density via\footnote{The full derivation is in \cite{schaye2001}.} 
\begin{equation}
    N_H\approx 1.6\times 10^{21} \mathrm{cm^{-2}} n_H^{1/2} T_4^{1/2} \left(\frac{f_g}{0.17}\right)^{1/2}\,.
\end{equation}
Two limiting cases can also be specified. If the gas is in the optically thick limit, $N_{H^0} \approx N_H$, and the above equation can also be used to predict the neutral column density. In the optically-thin limit, instead, 
\begin{equation}
    N_{H^0}\approx 2.3\times 10^{13} \mathrm{cm^{-2}} \left(\frac{n_H}{10^{-5}\mathrm{cm^{-3}}}\right)^{3/2} T_4^{-0.26} \Gamma_{-12}^{-1}\left(\frac{f_g}{0.17}\right)^{1/2}\,.
\end{equation}
Through these equations, the limiting cases of optically thin and optically thick absorbers can be computed numerically, bracketing the case of self-shielding around the critical volume density predicted by numerical solutions of the radiative transfer equation.

\subsection{Assumptions and limits of ionization corrections}\label{subsec:ionclimit}

 Applying ionization corrections to absorption line systems is a powerful tool to constrain the underlying physical conditions of the multiphase CGM. However, it is fundamental to remember that results are subject to the model assumed for the ionization conditions and the chemical composition of the gas. Although general parameters such as density and metallicity of a population are generally stable for reasonable variation of the model assumptions, the parameters inferred for individual systems may change considerably according to the adopted model \cite{Fumagalli2016}.  
 This Section reviews some of the primary sources of uncertainties that enter the model.

\subsubsection{Radiation field}

Calculating the photoionization rate requires an assumption on the radiation field impingent on the cloud. The redshift-dependent metagalactic UV background (UVB) is the first and often the only choice in this type of modeling. However, the UVB is a model-dependent quantity prone to uncertainties, and various formulations can be found in the literature \cite{Haardt2012,FaucherGiguere2020,Khaire2019}. Indeed, multiple models present non-negligible differences at some redshifts and in some energy ranges (see section 5.3 in \cite{simcoe2011}), leading to potential differences in the inferred properties of the CGM. 
For example, the impact on the metallicity inferred from absorption lines has been discussed at length in the literature (e.g., \cite{Lehner2022,Chen2017}) and can be substantial in some cases.  To prevent this indetermination, the UVB can be parameterized, and these parameters can be inferred from the data or marginalized (see appendix B in \cite{Crighton2015}). The quality with which parameters for the UVB can be constrained depends on the available ions in the observations. However, most data do not have enough information for robust UVB measurements.

The gas clouds under analysis, especially for the CGM, are also likely to be illuminated by an additional contribution from local sources, such as stars, active galactic nuclei (AGNs) \cite{Fumagalli2016}, or radiation arising from recombination \cite{Rahmati2013}. 
Including local sources adds many free parameters, as the intensity and the shape of the local radiation field must be specified. This leads to degeneracies between parameters that are hardly constrained by the information available in most sightlines where a limited set of ions is typically available. 

\subsubsection{Abundance and dust depletion}

The abundance pattern $A_i$ is a crucial ingredient of the photoionization modeling needed to predict the underlying abundance of the elements traced by observations for a given metallicity. 
Due to the fundamental degeneracy between the observed column density and the underlying abundance of each element, the abundance pattern cannot be trivially varied as a set of parameters but must be assumed.  
Most applications consider a solar composition for the gas, mainly because of the lack of other information on the actual gas composition. Leaving aside the intrinsic uncertainty of the Sun's abundances \cite{Asplund2009,Asplund2021}, questions arise on the applicability of the solar pattern to the CGM, especially at high redshift. Notable examples are metal-poor regions, where deviations from the standard pattern are observed \cite{Welsh2022}. In these cases, one can turn, for instance, to the abundance pattern derived from metal-poor stars in the Milky Way as done in the current literature \cite{Saccardi2023}.
Recent work has also investigated the ratio of iron to $\alpha$ elements ($Fe/\alpha$) \cite{Zahedy2017}. Supersolar $Fe/\alpha$ ratios have been identified in the inner halo of quiescent galaxies, suggesting enrichment from Type Ia supernovae at a level comparable to what is observed in the solar neighborhood or the intracluster medium.
Instead, an enhancement of $\alpha$ elements has been identified around star-forming galaxies in a typical pattern of core-collapse supernovae.

In addition to the assumed abundance pattern, dust depletion can modify the underlying element-by-element distribution by suppressing, e.g., the abundance of refractory elements compared to volatile ones \cite{Jenkins2009,Konstantopoulou2022}. While the specific depletion pattern of each element need not be universal, the collective dust depletion of all elements can be parametrized with a global depletion strength factor $F_{*}$ (``the Jenkins correction''), which can be constrained or marginalized with data. Unsurprisingly, the insertion of a dust depletion in the ionization model mainly affects the inferred metallicity above $10\%$ solar \cite{Fumagalli2016}, leaving the results for more metal-poor systems unchanged. 

\subsubsection{Treatement of multiple phases}

The coexistence of cold and dense gas moving inside a hot and diffuse gas is expected within the CGM, as it naturally develops inside galactic outflows \cite{Schneider2020}, or when cold accretion interacts with galaxy halos \cite{Mandelker2019}, or from thermal instabilities in collapsing large-scale structures \cite{Mandelker2019b}. 
A complex and dynamic temperature/density structure develops within the turbulent mixing layers created by the Kelvin-Helmholtz instability \cite{Fielding2020}, making detailed small-scale hydrodynamic simulations unavoidable for capturing the complicated physics of the multiphase CGM. 

As sightlines cross the CGM gas, ions from different phases often coexist in the same absorption system. Thus, it is crucial to account for all the physical conditions in different gas phases when computing ionization corrections. 
Due to the difficulties in solving the radiative transfer equation in complex geometries, most applications that rely on 1D radiative transfer code must decompose this problem into more tractable parts, where each phase is analyzed separately by modeling uniform slabs. This approach involves making assumptions about which ions belong to a given phase, which is especially problematic for ions with intermediate ionization stages such as \ion{C}{IV}, \ion{Si}{IV}, or even \ion{O}{VI} which can arise either from a cold and highly photoionized phase at low density, or denser but hotter gas \cite{Stern2016}. The results are, therefore, subject to the assumption made on which ion should be included when modeling a specific gas phase, and (systematic) errors can occur (see appendix A in \cite{Fumagalli2016}).

To overcome this limitation, multiphase models can be constructed by combining the results of two or more gas phases, each computed independently with 1D radiative transfer codes. This approach allows us to identify models that simultaneously satisfy observations of ions over an extensive range of ionization potentials (see, e.g., figure 8 in \cite{Zahedy2021} or \cite{Qu2022}). However, combining two or more phases becomes a non-trivial task due to the increase of free parameters in the model. With increasing computational capabilities, new approaches can be explored using full hydrodynamic models. Examples of this method rely on grids of hydrodynamic simulations of multiphase gas in turbulent media exposed to radiation backgrounds \cite{Buie2020}. These boxes are used to derive ion predictions that can be compared with observations as a function of physical parameters (e.g., UVB intensity or spectrum, turbulent velocities, etc..), thus obviating the need for explicit assumption on which gas phase to treat. Although quite computationally intensive, this novel direction represents a promising frontier for obtaining ionization corrections and is the focus of current research.

\subsubsection{Non-equilibrium effects}\label{subsec:nonequilibrium}

\begin{figure}[b]
\includegraphics[scale=.4]{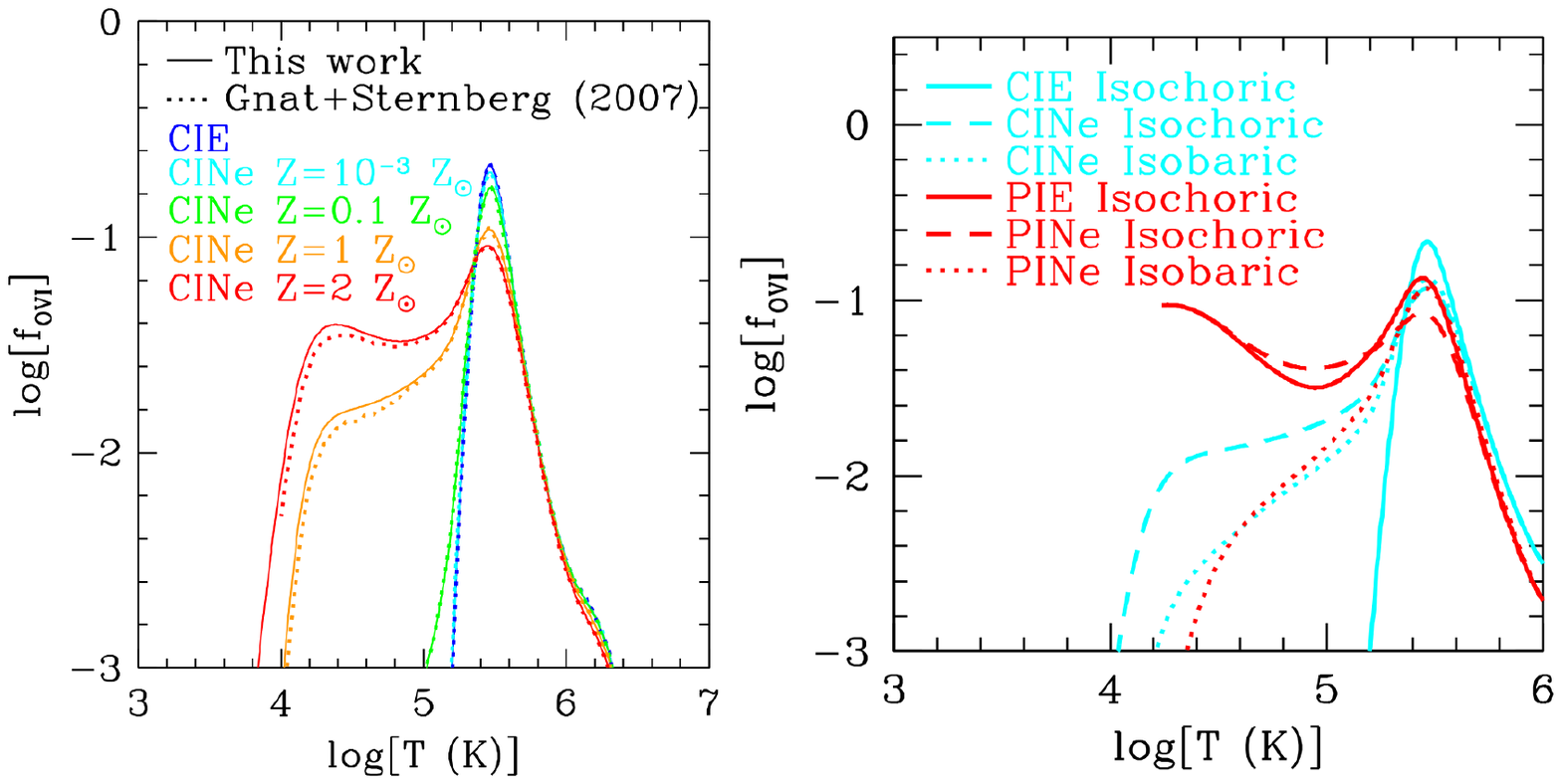}
\caption{Example of non-equilibrium effects of gas cooling rapidly from the calculations of \cite{Oppenheimer2013,Gnat2007}. Left:  Fraction of oxygen that is in \ion{O}{VI} 
in collisional ionization equilibrium (CIE, blue) and in the non-equilibrium cases (CINe) of a cloud that cools isocorically from $T=10^{7}$\,K to $T=10^{4}$\,K at different metallicities (as labeled). At progressively higher metallicities, the cooling timescale becomes shorter and shorter than the recombination timescale, leading to more pronounced non-equilibrium effects, and oxygen persists in higher ionization stages even at low temperatures. Right: Comparisons between calculations of the fraction of oxygen that is in \ion{O}{VI} for collisional ionization (CIE and CINe for the equilibrium and non-equilibrium cases) and the presence of radiation (PIE and PINe for the equilibrium and non-equilibrium cases). Two cases are considered: isochoric and isobaric transformations. Radiation effects are more prominent in isochoric transformations because, in isobaric cooling, the gas density rises, and the ionization parameter drops, leaving a solution similar to the collisional ionization case. {\it Credits: Oppenheimer B.D. and Schaye J., 2013, MNRAS, 434, 1043. Reproduced with permission.}}
\label{fig:nonequil}       
\end{figure}

The widely adopted formalism introduced in this Section assumes that gas is in equilibrium, i.e., the ionization and thermodynamic conditions of the gas can be computed iteratively till convergence is reached. In reality, if the ionization or recombination time of a species is long compared to other time scales (such as the dynamical time, the Hubble time, or the cooling time), non-equilibrium effects become relevant and must be accounted for. 
In a series of papers \cite{Oppenheimer2013,Gnat2007}, non-equilibrium effects have been studied considering gas cooling from $T\gtrsim 10^{6}$\,K to $T\approx 10^4$\,K, also accounting for effects of photoionization. 

In the case of gas cooling with no radiation, if the recombination time scale is longer than the cooling time scale, the gas will exhibit an ionization structure corresponding to the one of gas hotter than the actual temperature at any given time. The degree of lag between the temperature variation and the change in the ionization state will depend on the cooling time scale. Significant lags can occur for rapidly cooling gas, a condition that is easily encountered at high metallicity. For example, in solar or supersolar gas, a substantial fraction of \ion{O}{IV} can persist at temperatures approaching $10^{4}$\,K, which is well below the temperature for which this ion is collisionally ionized (see left panel of Figure~\ref{fig:nonequil}). Even at low metallicities, non-equilibrium effects can be generated in primordial conditions because hydrogen can experience lags in recombination. In the presence of elements such as oxygen, which are in charge exchange with hydrogen, collisionally-ionized ions can persist at temperatures below their typical ionization potential because of charge
transfer.

When including photoionization on top of collisional ionization, 
the effects of radiation are more apparent at lower densities, where the ionization parameter (the ionizing photon to gas density) is higher. This leads to a different behavior when considering isochoric or isobaric cooling, with photoionization non-equilibrium effects being more prominent in isochoric than in isobaric cases. This is because, in comparison to an isochoric transformation, gas cooling on an isobar increases its density as the temperature decreases, thus reducing the ionization parameter. Considering again the example of \ion{O}{IV} above, this implies that gas cooling isochronally in the presence of radiation remains more ionized compared to the pure collisional ionization (non)-equilibrium. In contrast, for an isobaric case, including radiation does not produce significant differences compared to the collisional ionization scenario (see the right panel of Figure~\ref{fig:nonequil}).

A further non-equilibrium effect arises when the radiation field varies on shorter timescales than the ones of recombination. An intriguing example discussed in \cite{Oppenheimer2013b} is a diffuse medium in the presence of the UVB and an AGN, which is in a bright state for a few million years. As soon as the AGN turns off, the medium settles in a new equilibrium state: the photoionization equilibrium with the UVB. Although hydrogen rapidly recombines when the AGN switches off, within the AGN proximity zone, ions recombine with various timescales, which for some ions can be longer than a few tens of millions of years (e.g., helium-like states such as \ion{C}{V}, \ion{O}{VII}, \ion{Ne}{IX} are the principal bottleneck in the recombination sequence). As a consequence, high-ionization states can appear enhanced. In contrast, low-ionization states (e.g., \ion{C}{IV}, \ion{O}{IV}, \ion{Mg}{II}) can appear suppressed compared to the equilibrium case. The non-equilibrium effects of absorbers near quasars are thus non-negligible under various AGN properties, especially for flickering sources.

\section{Observing the multiphase CGM in emission: hydrogen and helium lines}
\label{sec:Hemission}

The technique of mapping the CGM in absorption, reviewed in Section~\ref{sec:absorption}, is complemented by direct detection of the CGM in emission, a methodology rapidly expanding thanks to critical technological advancements at 8m class telescopes. 
This Section introduces the main techniques used to image the diffuse CGM (Section~\ref{sec:lowSB}). It reviews current studies that map the diffuse \lya\ emission from galaxies (Section~\ref{sec:lyagal}), quasars (Section~\ref{sec:lyaqso}), and the cosmic web (Section~\ref{sec:lyacw}). A brief discussion of the additional insight gained through observations of non-resonant transitions (mainly \ha\ and \heii, Section~\ref{sec:nonres}) is included. 

\subsection{Mapping low-surface brightness gas in emission}\label{sec:lowSB}

\begin{figure}[b]
\includegraphics[scale=.4]{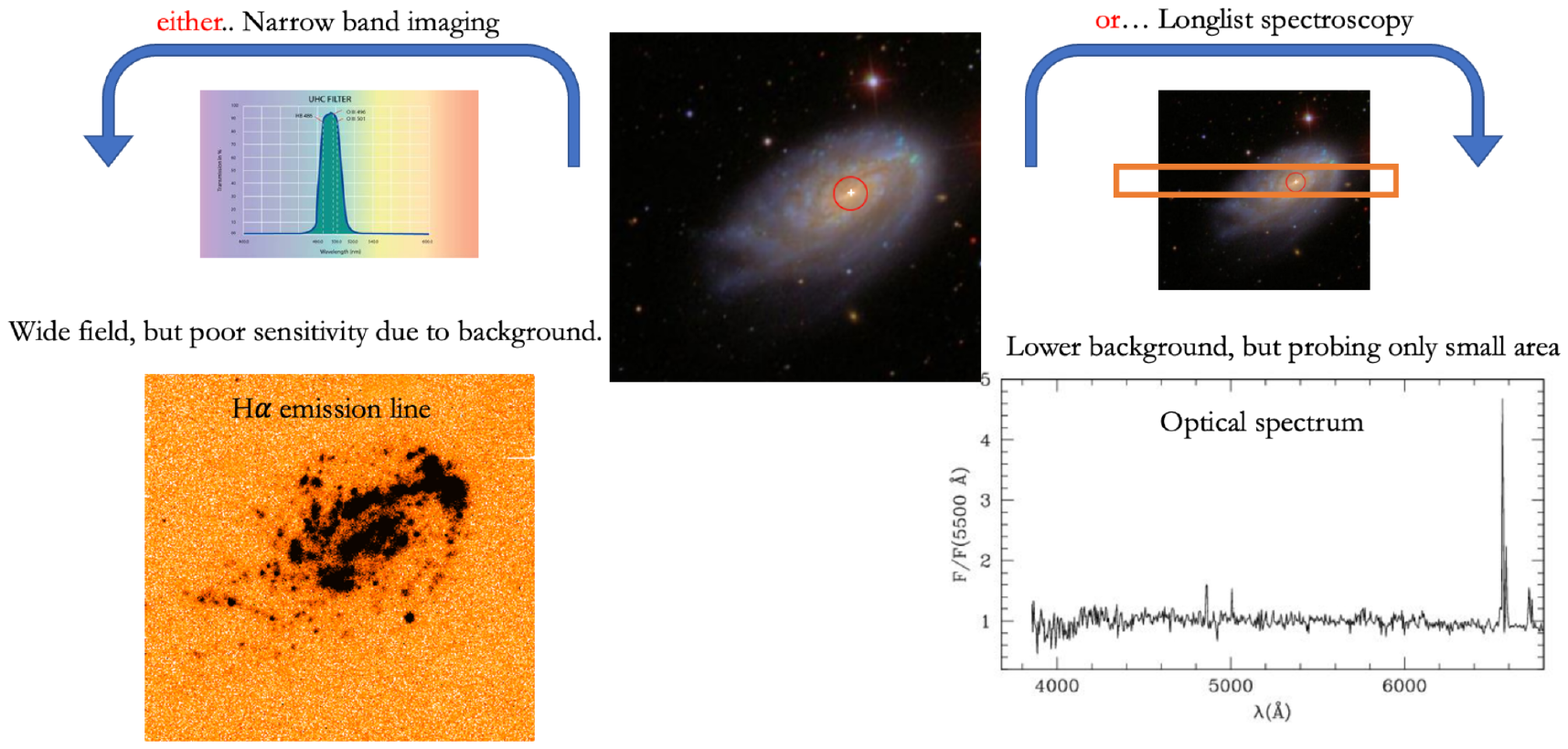}
\caption{Schematic illustration of the main techniques traditionally used to study gas in emission inside galaxies. On the left, galaxies can be imaged through narrow-band filters centered on emission lines, yielding a wide-field image. Filters and camera detectors have good nominal sensitivity, but most general-purpose filters are much wider than the targeted emission line (see an example transmission curve in the top-left panel). This causes a dilution of the signal by the background, leading to an overall low sensitivity for very faint emission lines. On the right, spectroscopy via slit (or fibers) can be used to target a portion of the galaxy. The light is dispersed, and one can analyze individual emission lines in wavelength intervals of the order of the resolution element. Therefore, the signal from the background in a resolution element is much less than in the entire passband of a filter, alleviating the problem of signal dilution typical of imaging cameras. The downside is an incomplete image reconstruction and slit/fiber losses that make it very challenging the study of extended low surface brightness emission. {\it Credit: GOLDMine website \url{http://goldmine.inaf.it/}}}
\label{fig:imgvsspec}       
\end{figure}

Imaging gas near and outside galaxies is challenging, especially at significant cosmic distances. This is because the emission expected from the CGM is a few orders of magnitude less than that of, e.g., photoionized regions inside galaxies, such as the \ion{H}{II} regions. 
The emissivity of an ionized gas that is recombining, for example, \ha, is given by 
\begin{equation}
    \epsilon_{\mha} = h\nu_{\mha} \alpha_{\mha}^{eff} n_p n_e
\end{equation}
where $h\nu_{\mha}$ is the photon energy, $\alpha_{\mha}^{eff}$ is the effective recombination coefficient leading to \ha, and $n_p$ and $n_e$
are the proton and electron density, respectively.
This equation illustrates the point: the gas inside the ISM has densities $\gtrsim 10$\,cm$^{-3}$, which is several orders of magnitude greater than the CGM typical density of $10^{-3}-10^{-1}$\,cm$^{-3}$, or the IGM at $10^{-6}-10^{-5}$\,cm$^{-3}$. Hence, the emissivity of the CGM and the IGM is many orders of magnitude lower than the typical signal observed inside galaxies. Traditionally, two main techniques have been used to map gas emission lines: imaging and spectroscopy (Figure~\ref{fig:imgvsspec}). 

The imaging approach is straightforward: it relies on narrow-band filters, i.e., with a limited passband $\Delta\lambda_f$ encompassing the emission line of interest. There are two main advantages to this technique. Firstly, CCD mosaics can be easily afforded, leading to a wide field of view ($\gtrsim 10$\,arcmin), which is more than adequate to encompass the extent of the CGM or IGM with high spatial resolution.  Secondly, imaging provides high nominal sensitivity. The optical design of an imaging camera is relatively simple, as light needs to go through the filter only before reaching the detector. The transmission efficiency of modern filters $T_{f}(\lambda)$ is usually very high ($\gtrsim 90-95$ percent), ensuring an end-to-end throughput $T_{eff}(\lambda)$ that is primarily limited by the detector quantum efficiency and the mirror reflection. As both of these are high for most leading observatories, $T_{eff}(\lambda)\approx 0.4-0.6$ (see figure 8 in \cite{Fumagalli2014} or figure 1 in \cite{Kikuta2023}).

The main disadvantage of the imaging technique is poor performance in background-limited observations. When the source signal $S(\lambda)$ is a faint emission line, the main source of noise comes from the sky background $B(\lambda)$. The signal-to-noise ratio for an exposure time $t$ in this limit becomes
\begin{equation}
    \frac{S}{N} \approx \frac{t\int S(\lambda) T_{eff} (\lambda)\Delta\lambda_f}{\sqrt{t\int B(\lambda)T_{eff} (\lambda)\Delta\lambda_f}}\,.
\end{equation}
Although narrow, filter band-passes are typically $\Delta \lambda_f\approx 100$\,\AA, much wider than the intrinsic emission line, of order $\Delta\lambda_l \approx 2-3$\,\AA. Thus, the signal-to-noise equation becomes 
\begin{equation}
    \frac{S}{N} \approx \frac{t\int S(\lambda)\Delta\lambda_l}{\sqrt{t\int B(\lambda)\Delta\lambda_f}}\,,
\end{equation}
where we have approximated $T_{eff}\approx 1$ for simplicity, and we have rewritten the numerator to leave only the non-zero part of the integral (i.e., the contribution of the line emission). In contrast, the background emits at all wavelengths, diluting the signal. As $\Delta \lambda_f \gg \Delta \lambda_l$, the signal-to-noise is significantly suppressed (that is, for every source photon, several background photons enter the filter), making the detection of faint signals very challenging with a narrow-band filter. This technique can be optimized by building bespoke narrow-band filters with $\Delta\lambda_f\approx 10$\,\AA, an approach that has led to some of the first imaging of the CGM around high-redshift radio-quiet quasars \cite{Cantalupo2014}. Nevertheless, manufacturing these filters for large telescopes is costly, accurate knowledge of the source redshifts is required, and the construction of different filters is necessary for different targets.

\begin{figure}[b]
\centering
\includegraphics[scale=.35]{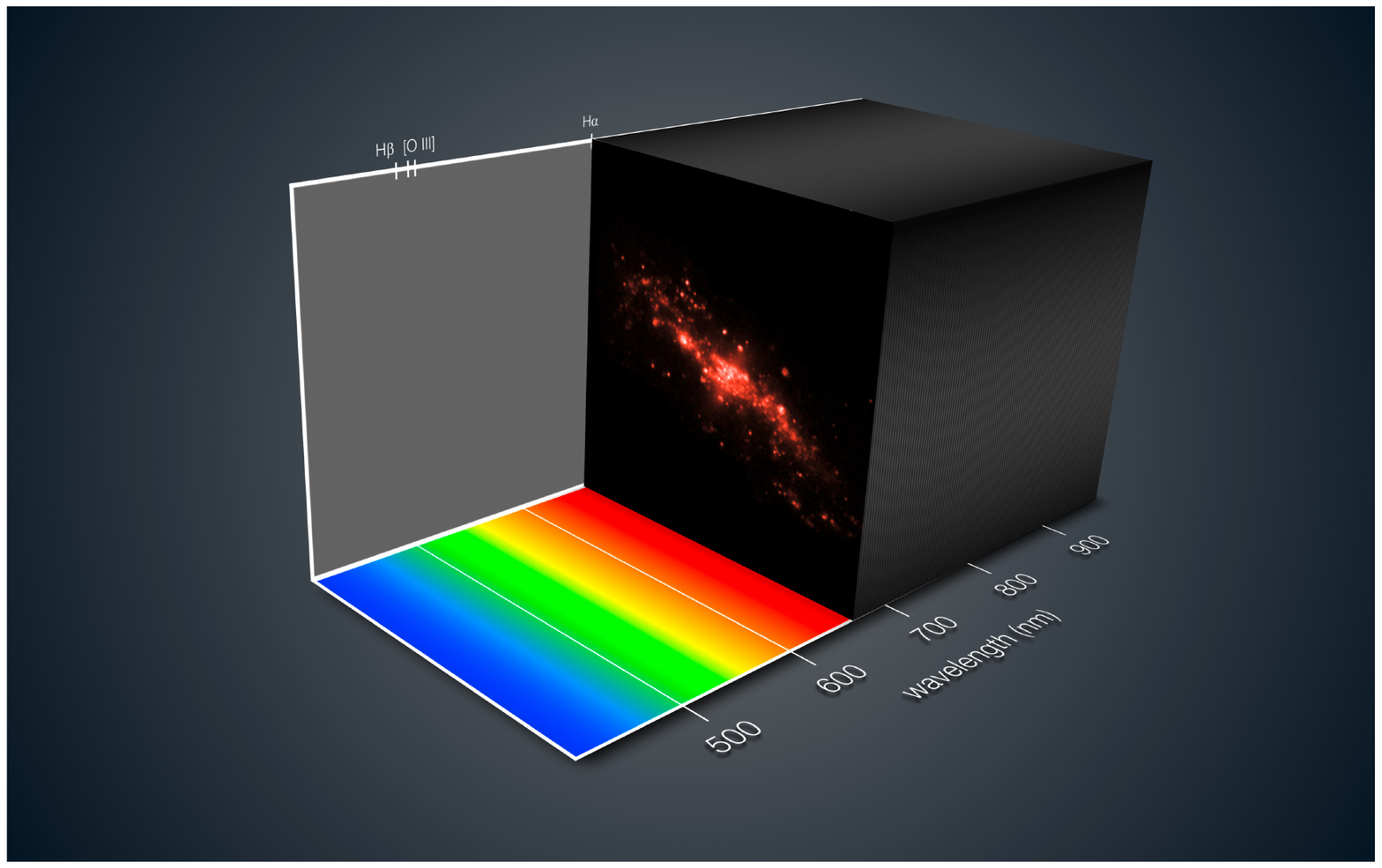}
\caption{Example of a MUSE datacube for the galaxy NGC4650A. IFSs decompose the field of view in various slices analyzed by one or more spectrographs. During data reduction, the spectra are combined in a 3D datacube, such that the image of the full field of view is reconstructed at each wavelength. This technique combines the advantages of imaging and spectroscopy, i.e., the background is optimally reduced around emission lines, and the full imaging capabilities are preserved.  {\it Credit: ESO/MUSE consortium/R. Bacon/L. Calc\c\\ada}}
\label{fig:ifs}       
\end{figure}

An alternative approach relies on spectroscopy. The essential advantage of this technique is that, by dispersing the light with sufficient resolution to resolve the line marginally, the background contribution is minimized in the resolution element (i.e., $\Delta \lambda_f \approx \Delta \lambda_l$ in the above equation), leading to an optimal $S/N$. However, there are downsides: slit or fiber-fed spectrographs map only tiny regions (of the order of a few arcseconds), leading to an incomplete view of the extended CGM and causing severe slit losses, which hamper the detectability of low surface brightness signals from extended regions. Moreover, the dispersion elements have lower throughput than filters, and spectrographs typically have more complex designs, including additional optics. In general, they have lower end-to-end throughput compared to imaging cameras.

A revolution in the study of the CGM in emission has been brought about by deploying high-throughput integral field spectrographs (IFSs) at 8m-class telescopes and, in particular, the Multi Unit Spectroscopic Explorer (MUSE) \cite{Bacon2010} at the Very Large Telescope (VLT) and the Keck Cosmic Web Imager (KCWI) at the Keck II telescope \cite{Morrissey2018}. The basic concept beyond an IFS is the ability to slice a large field of view into many small slitlets\footnote{IFSs composed of fiber bundles are also in operation, e.g., MaNGA, VIRUS, or WEAVE/LIFU.}, each dispersed by a series of spectrographs. During data reduction, the processed spectra can be repackaged in a three-dimensional (3D) datacube, where one axis is for wavelength and two axes are for the spatial extent (Figure~\ref{fig:ifs}). IFSs combine the best of the two worlds: data are acquired in spectroscopic mode, thus with an optimal $S/N$ over a wide field of view at high resolution so that images of the CGM can be reconstructed without slit or fiber loss. The throughputs are not as high as in imaging, e.g., between $\approx 0.25$ and $\approx 0.35$ for MUSE. The advantages offered by these instruments outweigh this limit, making IFSs the current instrument of choice for studying the CGM in emission.

\subsection{Extended \lya\ emission in galaxies}\label{sec:lyagal}

 \subsubsection{The size of \lya\ halos}

Early studies of the CGM in emission around galaxies employed narrow-band imaging techniques, including stacking analysis, to reach sufficient sensitivities to uncover the extended low-surface-brightness emission from the halo. At $\approx 2.6$, Steidel et al. \cite{Steidel2011} investigated a sample of $\approx 100$ Lyman break galaxies (LBGs), which were representative of the entire LBG population and selected without any known bias regarding \lya\ emission.   In deep continuum-subtracted \lya\ stacks, they detected \lya\ emission up to distances of $\approx 80$\,kpc with surface brightness profiles characterized by an exponential scale-length $\approx 5-10$ times larger than the UV continuum one. 

When dividing the sample accordingly to the \lya\ properties of the inner part of the galaxies (i.e., absorption and emission with varying equivalent widths), the radial profile of the halo does not change shape but correlates in normalization with the emission properties of the inner parts (see figure 9 in \cite{Steidel2011}). Brighter emitters have brighter surface brightness profiles in their halos. 
This behavior also applies to the most extreme examples of \lya\ blobs (LABs), which have similar surface brightness profiles, just rescaled to a higher normalization. 
Finally, despite their low surface brightness, the extended halos contribute substantially to the total integrated \lya\ luminosity ($\gtrsim 5$ times their ISM emission, or more for the case of weak-to-no \lya\ emission inside the galaxy). 
This result implies that virtually all galaxies are surrounded by an extended CGM that emits in \lya\ regardless of their ISM emission properties and that the characterization in classes of emission (e.g., LABs or LAEs, i.e., \lya\ emitters) depends on the surface brightness limit of the observations and on the radius within which the \lya\ fluxes are computed. 

With the advent of IFSs and, in particular, MUSE, observations of \lya\ around galaxies have become easier up to $\approx 10^{-19}$\,\sbunit, both in stacks and individual sources \cite{Wisotzki2016,Leclercq2017}, including systems that are generally continuum-faint ($m\gtrsim 27$) and detected exclusively through their \lya\ emission. 
Analyzing galaxies in this lower mass scale (with stellar masses of the order of $\gtrsim 10^{9}$\,M$_\odot$) reveals similar findings to those around LBGs. \lya\ halos appear ubiquitous with detection rates of $\approx 80$ percent. Adequate sensitivity appears to be the only limit to the detectability of extended halos. As in LBGs, the halos surrounding these LAEs represent a significant source of \lya\ photons, with fractions ranging from more than half to $\approx 90$ percent of the total \lya\ luminosity.   
Again, similarly to LBGs, the halos of LAEs are characterized by an exponential profile with a scale length of $\approx 4.5$\,kpc, an order of magnitude larger than the UV continuum.  
These scales imply that the \lya\ halos extend to over half the virial radius at this mass scale, and thus a significant fraction of the cool inner CGM is traced by this technique.  
The halo size depends weakly on UV or \lya\ flux properties and correlates more with the size of the stellar UV disk.
Individual detections in a sample of UV-selected galaxies (i.e., not identified based on \lya\ properties) confirm these trends \cite{Kusakabe2022}, making them general features of the galaxy population in this mass scale. 

IFS data also provide rich information on the kinematics of the \lya\ line \cite{Verhamme2018}, a diagnostic that can be combined with radiative transfer calculations to learn about the underlying physics of the CGM and the powering mechanisms. The literature on the subject is extensive but not discussed in this contribution, as is reviewed in another chapter of this book.

\begin{figure}[b]
\centering
\includegraphics[scale=.4]{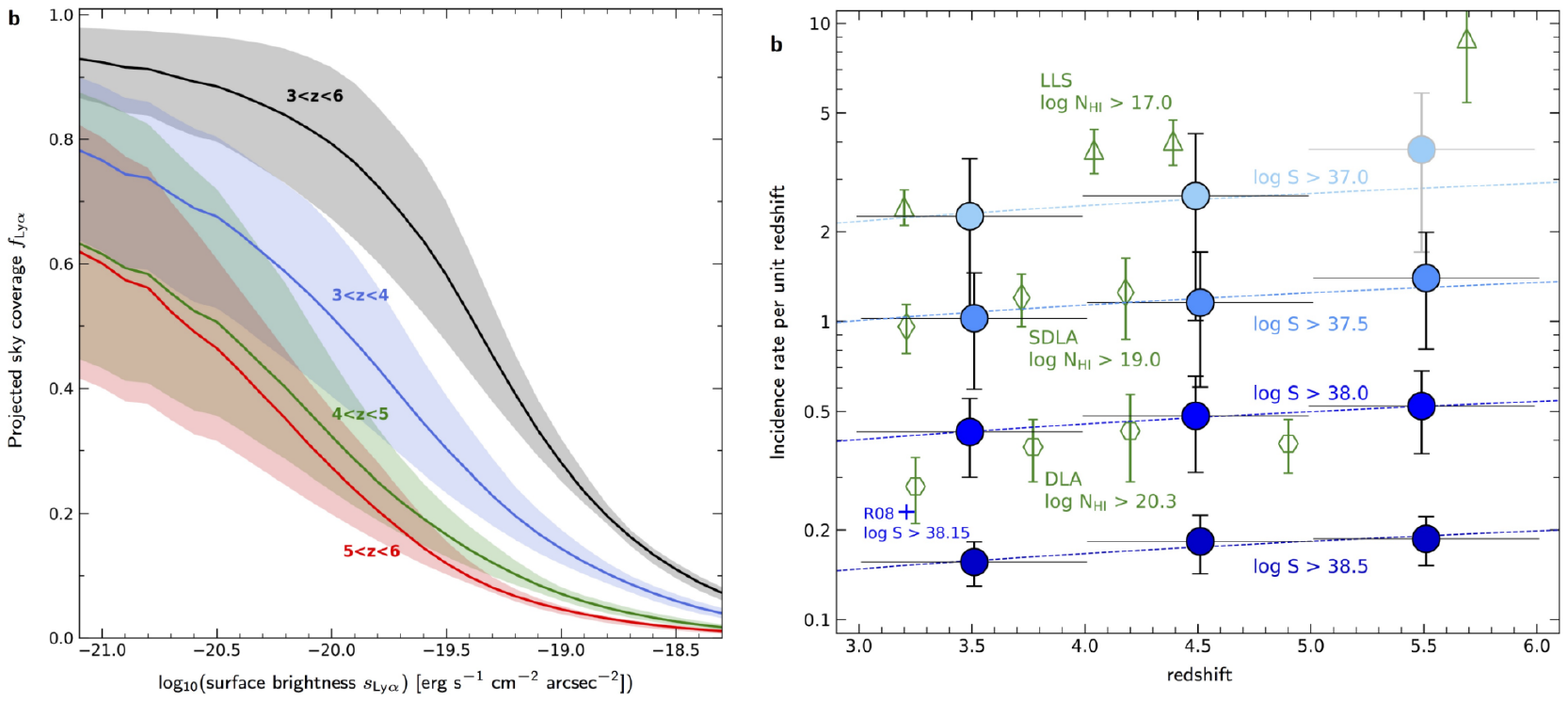}
\caption{
Left: the projected sky coverage of \lya\ halos as a function of redshift (as labeled) and limit surface brightness. The cross-section of individual LAEs increases for increasing sensitivity, yielding more significant fractions of the sky being covered by \lya\ emission when projected along a given redshift interval. 
Right: The incidence rate per unit redshift of \lya\ emitting halos as a function of redshift and surface brightness threshold (colored circles, as labeled). For comparison, the incidence rate per unit redshift of classes of strong absorption line systems is also shown with green empty symbols. 
Through this comparison, the cross-section of halos at $\Sigma \gtrsim 10^{38}$\,erg~s$^{-1}$~kpc$^{-2}$ appears sufficient to reproduce all DLAs, while LLSs could be reproduced at $\Sigma \gtrsim 10^{37}$\,erg~s$^{-1}$~kpc$^{-2}$. However, the distribution of optically thick hydrogen in the halo is not uniform since it has a covering factor of $f_c\approx 0.25$. This implies a non-unique correspondence between emitting halos and gas in absorption. 
{\it Credits: Wisotzki L. et al., 2018, arXiv:1810.00843. Reproduced with permission.}}
\label{fig:coverfraclya}       
\end{figure}

 \subsubsection{The sky covering factor of \lya\ halos}

The presence of a significant extent of \lya\ emitting gas near low-mass systems that are distributed with a high number density, of the order of $n_{gal}\approx 10^{-2}$\,cMpc$^{-1}$, has fascinating implications for estimating the fraction of sky covered by \lya\ emission and optically-thick gas \cite{Kusakabe2022,Wisotzki2018}.
For a population of LAEs with surface brightness ($\Sigma$) dependent on the cross section $C = \pi r^2(\Sigma)$, \lya\ halos projected in a redshift range $3\lesssim z \lesssim 6$ cover $\gtrsim 80$ percent of the sky for $\Sigma \lesssim 10^{-20}$\,\sbunit\ (see Figure~\ref{fig:coverfraclya}). The totality of the sky is thus glowing in \lya. 
The sky coverage can be recast in the incidence rate $\dv{n}{z}$, expressing the probability of intersecting a region covered by \lya\ across a given redshift path. 
For a given population with number density $n_{gal}(z)$ observed in a volume $V_{p}$ defined by an area $A_{p}$ and a redshift window $\Delta z_{p}$, the incidence is expressed by 
\begin{equation}
    \dv{n}{z} (\Sigma) = n_{gal} \pi r^2(\Sigma) \frac{V_p}{A_p \Delta z_p}\,.
\end{equation}
The number density of objects is a priori not fully known and must be derived, e.g., by integrating the luminosity function $\phi(L)$ between a chosen minimum and maximum. 
When allowing for a flux-dependent size of the emitting halos, the incidence becomes
\begin{equation}\label{eq:lyahinc}
    \dv{n}{z} (\Sigma) = \frac{V_p}{A_p \Delta z_p} \int_{L_1}^{L_2} \pi r^2(\Sigma,L) \phi(L) \dd L\,.
\end{equation}
Although the calculation is relatively insensitive to the choice of $L_2$ for an exponentially declining luminosity function, extrapolations below the sensitivity limit through the parameter $L_1$ can dramatically affect the final result. 
For LAEs at $3\lesssim z \lesssim 6$ and $L_1 = 10^{41}$\,erg~s$^{-1}$,  $\dv{n}{z}\approx 1$ is reached for $\Sigma \approx 10^{37.5}$\,erg~s$^{-1}$~kpc$^{-2}$, which is the typical surface brightness encountered at $40$\,kpc from the LAEs (Figure~\ref{fig:coverfraclya}; see \cite{Kusakabe2022} for similar estimates starting from UV selected galaxies).

The incidence of \lya\ halos can be compared with the incidence of optically thick absorption lines, including LLSs with $N_{HI}\gtrsim 10^{17}$\,cm$^{-2}$ and damped \lya\ systems (DLAs) with $N_{HI}\ge 10^{20.3}$\,cm$^{-2}$. Figure~\ref{fig:coverfraclya} shows that \lya\ emitting halos around LAEs can comfortably account for DLAs for a surface brightness limit of $\Sigma \gtrsim 10^{38}$\,erg~s$^{-1}$~kpc$^{-2}$, while $\Sigma \gtrsim 10^{37}$\,erg~s$^{-1}$~kpc$^{-2}$ is needed to reproduce the incidence of LLSs. At face value, this result implies that LAE halos are responsible for the observed population of DLAs and LLSs in quasar surveys within radii of $\approx 30$\,kpc and $\gtrsim 60$\,kpc, respectively. 
Further insight into this association can be gained by performing the opposite experiment, i.e., searching LAEs around the location of known LLSs. In the MUSE Analysis of Gas around Galaxies (MAGG), Lofthouse et al. \cite{Lofthouse2023} find that LAEs are often encountered at distances of $>100$\,kpc from LLSs, including in filamentary structures connecting galaxies. Thus, not all the \lya\ emitting gas gives rise to optically thick absorbers, but only a fraction of the \lya\ halos to distances $\lesssim 100$\,kpc appears to be covered by LLSs. Lofthouse et al. estimate this fraction in the order of $f_c\approx 0.25$.

\subsubsection{Emission mechanisms}

Although the demographics and distribution of \lya\ emitting gas are now well characterized in sufficiently large samples, the mechanisms powering the observed extended halo emission are less constrained \cite{Momose2016}. 

Most analyses \cite{Steidel2011,Wisotzki2016,Leclercq2017} agree that 
a viable mechanism is the scattering of \lya\ photons produced inside the galaxy's \ion{H}{II} regions. Photons then diffuse outward in the CGM, through scattering on the surface of neutral gas clouds. The extend and covering factor of the cool and neutral gas in the halo are, under this scenario, the main parameters that regulate the observed emission.  Modeling this process is challenging, as detailed radiative transfer calculations in complex (and often unknown) geometries are required. From an energetic point of view, scattering is highly plausible, as the observed \lya\ fluxes are well below the theoretical limit arising from hydrogen recombination for a given observed UV flux unless dust absorbs a large amount of \lya\ photons that do not reach the observer. 

As the data do not exclusively indicate scattering as the unique process at play, other powering mechanisms cannot be excluded \cite{Leclercq2017}. 
As the gas accretes onto dark matter halos, gravitational energy is converted into thermal energy, giving rise to collisionally excited \lya\ emission (the so-called cooling radiation) in the CGM. 
As gas needs to be accreted to sustain the activities of galaxies, some level of cooling radiation must be present. However, there are reasons to believe this is not the dominant contribution observed in the halos of LAEs. Firstly, models predict luminosities from cooling radiation at the typical mass scale of these galaxies below the observed ones, making collisionally-excited \lya\ a sub-dominant powering mechanism. Moreover, cooling radiation should manifest itself with blue peaks in spectra, a feature that is observed only in a small fraction of the sample.   

Another powering mechanism is the production of \lya\ photons in the halo through recombination, a process that is often defined as \lya\ fluorescence. The main difference from scattering is the origin of the photons, which is {\it in-situ} rather than arising from the central UV bright regions of the galaxy. Fluorescence of the UVB, by which photons from the metagalactic background photoionize the halo gas directly, can be discarded as the primary mechanism since the observed surface brightness can be up to a factor $\approx 10$ higher than expected from current UVB models. \lya\ fluorescence is nevertheless a plausible contributor to the emission in the inner halo, where modest fractions ($\lesssim 5$ percent)  of ionizing photons escaping from the ISM are enough to power low-surface brightness halos.

\begin{figure}[b]
\includegraphics[scale=.28]{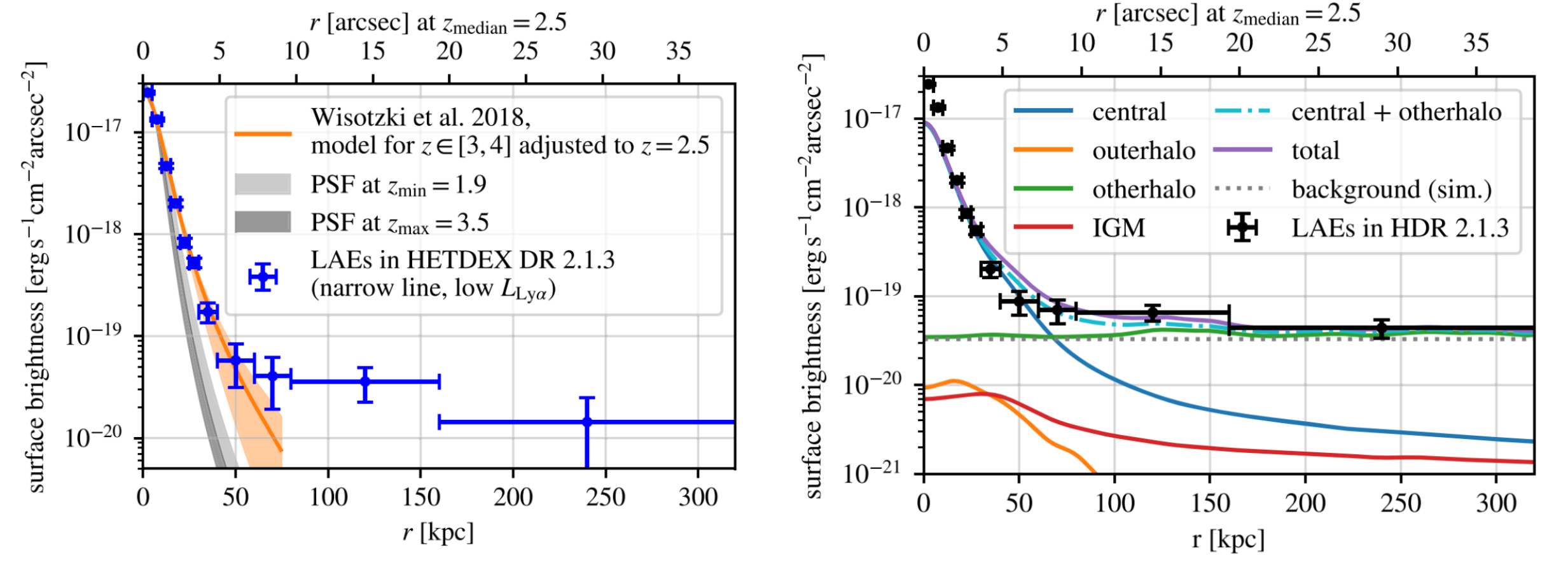}
\caption{Left: Comparison between the \lya\ surface brightness profiles around LAEs from the MUSE observations (orange line) and the HETDEX survey (blue points). Both profiles are more extended than typical point sources (grey bands) and are consistent in the inner parts of the halos on scales $\gtrsim 50-60$\,kpc. HETDEX observations probe larger distances from LAEs, revealing a flattening in the profile. Right: The observed profile from HETDEX (black points), compared to predictions of numerical models \cite{Byrohl2021}. The central galaxy (blue line) with its halo (orange) contributes enough photons only for the inner part of the profile. The contribution of additional halos clustered in the vicinity (green) explains the observed flattening at a large distance, as the IGM contribution (red) appears insufficient. {\it Credits: Lujan Niemeyer M. et al., 2022, ApJ, 929, 90. Reproduced with permission.}}
\label{fig:poweringmec}       
\end{figure}

A third mechanism that can be invoked to power halo emission is the superposition of \lya\ photons from UV-faint (hence undetected) satellites surrounding the central galaxy. Unless the number of these satellites is exceptionally high, the signal is expected to be clumpy and asymmetric, a characteristic not observed in the data for small separations from the central galaxy. Moreover, although the satellites are intrinsically UV faint if the halo is powered by these unresolved sources in stacked profiles, one should expect more similarities between the UV and \lya\ emission than currently observed. 
Although a superposition of multiple galaxies does not seem suitable to explain emission on scales of $\lesssim 50-60$\,kpc, this scenario becomes relevant for distances $\gtrsim 50$\,kpc \cite{HerreroAlonso2023}. For example, the large-scale Hobby-Eberly Telescope Dark Energy Experiment (HETDEX) Survey \cite{Gebhardt2021} has provided a mapping of the radial profile in bright LAEs to distances of $\approx 250$\,kpc \cite{Niemeyer2022}. By stacking $\approx 1000$ LAEs at $1.9 \le z \le 3.5$ with luminosity $\approx 10^{42.5}$\,erg~s$^{-1}$, Lujan Niemeyer et al. \cite{Niemeyer2022} find an exponentially decreasing profile in agreement with the MUSE observations for $\lesssim 100$\,kpc. At more considerable distances, the profile flattens considerably. Radiative transfer simulations \cite{Byrohl2021} suggest that resonant scattering can power the inner halo, but emission from other halos becomes critical to powering the halo at $r \geq 100$\,kpc (Figure~\ref{fig:poweringmec}).

\subsection{Extended \lya\ emission in quasars}\label{sec:lyaqso}

\subsubsection{Detections and properties of \lya\ nebulae}

Although typical LAE and most LBG reside in halos $\lesssim 10^{12}$\,M$_\odot$ at $z\approx 3$, the CGM of more massive halos with $\approx 10^{12}$\,M$_\odot$ or greater can be studied by targeting quasars\footnote{An in-depth review of the physics of extended nebulae in quasars is in \cite{Cantalupo2017}.}.
Extended emission from radio-loud quasars has been known since the 1990s \cite{Heckman1991}, similarly to the detection of extended emission in radio galaxies \cite{McCarthy1987}.  
From the start, these nebulae were interpreted as arising from the CGM photoionized by quasars. Still, anisotropies in the emission in the direction of the radio jet indicate a more complex physics peculiar to these radio-loud systems. 
The link between bright emission and outflows in radio-loud systems is indeed confirmed by MUSE observations \cite{Kolwa2019}, where significant velocity shifts are observed in the emission spectra, reaching up to $\approx 1000$\,km~s$^{-1}$. These signatures are interpreted as perturbations in the gas dynamics along the radio axis, a signature of an outflow via jets. $Ly\alpha$ absorption profiles in radio-loud quasars often show multiple absorption components, usually interpreted as material entrained in outflows \cite{Swinbank2015}. 

Observations of the CGM in less extreme environments traced by radio-quiet quasars remained elusive. The first systematic detections of the extended CGM in emission in high-redshift radio-quiet quasars were obtained using the narrowband imaging technique with custom-made filters targeting some of the brightest $z\gtrsim 2$ quasars \cite{Cantalupo2014,Hennawi2015}. Nebulae extending beyond $\approx 200$\,kpc and up to $500$\,kpc, beyond the typical scale of the virial radius of quasar hosts, were identified with surface brightness $\gtrsim 10^{-17}$\,\sbunit. Multiple active sources were also found in these nebulae, further highlighting these systems' remarkable and rare nature. These enormous \lya\ nebulae (ELAN) are now understood to be the tip of the iceberg, $\lesssim 1$ percent of a more ubiquitous population of nebulae near quasars.

In fact, IFS follow-up observations of large samples of radio-quiet quasars at $z\approx 3-6$ have revealed a $\approx 100$ percent detection rate \cite{Borisova2016,ArrigoniBattaia2019,Fossati2021,Farina2019}
for nebulae with typical sizes $\lesssim 100$\,kpc when measured at a reference intrinsic surface brightness \cite{ArrigoniBattaia2023}.
These observations trace a substantial cool ($T\approx 10^4$\,K) and likely dense gas phase in quasar hosts.
Nebulae around quasars are generally more extended ($\approx 3$ times) than the \lya\ halos in other galaxy populations. This property is attributed to the more extended CGM in these massive halos and brighter central ionizing sources. The brightest nebulae are also the most outstretched, with size and luminosity linearly correlated in log-log space \cite{ArrigoniBattaia2023}. Unlike ELAN, typical nebulae also appear more symmetrical and round. Spectrally, the extended \lya\ emission is characterized by velocity dispersion $\sigma \lesssim 400$\,km~s$^{-1}$, which is comparable to the expected motion in the potential of halos with masses $10^{12.5}$\,M$_\odot$.

\subsubsection{The redshift evolution of \lya\ profiles}

\begin{figure}[b]
\centering
\includegraphics[scale=.26]{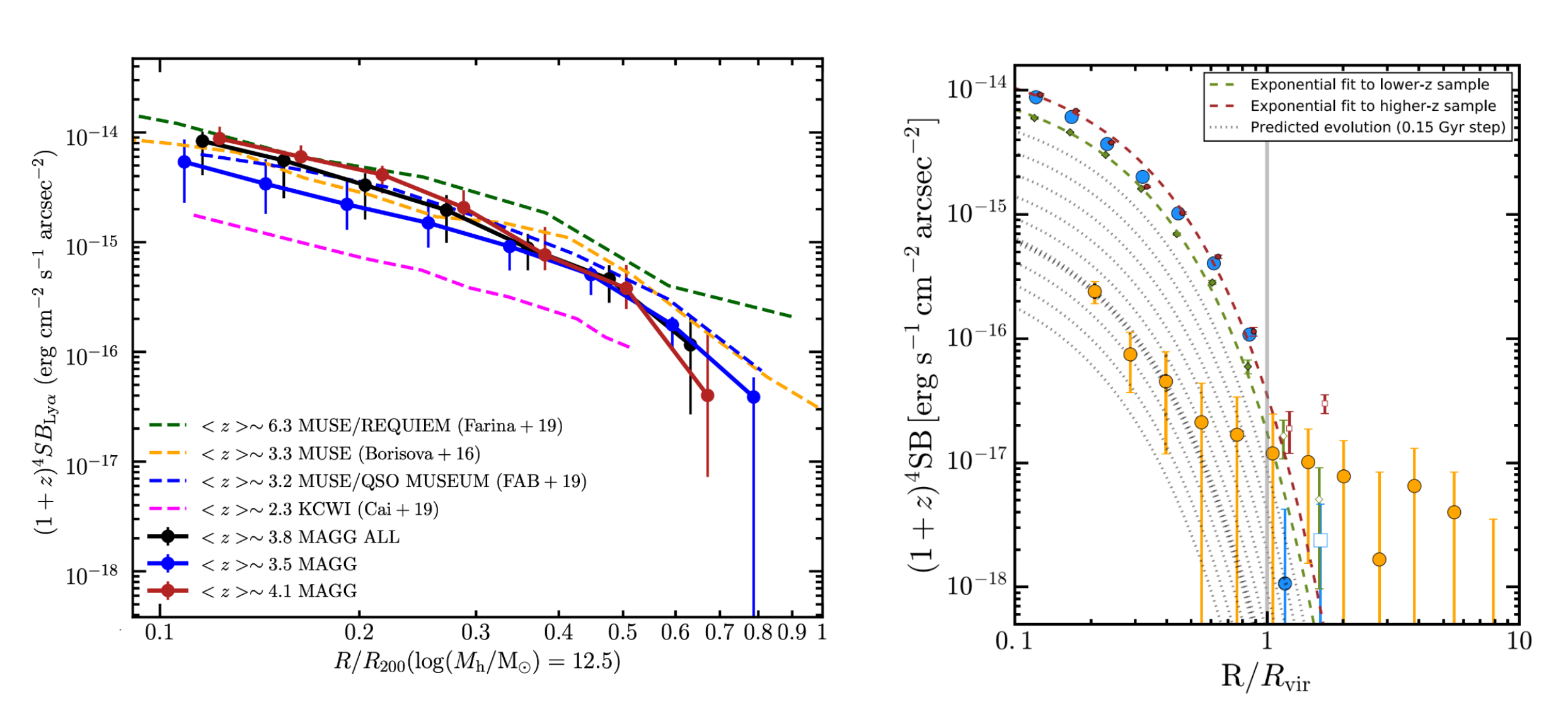}
\caption{Left: the cosmologically-dimmed corrected \lya\ surface brightness profile as a function of radius scaled for the virial radius of a $10^{12.5}$\,M$_\odot$ halo. A comparison of various observations from $z\approx 2.3$ to $z\approx 6$ (as labeled) is presented. The profiles share a similar shape and present a modest redshift evolution except for the $z\approx 2.3$ measurement. Right: Comparison of the $z\approx 3$ (blue) and $z\approx 2$ (yellow) profiles with simple evolution models (dotted lines, in steps of 0.15\,Gyr) based on the observed evolution around $z\approx 3$. A simple mode in which the gas profiles evolve with decreasing mean density as the Universe expands is insufficient to explain the rapid evolution between $z\approx 3$ and $z\approx 2$. A change in the physical properties of the CGM, for instance, induced by a transition between a hot and a cool halo, must be invoked. {\it Credits: Fossati M. et al., 2021, MNRAS, 503, 3044 (left) and Arrigoni Battaia F. et al., 2019, MNRAS, 482, 3162 (right). Reproduced with permission.}}
\label{fig:qsoprofile}       
\end{figure}

Stacks of surface brightness profiles obtained from quasar nebulae at $z\approx 3-4$ can be modeled by exponential functions with scale-length $\approx 15.5$\,kpc \cite{ArrigoniBattaia2019}. Once corrected for the redshift-dependent cosmic dimming ($\propto (1+z)^4$) and rescaled to the expected size of the viral radius ($R_{vir}$), stacks of profiles in different redshift bins at $z\gtrsim 3$ overlap, including the highest-redshift nebulae observed at $z\approx 6$. The outliers of this trend are the nebulae at $z\approx 2$ (\cite{Cai2019}; see Figure~\ref{fig:qsoprofile}, left). To interpret this finding, Arrigoni Battaia et al. \cite{ArrigoniBattaia2019} developed a simple scaling relation based on the formalism by Hennawi \& Prochaska \cite{Hennawi2013} for the evolution of the \lya\ profile (see also \cite{Cai2019}).
Starting from the ansatz that the gas is highly ionized and hence optically thin, the intrinsic surface brightness scales as 
\begin{equation}
    \Sigma \propto n_H N_H    
\end{equation}
where 
\begin{equation}
   N_H \propto n_H R_{vir}\,.   
\end{equation}
As the Universe expands, the mean density decreases with time as 
\begin{equation}
   n_H \propto (1+z)^3\,;   
\end{equation}
likewise, for a population of quasars with comparable halo masses at the redshifts probed, the cosmic expansion causes an increase in the virial radius 
\begin{equation}
   R_{vir} \propto (1+z)^{-1}\,.   
\end{equation}
Given this, the intrinsic surface brightness is expected to evolve as
\begin{equation}
    \Sigma \propto n_H N_H = n_H^2 R_{vir} \propto (1+z)^5\;.    
\end{equation}
The expected evolution between $z\approx 3.1$ and $z\approx 2.3$ is only a factor $\approx 3$, which falls short of the observed variation.  A mere drop in the cosmic density with time is insufficient to explain the observed trend, but a more rapid time evolution due to a change in the physical conditions of the CGM gas is required. 
Intriguingly, between $z\approx 3$ and $z\approx 2$, it is expected that the halo mass for which stable shocks can be sustained without infalling cold gas vary from $\approx 10^{13}$\,M$_\odot$ to $\approx 10^{12}$\,M$_\odot$  \cite{dekel2006}, thus crossing the characteristic halo mass of quasars at $\approx 10^{12.5}$\,M$_\odot$. Therefore, the CGM of higher redshift quasars is expected to be more dominated by cool gas, while lower redshift quasars have their halos preferentially filled by hot gas. This change in the CGM structure is likely to be reflected in the rapid variation observed for the \lya\ surface brightness.

\subsubsection{The powering mechanisms}

Because of the presence of a bright and highly ionizing central source, it is conceivable that recombination radiation is a primary powering mechanism for the extended quasars' nebulae. 
In fact, for typical photoionization rates of the order of $\Gamma_{H^0}\approx 10^{-7}-10^{-6}$\,s$^{-1}$ (several orders of magnitude above the UVB value), the neutral hydrogen fraction in a medium with density $\approx 1$\,cm$^{-3}$ is of the order of $x_{H^0}\approx \alpha/\Gamma_{H^0} \approx 10^{-7}$, where $\alpha$ is the recombination coefficient. Cooling radiation (collisionally-excited emission) can thus be considered sub-dominant when the gas is fully ionized. This is because the emissivity of \lya\ from collisional excitation scales as the product $n_{H^0}n_e$, while recombination radiation scales with the $n_{H^+}n_e$ (see section 5.3 in \cite{Pezzulli2019}).
However, the role of scattering from the central broad-line region is more debated. Due to the bright \lya\ emission from the quasar's broad line region, it is reasonable to ask whether fluorescence dominates over scattering or whether scattering is a key (perhaps even dominant) emission source. 

Some insight into the emission mechanisms can be gained by examining possible correlations between the emission properties of the \lya\ nebulae and the quasar properties, including their UV luminosity and the \lya\ emission from the broad line region.
The analysis of these trends in bright quasar samples \cite{ArrigoniBattaia2019} does not reveal strong correlations between the nebula surface brightness and the quasar UV magnitude or peak \lya\ luminosity. 
The lack of clear correlations with the quasar magnitude can constrain whether the medium is optically thin or thick. In the latter case, some degree of correlation between the nebula and the UV flux of the quasar is expected, as optically thick gas around the quasar would behave as an \ion{H}{II} region where ionization and recombinations are directly proportional. In the case of fully ionized gas, the \lya\ emission properties do not correlate directly with the ionizing photon flux, as they also depend on the properties of the ionized material, such as its density distribution. 

When extending this analysis to quasars fainter by $\approx 4-5$ magnitudes \cite{Mackenzie2021}, correlations emerge between the nebula \lya\ and the quasar properties. In particular, the increased \lya\ luminosity of the nebula as a function of \lya\ peak emission in the quasar could hint at scattering as a powering mechanism: the brighter the broad line region, the brighter the nebula is as \lya\ photons are scatter outward. However, the existence of a competing effect should also be noted. As the quasar ionizing luminosity increases, the neutral fraction of the nebula decreases ($x_{H^0}\approx \alpha/\Gamma$), reducing the neutral atoms required in the first place for scattering to occur. In the end, both scattering and recombination radiation likely act jointly as power sources of the \lya\ nebulae, but the degree to which each mechanism contributes remains unclear. Further insight into this problem can be gained by analyzing non-resonant lines, as discussed in Section~\ref{sec:nonres}. 

\subsection{Extended \lya\ emission in the cosmic web}\label{sec:lyacw}

\begin{figure}[b]
\sidecaption
\centering
\includegraphics[scale=.34]{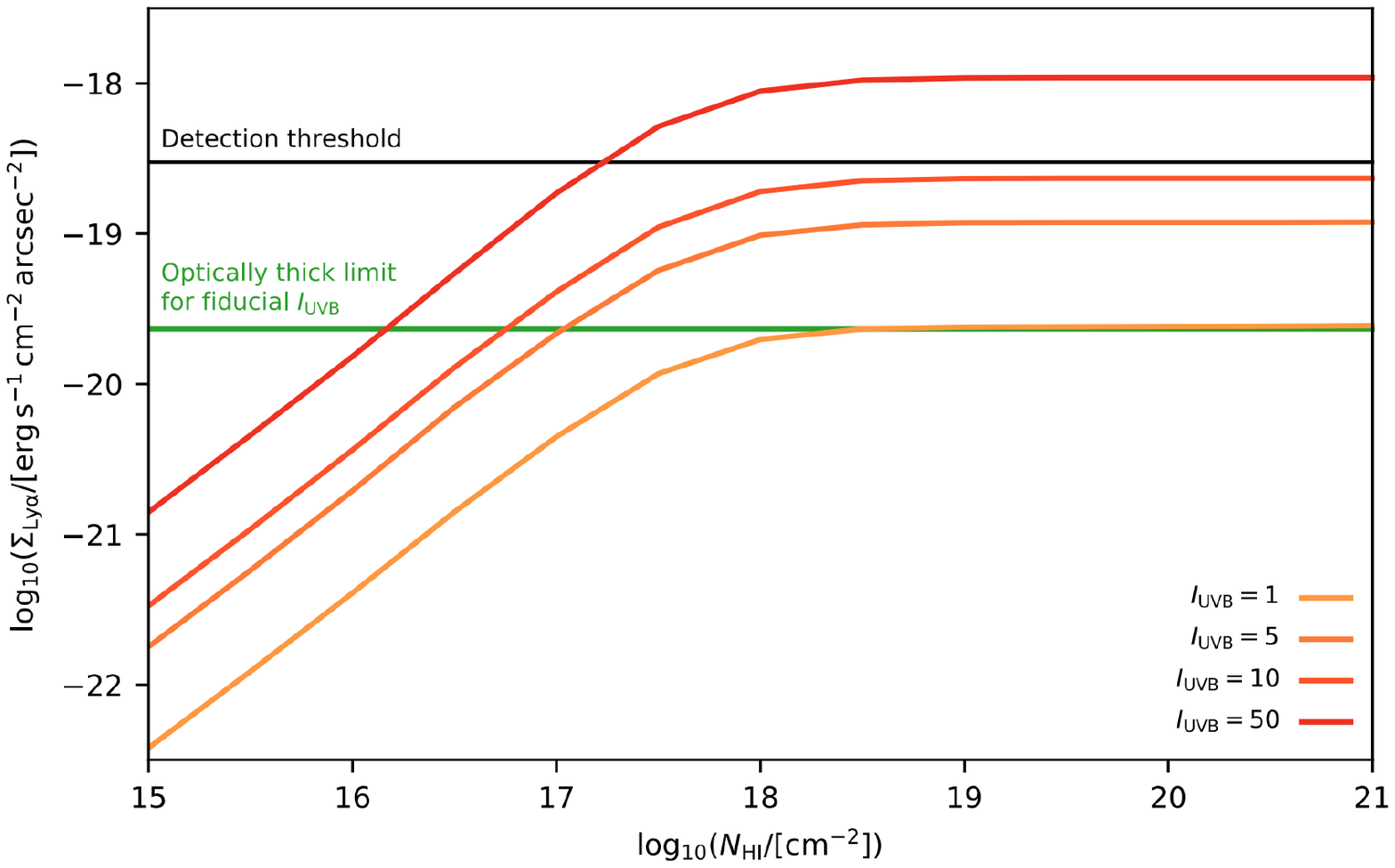}
\caption{Expected surface brightness values as a function of the neutral hydrogen column density of gas slabs illuminated by a UVB with varying intensity, obtained by scaling a reference UVB spectrum at $z\approx 3$ \cite{Haardt2012} with multiplicative factors (as labeled). Two regimes are visible. For optically thin gas, the surface brightness scales linearly with the column density and the radiation field intensity. For optically thick gas, at $N_{HI}>10^{18}$\,cm$^{-2}$, the gas converts about 60 percent of the ionizing photons in \lya\ radiation, regardless of the column density value. The expected surface brightness plateaus. The green line marks this value for the chosen reference UVB, while the black line marks a typical sensitivity of moderate-depth MUSE observations. A significant gap exists between current depths and expected surface brightness. Detecting the IGM in emission is thus especially challenging. {\it Credits: Umehata H. et al., 2019, Science, 366, 97. Reproduced with permission.}}
\label{fig:uvbfluo}       
\end{figure}

\subsubsection{Expected emission levels}

The ability to image the diffuse gas outside galaxies at even more considerable distances, reaching the IGM, would provide a complete view of how galaxies assemble within cosmic structures and how gas is exchanged between the cosmic web and the CGM, a critical component for comprehensive mapping the baryon cycle. 
However, directly imaging the cosmic web is less trivial than the CGM, as the densities involved are even lower than for the CGM, reaching the mean cosmic density ($\approx 10^{-5.5}$\,cm$^{-3}$ at $z\approx 3$). Moreover, the primary emission mechanisms are fluorescence from the UVB and cooling radiation. This poses notable challenges because of the expected low surface brightness.

If the gas is optically thin, the \lya\ emissivity will depend on the balance between recombination and ionization, and the observed photon flux will be \cite{Gould1996} 
\begin{equation}
    \phi = (1+z)^{-3}\eta_{thn} N_{H^0} \int_{\nu_0}^\infty \frac{J(\nu)}{h\nu}\sigma_\nu\dd\nu\,,
\end{equation}
where the term $(1+z)^{-3}$ accounts for cosmic dimming, and $\eta_{thn}\approx 0.42$ describes the effective number of recombinations leading to \lya\ photons. 
As the gas column density progressively rises, so does the \lya\ photon flux till the gas cloud becomes optically thick. In this limit, the gas acts as a perfect converter of ionizing photons into line emission, leading to an observed photon flux of
\begin{equation}
    \phi = (1+z)^{-3}\eta_{thk}\int_{\nu_0}^\infty \frac{J(\nu)}{h\nu}\dd\nu\,,
\end{equation}
where $\eta_{thk}\approx 0.62$ describes the effective number of \lya\ photons emitter per incident number of ionizing photons. In the optically thick regime, the observed photon flux is thus solely set by the amplitude of the incident radiation field (Figure~\ref{fig:uvbfluo}).
Further, in an optically thick cloud, collisions convert about half of the ionizing background energy into \lya\ photons. 
Altogether, the emission from the cosmic web is at most of the order set by the amplitude of the UVB, which is considerably low ($\approx 10^{-20}$\,\sbunit) for high redshifts because of the strong suppression induced by cosmological dimming.

\begin{figure}[b]
\includegraphics[scale=.36]{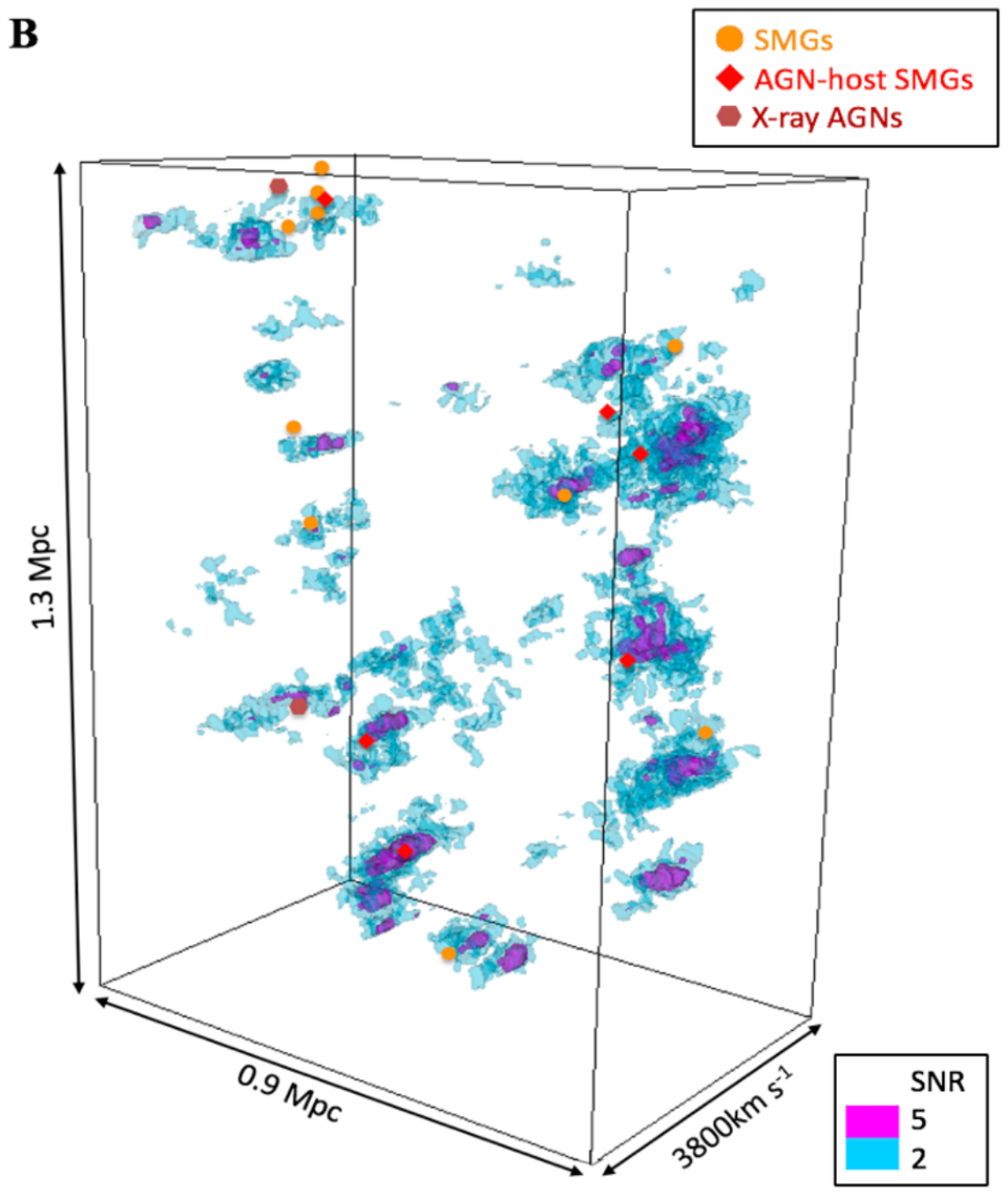}
\sidecaption
\caption{Mpc-scale \lya\ cosmic web in the overdense protocluster SSA22 at $z\approx 3.1$. The color coding (purple and cyan) reflects the $S/N$ of the detection. The analysis of the MUSE cube allows us to determine the correlation between the filaments and the active galaxies (sub-mm galaxies with yellow circles, X-ray bright AGNs with brown hexagons, and sub-mm galaxies hosting AGNs with red diamonds). When analyzed in velocity space, galaxies appear embedded in filamentary structures that connect them on large scales, fuelling the observed intense activity in this Mpc$^{3}$ region of the Universe. {\it Credits: Umehata H. et al., 2019, Science, 366, 97. Reproduced with permission.}}
\label{fig:ssa22}       
\end{figure}

\subsubsection{Status of current observations}

One way to make progress in imaging the cosmic web is to focus on overdense regions of the Universe, where gas in the IGM might be more easily above the density for self-shielding, and where galaxies and AGNs could locally enhance the metagalactic background compared to average regions of the Universe. This approach was taken by Umehata et al. \cite{Umehata2019}, who used MUSE to target the core of SSA22, a rich protocluster at $z\approx 3.1$. In a series of moderately deep exposures, extended emission was detected at low surface brightness along two filaments on scales of $\gtrsim 1$\,Mpc (Figure~\ref{fig:ssa22}). Through spectral analysis of the MUSE cube, it is apparent that these filaments connect several galaxies that are fully embedded in these structures. These observations represent some of the clearest examples of intergalactic gas connecting multiple halos on scales much larger than the virial halos.

Compared to the expectation of UVB fluorescence, the detected signal is $\gtrsim 10$ times brighter. This can be explained by the overdensity of highly star-forming galaxies and AGNs in the region, which contribute to a cumulative ionizing photon rate of $\approx 10^{57}$\,s$^{-1}$, well above the required number of $\approx 10^{55}$\,s$^{-1}$ needed to power the observed emission. Therefore, a modest escape fraction of ionizing radiation from these galaxies ($\lesssim 1\%$) is enough to close the photon budget needed to explain the observations. From Figure~\ref{fig:ssa22}, it is clear that these observations also probe the CGM in proximity to galaxies, where other powering mechanisms are likely to operate (including, e.g., shocks from outflows). This distinction is also apparent spectroscopically, where regions close to galaxies show wider \lya\ profiles than regions of lower surface brightness far from galaxies (see figure S4 in \cite{Umehata2019}). 

A second approach to the search for diffuse emission from the cosmic web is directly detecting very low-surface brightness features in ultradeep IFS observations. The sensitivity reached by MUSE in deep fields ($\gtrsim 100$\,h in the MUSE Extremely Deep Field or MXDF \cite{Bacon2021} and in the MUSE Ultra Deep Field or MUDF \cite{Fossati2019}) is of the order of $\approx 5\times 10^{-20}$\,\sbunit, which starts approaching the levels required for the detection of UVB fluorescence.
In the MXDF, Bacon et al. \cite{Bacon2021} searched for extended emission in known groups of LAEs using a wavelet analysis, which helps enhance faint signals on different scales. 
This technique identified extended emission in $\approx 60$ percent of the studied LAE groups, including examples of extended filamentary emission connecting LAEs. Although on much smaller scales ($\approx 200-300$\,kpc) than the filaments identified in SSA22, these detections are the first images of the cosmic web in representative regions of the Universe, far from substantial overdensities or bright quasars (see also a recent detection of a cosmic filament connecting two quasars \cite{Tornotti2024}).

By separating compact emission (from the ISM and CGM) and extended emission, the authors found that the extended filaments account for between $60-80$ percent of the total \lya\ emission. 
Based on the most recent models, UVB fluorescent \lya\ emission is not the dominant powering mechanism, although it can account for $\approx 1/3$ of the detected signal at $z\approx 3$.
In groups with AGNs, an additional source of \lya\ photons is fluorescence from leaking ionizing radiation, which is not considered the norm for all groups. To account for the missing fraction of \lya\ flux, these authors have proposed a model in which a fraction of the observed signal arises from a population of undetected LAEs, which contribute to the diffuse surface brightness below the luminosity for individual detections $L_{det}$ as
\begin{equation}
    SB(<L_{det}) = \Delta l \zeta \int_0^{L_{det}} \delta L \phi(L) \dd L\,,
\end{equation}
where $\Delta l$ describes the depth of the volume considered, $\zeta$ is a factor that models the autocorrelation of LAEs, and $\delta$ is the LAE overdensity. This unresolved population can account for between $\approx 50-80$ percent of the observed signal. 
As this estimate rests on the extrapolation of the LAE luminosity function, significant uncertainties on the actual fraction contributed by this emission mechanism remain.  

These earlier detections show that current instruments are now on the verge of mapping \lya\ emission from the cosmic web on large scales.
A particularly promising avenue for future progress is to move from $z\approx 3$ to $z\approx 2$, where the reduced cosmic dimming will make the expected signal brighter by a factor $(4/3)^4\approx 3.2$, and hence observations $\approx 10$ faster. The design of a blue-sensitive large field-of-view IFS for VLT, BlueMUSE, is currently in development \cite{richard2019}.

\subsection{The value of non-resonant lines: \ha\ and \heii}\label{sec:nonres}

A fraction of $\approx 2/3$ of all recombinations leads to \lya\ photons. Therefore, this line is the brightest transition that can be targeted for studying the CGM in emission. In addition to being intrinsically bright, the \lya\ line is also conveniently detected at optical wavelengths for a considerable range of redshifts, between $z\approx 2-6$. 
Being a $n=2\rightarrow1$ transition, \lya\ is a resonant line. This is because a significant fraction of electrons are in the ground state under typical conditions for astrophysical environments, and the emitted \lya\ photons can be easily absorbed by electrons that are excited again to $n=2$.
This property makes it more difficult to constrain the underlying kinematics of the gas as photons scatter and diffuse in frequency before escaping the cloud. It also creates ambiguity in the emission source that powers the CGM due to the considerable contribution that scattering can take in the proximity of central bright \lya\ sources such as LAEs or AGNs.

\begin{figure}[b]
\centering
\includegraphics[scale=.26]{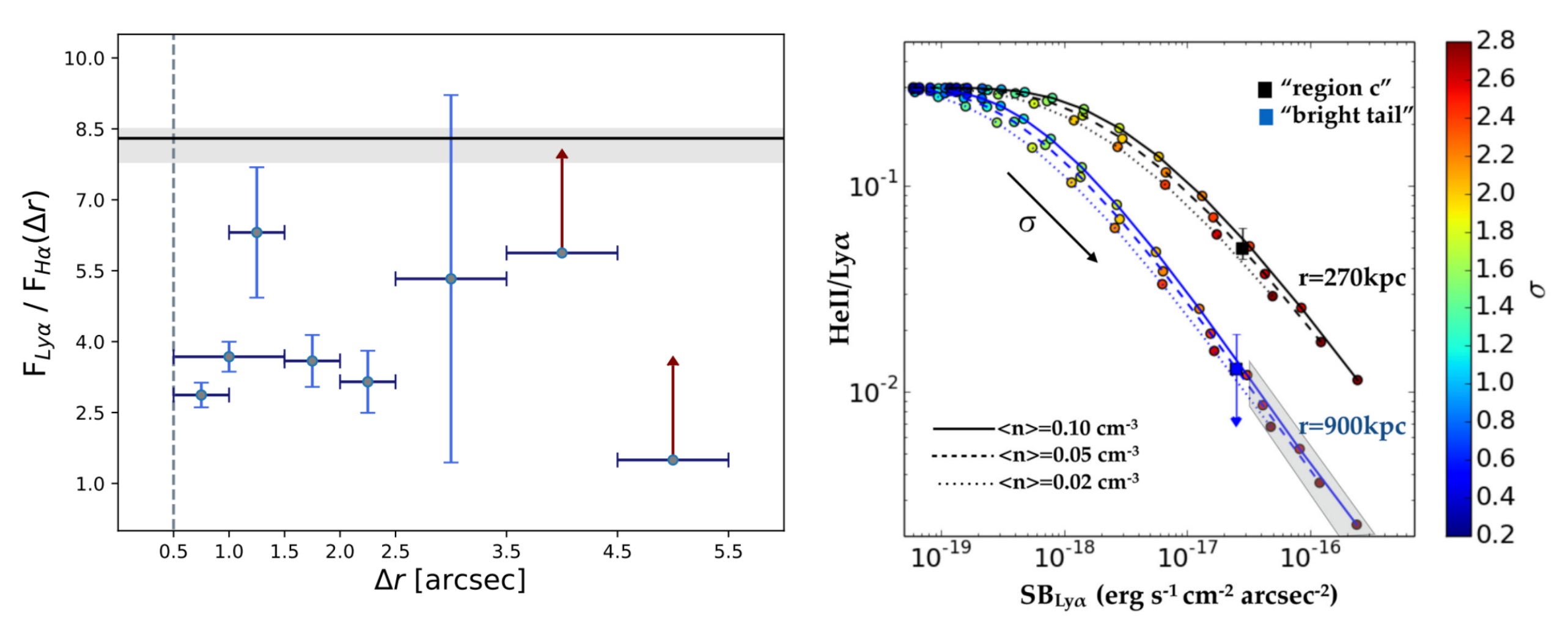}
\caption{Left: the observed \lya-to-\ha\ flux ratio in the nebula of the quasar J001005$+$061720. Data points (blue, with error bars) or limits (upward arrows) mark the flux ratio of the lines in different spatial regions once integrated over optimal velocity windows. The black horizontal line marks the expected ratio for case B recombination in low-density gas at a temperature of $\approx 10^4$\,K. The lack of substantial \lya\ fluxes compared to \ha\ indicates that prominent scatter from the central source is absent in this system, as radiative transfer effects are expected to boost the \lya\ fluxes above the expected case B recombination cases. Right: \ion{He}{II}-to-\lya\ fluxes in photoionization calculations of regions characterized by a density distribution with different mean densities (solid, dashed, and dotted lines) and dispersion parameter $\sigma$ (color bar). Two different distances from the central ionizing source are adopted (270\,kpc and 900\,kpc). Observational constraints (data point and limit) from the Slug Nebula are also shown. Models can reproduce the data for relatively large distances and high velocities typically seen in the turbulent ISM. This analysis suggests a broad density distribution and clumpy structure for 
the emitting gas. {\it Credits: Langen V. et al., 2023, MNRAS, 519, 5099; Cantalupo S. et al., 2019, MNRAS, 483, 5188. Reproduced with permission.}}
\label{fig:nonreslines}       
\end{figure}

Non-resonant lines, such as \ha\ in the Balmer series ($n=3\rightarrow2$) or \ion{He}{II} lines (e.g., the bright $\lambda = 1640$\,\AA\ transition), become valuable tools to discern the physical properties of the gas. However, they are harder to image because of their lower luminosity or because they are redshifted at near-infrared (NIR) wavelengths for $z\gtrsim 2$ systems. 
Non-resonant lines are helpful in three ways: 1) to measure the gas kinematics, 2) to learn about the powering mechanisms, and 3) to constrain the density structure. 

The value of non-resonant lines for resolving gas kinematics is clear. As \lya\ is scattered, not only its line profile can be altered, but also its centroid can shift relative to the systemic velocity, leading to uncertainties in the estimate of the system's rest frame from measurements of \lya\ alone. Non-resonant lines are required to infer the underlying systemic velocity (see figure 2 in \cite{Steidel2010}). In turn, a solid anchor point for the galaxy's rest frame becomes invaluable for inferring the motion of the CGM gas probed by \lya, i.e., constraining the presence of inflowing or outflowing material based on blue or red components in the \lya\ line \cite{Gronke2017}. Knowledge of the intrinsic line profile also enables the study of absorption lines imprinted on the top of the \lya\ emission, as done in bright radio-loud quasars \cite{Kolwa2019,Swinbank2015}, and ultimately to assess the clumpiness and structure of the CGM through radiative transfer models \cite{Gronke2016}.

The comparison of \lya\ to non-resonant lines, and in particular \ha, is a powerful tool to learn about the powering mechanisms in the CGM, particularly in disentangling the contribution of scattering and recombination radiation \cite{Mas-Ribas2017}. 
At the one limit, in the presence of recombination radiation only for the low-density regime of the CGM at a temperature of $T= 2\times 10^{4}$\,K, the expected \lya-to-\ha\ flux ratio equals 8.3 in case B, and 12.7 in case A. This value can range between 7.8 and 8.5 for temperatures in the interval $T= 5\times 10^{3}$\,K and $T= 4\times 10^{4}$\,K for case B, while between 10.5 and
13.7 in case A \cite{Langen2023}. Considering instead the opposite limit of a \lya\ emitting source embedded in a medium that scatters radiation, no net production of \ha\ is expected {\it in situ} within the CGM.  All in-between cases will be characterized by an enhanced ratio compared to case A or B expectations, as additional \lya\ photons are scattered from other regions on the top of the \lya\ and \ha\ emission produced {\it in situ}.

Although detections of \ha from the halo of LAEs or even LBGs are currently very challenging, this technique can be applied more readily to quasar nebulae, where \lya\ surface brightness  $\gtrsim 10^{-17}$\,\sbunit\ can be followed up with NIR spectrograph to sensitivities of $10^{-18}$\,\sbunit. In fact, \ha has been detected in high surface brightness regions inside quasar nebulae \cite{Leibler2018,Langen2023} with \lya-to-\ha\ flux ratios in the range $3-6$ (Figure~\ref{fig:nonreslines}, left). These ratios are below the expected theoretical values, but some effects can contribute to suppressing the signal. Firstly, dust can reduce the \lya\ flux more than the \ha\ one, leading to reduced ratios. Moreover, different techniques used for \ha\ and \lya\ observations may lead to errors when matching long-slit spectroscopic observations with IFS data. However, these values are relatively close to the theoretical ratio for recombination radiation, and the lack of substantial \lya\ fluxes over \ha, over $>11-12$, points to recombination radiation as the primary powering mechanism in the targeted regions. The {\it James Webb Space Telescope} (JWST) and the future 30m-class telescope, with their NIR capabilities, will make a decisive contribution in this field. 

Lastly, the non-resonant \heii\ line is useful to constrain the density structure of the CGM when compared to \lya, under the assumption that both lines arise from a two-body process such as recombination or collisional excitation. Following the derivation in \cite{Cantalupo2019}, the observed flux ratio averaged over a given region is, for recombination radiation,  
\begin{equation}
    \frac{\mean{F_{\heii}}}{\mean{F_{\mlya}}} = \frac{h\nu_{\heii}\alpha^{eff}_{\heii}(T)\mean{n_en_{\heiii}}}{h\nu_{\mlya}\alpha^{eff}_{\mlya}(T)\mean{n_en_p}}\,
\end{equation}
with $\alpha^{eff}$ the effective recombination coefficients and $n_{\heiii}$ the $He^{++}$ number density. 
For gas with helium primordial abundance, $n_{\heiii}=0.087 x_{He^{++}}n_H$. Considering hydrogen as the main source of free electrons, $n_e \approx n_H x_{H^+}$. By definition, $n_p = n_H x_{H^+}$, and $x_{H^+}$ and $x_{He^{++}}$ are the fractions of ionized hydrogen and doubly ionized helium.
With these relations, the above equation can be rewritten as 
\begin{equation}
    \frac{\mean{F_{\heii}}}{\mean{F_{\mlya}}} \approx 0.087 \frac{h\nu_{\heii}\alpha^{eff}_{\heii}(T)\mean{x_{H^+}x_{He^{++}}n^2_H}}{h\nu_{\mlya}\alpha^{eff}_{\mlya}(T)\mean{n_H^2 x_{H^+}^2}} = R_0(T)\frac{\mean{x_{H^+}x_{He^{++}}n^2_H}}{\mean{n_H^2 x_{H^+}^2}}\,.
\end{equation}
For a highly-ionized gas (e.g., in the presence of a strong radiation field such as in quasars) $x_{H^+}\approx 1$, and the above equation can also be rewritten substituting the volume average over a region with the explicit dependence on the density distribution $p(n_H)$
\begin{equation}\label{eq:heclumps}
    \frac{\mean{F_{\heii}}}{\mean{F_{\mlya}}} \approx  R_0(T)\frac{\int x_{He^{++}}n^2_Hp(n_H)\dd n_H}{\int n_H^2 p(n_H)\dd n_H}\,.
\end{equation}
Suppose the distance from the ionizing source is known. In that case, both the He photoionization rates and the temperature will be known, making the expression above vary only via the density-dependent $x_{He^{++}}$, which is weighted by the square of the gas density. 
Through Equation~\ref{eq:heclumps}, the observed flux ratio between \heii\ and \lya\ becomes a function of the density distribution. 
Indeed, if the gas is at constant density, Equation~\ref{eq:heclumps} approaches a given maximal value. Conversely, for a density distribution, the flux ratio will decrease as  $x_{He^{++}}$ decreases with increasing density, and this contribution is weighted by $n_H^2$.

Cantalupo et al. \cite{Cantalupo2019} have applied the above formalism to the case of quasar \lya\ nebulae where \heii\ was also detected. Under some assumptions, they inferred the distance of the observed gas from the ionizing source. They then placed the observed line ratios on a diagram of $\frac{\mean{F_{\heii}}}{\mean{F_{\mlya}}}$ versus \lya\ surface brightness (Figure~\ref{fig:nonreslines}, right).
They also derived numerical estimates of the expected line ratio and \lya\ surface brightness in a turbulent medium that is characterized by a log-normal density distribution with average volume density $\mean{n_c}$ and dispersion $\sigma$, which is linked to the gas Mach number.
This step requires some additional assumptions about the geometry of the system. These authors found that the line ratios observed in the nebula could be modeled by gas at low densities ($<0.1$\,cm$^3$) and high-velocity dispersion ($\sigma \approx 2.5$), a condition not too dissimilar from what is found in the ISM. The emitting gas is, therefore, likely to have a broad density distribution, be clumpy, and multiphase. 
Unlike the ratio of \ha\ to \lya, this technique is based on several assumptions that are difficult to verify in observations. The results should, therefore, be taken with some caution. However, additional observations of \heii\ lines provide an avenue for further, albeit indirect, constraints on the nature of the emitting gas.

\section{Observing the multiphase CGM in emission: UV metal lines}
\label{sec:Zemission}

Although hydrogen lines, and in particular \lya, are ideal for mapping the structure and extent of the CGM up to high redshift, metal emission becomes a precious tool for studying the physics of the CGM and linking the metal production of galaxies with the metal content of their halo gas. 
Most bright metal transitions lie in the spectrum's near/far UV rest-frame region, extending to the X-ray part. The outlook for X-ray imaging and spectroscopy of the CGM is promising, with planned and concept missions (e.g., ATHENA or Lynx \cite{Barret2020,Lynx2018}) that will build on results from eROSITA \cite{Predehl2021}. 

This contribution focuses on rest-frame UV transitions, while future $X-$ray surveys will be reviewed in another chapter of this book.
 After a brief overview of the physics of the UV emission lines (Section~\ref{sec:uvlphys}), forecasts from numerical simulations are discussed (Section~\ref{sec:uvlsims}) before turning the attention to early results in observations (Section~\ref{sec:uvlobs}) and more recent studies which have assembled larger samples (Section~\ref{sec:uvlstats}). Finally, Section~\ref{sec:uvlemabs} discusses the promising approach of combining absorption and emission in individual systems.

\subsection{The physics of UV emission lines}\label{sec:uvlphys}

The emission properties of a gas parcel are described by the radiative transfer equation (Equation~\ref{eq:RTeq}) through the emission coefficient $j_\nu$ that represents the energy emitted over time in a unit volume $V$, per unit frequency $\nu$, and solid angle $\Omega$:
\begin{equation}
    j_\nu \equiv \frac{\dd E_\nu}{\dd V \dd \Omega \dd t \dd \nu}\,.
\end{equation}
The emission coefficient is related to the intrinsic emissivity $\epsilon_\nu$ of the material through its mass density $\rho$ as 
\begin{equation}
j_\nu = \frac{\epsilon_\nu\rho}{4\pi}\,.
\end{equation}
For line emission, $j_\nu$ can be computed explicitly as 
\begin{equation}
j_\nu = \frac{1}{4\pi}n_u A_{ul}h\nu \phi(\nu)\,,
\end{equation}
where $n_{u}$ is the number density of electrons in the state $u$, $A_{ul}$ is the Einstein A coefficient for spontaneous emission between states $u$ and $l$, and $\phi(\nu)\dd \nu$ expresses the probability that the emitted photon will be in the frequency interval $(\nu, \nu+\dd \nu)$.

\begin{figure}[b]
\centering
\includegraphics[scale=.18]{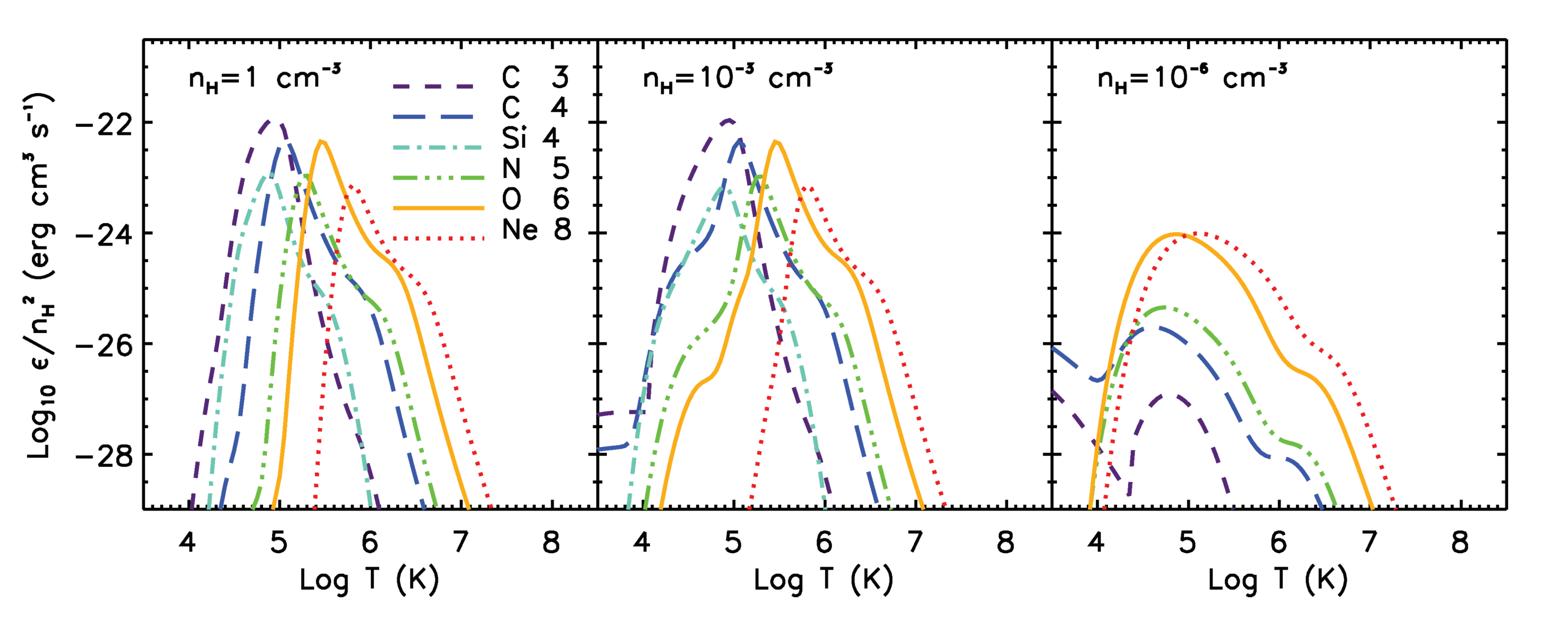}
\caption{Density-normalized gas emissivity at solar abundance for selected UV transitions as a function of temperature. The three panels reproduce results at different densities, as labeled. In the regime dominated by collisions, transitions have a characteristic peak at well-defined temperatures, and the emissivity scales as the square of the gas density. For low densities (middle and left panels), radiation can keep low-to-moderate transitions ionized (e.g., \ion{C}{III} or \ion{C}{IV}) even at low temperatures below $T\lesssim 2-3\times 10^{4}$\,K. {\it Credits: Bertone S. et al., 2010, MNRAS, 408, 1120. Reproduced with permission.}}
\label{fig:emissivity}       
\end{figure}

One can predict the emergent line flux from a gas cloud through the above equation specified for the transition of interest. Similarly to the case of ionization correction for absorption line systems, this calculation requires a detailed knowledge of the ionization conditions of the gas, which in turn is linked to a complex radiative transfer problem. As seen in Section~\ref{sec:absorption}, this problem becomes tractable numerically under some simplifying assumptions, such as solving for the equilibrium gas emissivity in optically thin gas ionized by photons and collisions. These calculations can be derived, e.g., using the {\sc Cloudy} code \cite{cloudy} and have also been published in the literature \cite{Bertone2010}.

As shown in Figure~\ref{fig:emissivity}, the emissivity shows characteristic shapes with well-defined peaks at various temperatures according to the specific ion. These originate from collisions that are the dominant ionization mechanism at moderate-to-high number densities and leave elements in a given ionization state over a well-defined temperature interval: at too-low temperatures, ions cannot be produced as there is not enough energy to excite the transition; at too-high temperatures, the ionization fraction declines exponentially as higher energy transitions become more favorable. In the regime where collisions dominate the ionization, the emissivity scales with the density as $\epsilon_\nu \propto \rho^2$.  
At sufficiently low densities, collisions become less significant, and the gas can remain ionized by radiation at lower temperatures (right panel of Figure~\ref{fig:emissivity}). In the case of a gas illuminated by radiation, the emissivity remains substantial even for cool ($T\lesssim 10^4$\,K) gas. 

Figure~\ref{fig:emissivity} encapsulates the prominent trends that make emission lines valuable tracers of the multiphase CGM. The mere detection of a line, especially from high-ionization state ions, constrains the likely range of gas temperatures. Moreover, the line brightness constrains the metallicity (more metals equals more ions and brighter emission) and density (through the density dependence of the emissivity discussed above), although the two quantities are degenerate.

\begin{figure}[b]
\sidecaption
\includegraphics[scale=.2]{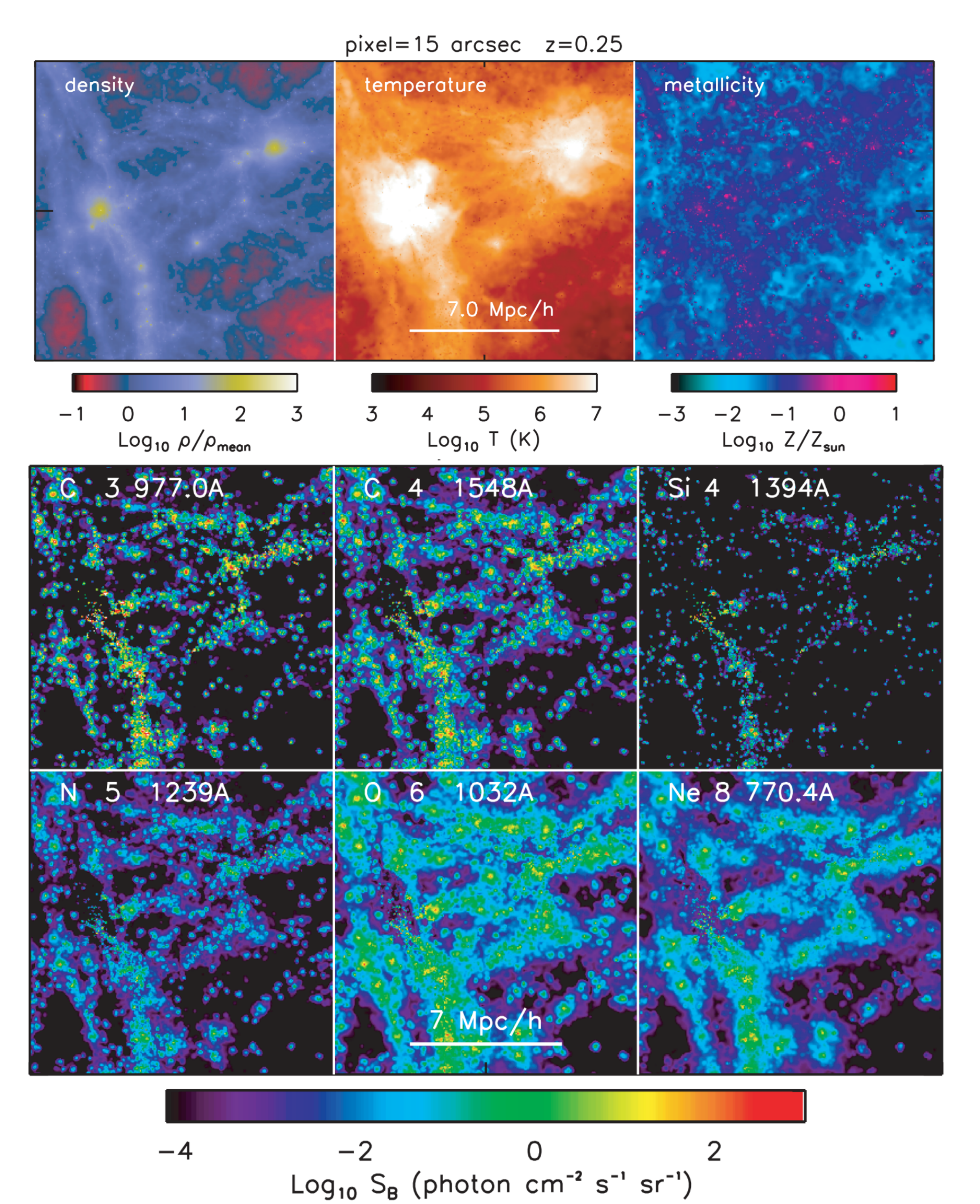}
\caption{Predictions of UV emission lines at $z\approx 0.25$ in the OWLS simulations. Top: map of density, temperature, and metallicity from a 14\,$h^{-1}$\,Mpc region of the simulation box. Bottom: Surface brightness from the CGM and IGM. UV metal emission is challenging to detect with surface brightness below $\approx 10^{-19}$\,\sbunit. Ions of low and intermediate ionization (\ion{Si}{IV}, \ion{C}{III}, \ion{C}{IV}) trace the cooler ($T\lesssim 10^{5}$\,K) and clumpy CGM of low mass halos, while ions of higher ionization potential better trace the more massive halos and the warm gas in the filament. The most massive halos are not visible in the UV, as the virial temperatures shift their peak emissivity in the X-ray. {\it Credits: Bertone S. et al., 2010, MNRAS, 408, 1120. Reproduced with permission.}}
\label{fig:owlsUVline}       
\end{figure}

\subsection{Predictions from numerical simulations}\label{sec:uvlsims}

The formalism described above can be applied to cosmological numerical simulations to predict the expected surface brightness maps of metals in the CGM and IGM at different redshifts \cite{Bertone2010,Bertone2010a,Bertone2012,vandevoort2013}. Specifically, the luminosity of a simulation particle or cell in a given emission line can be computed using a grid of emissivity like the one in Figure~\ref{fig:emissivity} as a function of temperature, density, and redshift after rescaling for the element abundance respective to solar and renormalizing by the volume of the emitting region.

Considering UV emission lines at $z<1$, the analysis by Bertone et al. \cite{Bertone2010} provides one of the first models for emission lines in recent cosmological simulations (see also \cite{vandevoort2013}). Using the OverWhelmingly Large Simulations (OWLS) project \cite{Schaye2010}, these authors predicted the expected rest-frame UV signal for the cooler ($T\lesssim 10^{6}$\,K) warm-hot intergalactic medium (WHIM) at densities $n_H <0.1$\,cm$^{-3}$, below which star formation does not occur in the simulation. Figure~\ref{fig:owlsUVline} shows the surface brightness maps of key UV metal lines.

Several results can be inferred from this analysis.  
First and foremost, UV metal lines are challenging to detect, as the surface brightness is consistently below $\approx 10^{-19}$\,\sbunit, with the bulk of the diffuse emission below $\approx 10^{-21}$\,\sbunit. 
Two units are conventionally used: one is based on energy, and the other is based on the photon number. 
To convert between these two scales of surface brightness, one can use the relation 
\begin{equation}
10^{-19}\textrm{\sbunit}=270 \left(\frac{\lambda}{10^3\textrm{\AA}}\right)\left(\frac{1+z}{1.25}\right)\textrm{photons~s$^{-1}$~cm$^{-2}$ sr$^{-1}$}\,.
\end{equation}
These values are orders of magnitude below the sensitivity of current instruments operating in the near UV \cite{Augustin2019,Hamden2020}, making the detectability of the CGM at $z<1$ a challenging goal of future larger missions, such as the Large UV/Optical/IR Surveyor (LUVOIR) \footnote{https://asd.gsfc.nasa.gov/luvoir/}.

Another characteristic that emerges from these numerical simulations is that emission from intermediate ionization stage ions (\ion{Si}{IV}, \ion{C}{III}, \ion{C}{IV}) appear more clumpy and trace the cooler ($T\lesssim 10^{5}$\,K) CGM of low mass halos. 
Of these ions, \ion{C}{III} is predicted to be the brightest by about one order of magnitude compared to \ion{C}{IV}, being carbon an abundant element that is easily found in its second ionization stage. In contrast, ions of high ionization stages (\ion{O}{IV}, \ion{Ne}{VIII}) are better suited to trace the diffuse and filamentary WHIM at $T\gtrsim 10^{5}$\,K, with \ion{O}{IV} predicted to be the brightest UV tracer of cosmic filaments.
Invisible to UV lines is instead the CGM of the most massive halos, where virial temperatures exceed $T\approx 10^{6}$\,K. At these temperatures, the gas is dominated by ions in very high ionization stages, which emit in the X-ray \cite{Bertone2010a,vandevoort2013}. 
Thus, X-rays observations are highly complementary to the ones in the UV, probing a different part of the phase diagram. 

As the UV line brightness scales with metallicity and the square of density (for collisions), images of the CGM and IGM will produce a biased view of the denser parts of the WHIM and the CGM, with an additional modulation induced by the temperature given that different UV lines trace different temperature intervals as discussed above.  Therefore, UV-metal lines (and X-rays) are unsuitable for imaging the bulk of the mass in the low-density and more metal-poor IGM. Ly$\alpha$ emission remains the best tool for these regions.

Taking advantage of the multiple physical models within the OWLS suite, Bertone et al. \cite{Bertone2010} also studied the impact of physical prescriptions on the predicted UV metal emission signal, finding that the expected surface brightness is, to first order, insensitive to reasonable variation of the model, thus making the predictions derived from the OWLS simulations generally robust concerning the adopted sub-grid physics. 
The main differences are observed in simulations without cooling, without supernova feedback, or when adding AGN feedback.
Without cooling, the gas thermodynamics is significantly altered, resulting in more regions with a favorable temperature to high-ion emission, such as \ion{O}{VI}.
Metals are retained closer to star-forming regions without supernova feedback, reducing the outer CGM and IGM emission. This results in clumpier surface brightness maps, mainly dominated by the inner CGM. Finally, the AGN feedback implemented in OWLS decreases the gas density and the metallicity in the central parts of massive halos, suppressing the fluxes in the highest-density features of the CGM. 

At higher redshift, $z>2$, despite the reduced observed surface brightness due to $(1+z)^4$ dimming, observations of metals in the CGM and IGM are starting to be within reach thanks to the sensitivity of IFSs like MUSE, both in stacks and deep exposures. Compared to lower-redshift predictions, the analysis of the OWLS simulations at $z>2$ \cite{Bertone2012}
shows how most of the emission arises from the gas inside or near collapsing objects, given that massive structures have not been assembled yet. Moreover, the evolution in the metallicity distribution is responsible for most of the differences observed for the lower redshift.

\begin{figure}[b]
\centering
\includegraphics[scale=.28]{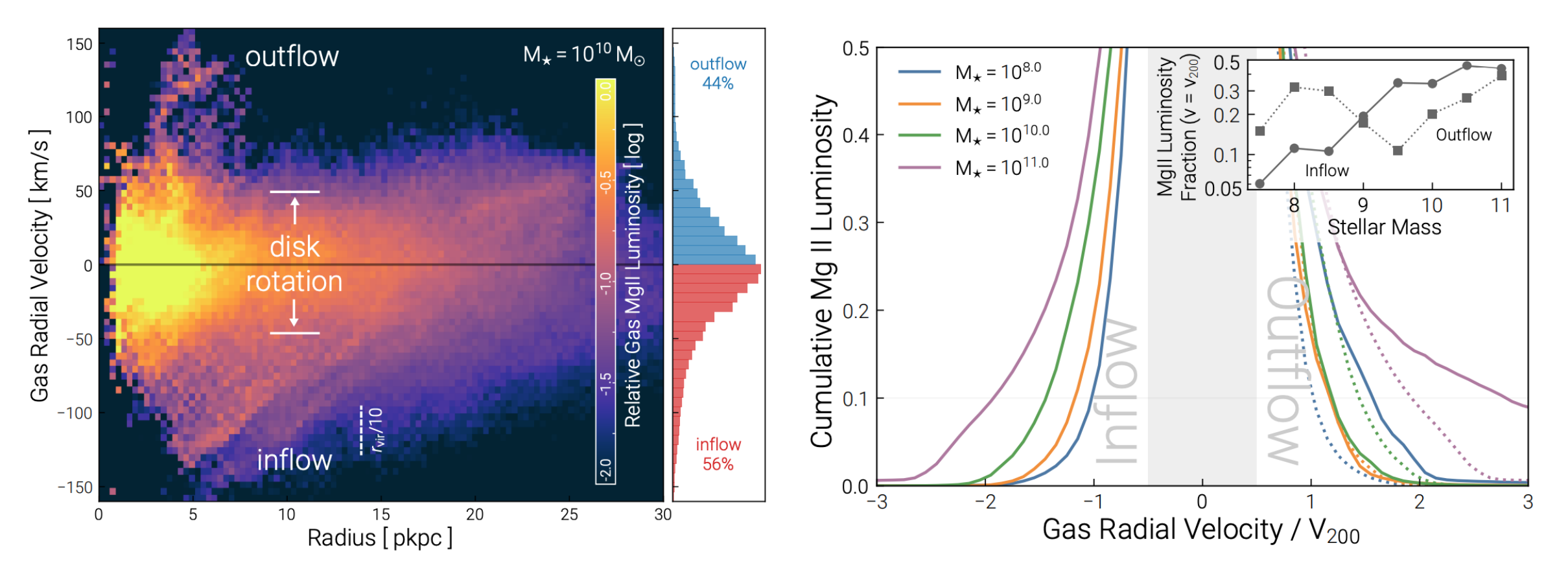}
\caption{Inflow and outflow contribution to the \ion{Mg}{II} emission in the TNG-50 simulation. Left: Distribution of \ion{Mg}{II} luminosity as a function of radius and radial velocity relative to central galaxies with stellar mass $\approx 10^{10}$\,M$_\odot$. The disk rotation dominates the inner part. Outflows and inflows, which  contribute equally to the \ion{Mg}{II} emission, become apparent for $v\gtrsim 60-80$\,km~s$^{-1}$. Right: Cumulative MgII luminosity of halos of different mass as a function of the radial velocity. Outflows dominate at low halo masses, while inflow dominates in massive halos (see inset). {\it Credits: Nelson D. et al., 2021, MNRAS, 507, 4445. Reproduced with permission.}}
\label{fig:tngmgii}       
\end{figure}

The \ion{Mg}{II} doublet at $\lambda\lambda 2796,2803$\,\AA\ represents an exciting transition for imaging the CGM. Although magnesium is not a particularly abundant element, with a number density $\approx 1$\,dex lower than carbon or oxygen, the possibility of detecting \ion{Mg}{II} at optical wavelengths at $z\approx 0.5-1.5$ makes this ion a powerful tool for studying the CGM at moderate redshifts with IFSs. In a recent study based on the high-resolution TNG50 simulation \cite{Pillepich2019}, Nelson et al. \cite{Nelson2021} predicted the emission from \ion{Mg}{II} halos in UVB-photoionized gas, without the inclusion of resonant scattering, but accounting for gas self-shielding. 
A simple prescription for dust depletion is also included by removing some of the magnesium from the gas phase.  
These authors followed a methodology similar to that in Bertone et al. \cite{Bertone2010} and predicted the \ion{Mg}{II} luminosity by interpolating a grid of tabulated emissivity onto the simulation. However, the superior resolution of TNG50 allows for a more in-depth study of the CGM properties. 

Through this analysis, Nelson et al. \cite{Nelson2021} predicted that \ion{Mg}{II} should be ubiquitous around star-forming galaxies at all the redshifts analyzed ($z\approx 0.3-2$), with surface brightness profiles at $10^{-19}$\,\sbunit\ extending up to $\approx 10-20$\,kpc for galaxies with stellar mass $\gtrsim 10^{9.5}$\,$M_\odot$ at $z\lesssim 1$. More extended halos, between $\approx 20-50$\,kpc, are predicted for surface brightness limits of $10^{-21}$\,\sbunit.
The extent of the \ion{Mg}{II} halo increases with increasing stellar masses, with a steeper increase above $10^{10}$\,$M_\odot$. More massive galaxies have, on average, more luminous halos. The scatter observed about the mean trends is driven by the accretion history of the halo and is also modulated by the galaxy environment.

Through the use of simulation data, it is also possible to study whether the \ion{Mg}{II} emission traces primarily the inflow on or outflow from galaxies (or both). The result of this analysis is presented in Figure~\ref{fig:tngmgii}. 
At all radii, the contribution of inflow and outflow is approximately equal, with no net preference for \ion{Mg}{II} emission to trace outflows. When considering gas moving at virial velocity, a higher fraction of \ion{Mg}{II} luminosity can be attributed to the outflowing gas. At the same time, inflow dominates for more massive systems, with equality reached at $M_*\approx 10^{9}$\,$M_\odot$.

\begin{figure}[b]
\centering
\includegraphics[scale=.3]{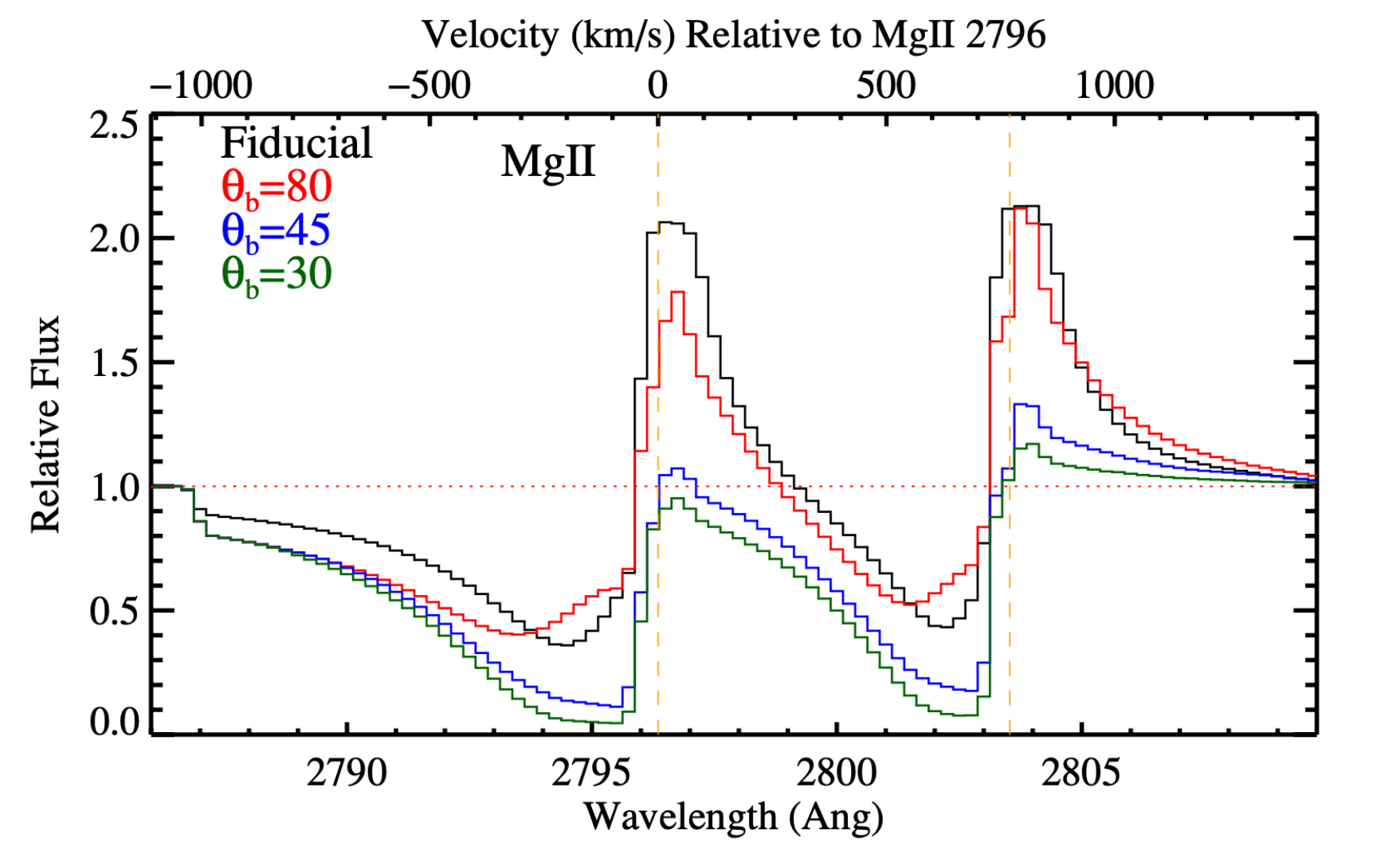}
\caption{Velocity profiles for the \ion{MgII}\ doublet in a wind model. The fiducial model (black lines) is for an isotropic wind covering the source. A characteristic double-peaked P-Cygni profile is visible, with redshifted emission from the back-scattered radiation in the receding part of the wind and the blueshifted absorption from the approaching side of the wind. Different models of biconical outflows with varying opening angles viewed along the rotational axis of symmetry are shown with color lines. The profile's shape depends on several parameters, including velocity, geometry, and dust. {\it Credits:  Prochaska J.~X. et al., 2011, ApJ, 734, 24. Reproduced with permission.}}
\label{fig:pcygni}       
\end{figure}

Additional information on halo kinematics can be obtained beyond imaging with spectral observations of the extended \ion{Mg}{II} emission. This doublet is resonant as the \ion{Mg}{II} lines arise from recombinations from the $^2P$ to the $^2S$ state. Hence, the emerging radiation from a galaxy is modulated by the halo gas kinematics \cite{Prochaska2011}. Radiative transfer calculations are therefore needed to understand this problem. 

Considering a standard simple geometry, such as an outflowing shell within which a source is embedded, the emergent spectrum will present a characteristic double-peaked P-Cygni profile (Figure~\ref{fig:pcygni}), which is characterized by absorption in the blue to systemic velocity and emission in the red part of the spectrum. The absorption profile arises from the front of the shell, which is moving towards the observer and absorbs the source radiation in its rest frame. Due to the strong nature of the \ion{Mg}{II} transition (with oscillator strengths of $\approx 0.3$ and $\approx 0.6$), this absorption component is often saturated. Thus, the profile shape constrains the gas velocity distribution more than its column density. The emission peak, found in the red part of the spectrum, arises instead from gas moving away from the observer, on the far side of the shell, that can back-scatter photons toward the observer. Only the redshifted photons (in the rest frame of the receding shell) can be scattered; hence, this radiation can travel unperturbed back through the front of the shell without being absorbed and reaching the observer. 

While rich in information on the motion of the gas, due to the scattering of radiation, the analysis of the kinematics profile is nontrivial, as it also depends on parameters such as the geometry of the wind and the viewing angle. Moreover, scatter radiation can also fill in the absorption component close to systematic velocity, leading to a possible misinterpretation of the data if this contribution is not considered (see details in \cite{Prochaska2011}). However, extended spectroscopy of \ion{Mg}{II} remains a promising avenue for gaining more insight into the CGM kinematics.

\subsection{Early examples of metal line emission from the CGM}\label{sec:uvlobs}

Observing the UV metal line emission from the CGM and IGM requires reaching a low surface brightness outside the sensitivity achieved routinely by imaging or long-slit spectrographs. Obtaining detailed maps of the metal content of the CGM is, therefore, the realm of IFSs, although this is still challenging. Nevertheless, examples in the literature can be found of the first glimpse into the metal emission of the diffuse gas around highly starburst galaxies, using both narrow-band filters and spectrographs. Some examples are reviewed next.

\begin{figure}[b]
\includegraphics[scale=.36]{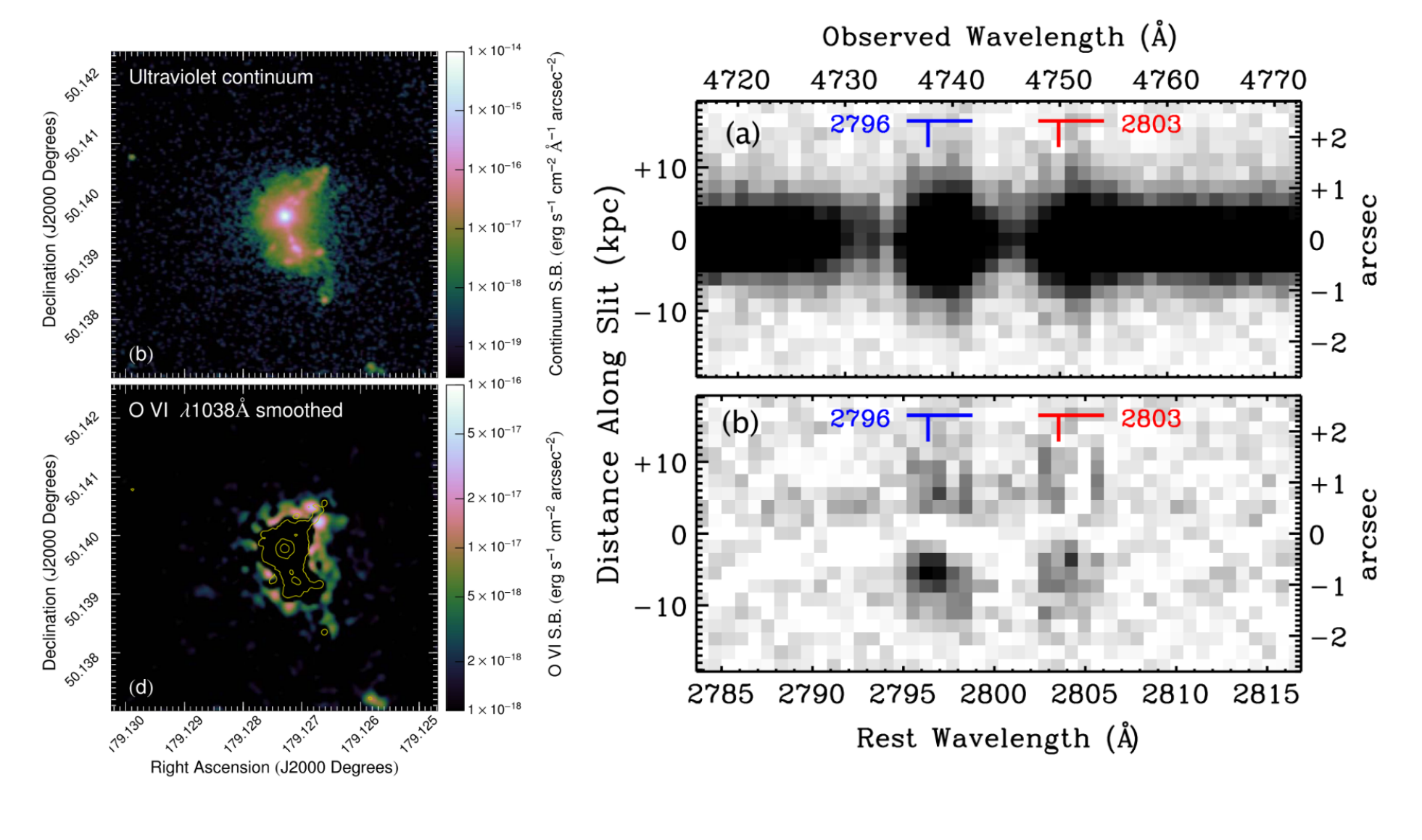}
\caption{Examples of early observations of metal emission from the CGM using spectroscopy and imaging. Left: A ring of \ion{O}{IV} emitting gas (bottom) detected outside the brightest UV components (top) of a starburst galaxy at $z\approx 0.235$. This gas is likely tracing $T\approx 10^{5.5}$\,K gas cooling from a hotter phase. Right: 2D long-slit spectroscopy of a $z \approx 0.694$ galaxy. A P-Cygni profile, characteristic of an outflow, is visible in the \ion{Mg}{II} line (top). Scattered continuum radiation in the wind traces the spatial extent of the outflow out to $\approx 7$\,kpc (bottom). {\it Credits:  Hayes M. et al., 2016, ApJ, 828, 49. Rubin K.~H.~R. et al., 2011, ApJ, 728, 55. Reproduced with permission.}}
\label{fig:earlyZimg}       
\end{figure}

Hayes et al. \cite{Hayes2016} reported on the imaging of the \ion{O}{VI} doublet at $\lambda\lambda 1032, 1038$\AA\ in the CGM of a highly starburst galaxy at $z\approx 0.235$ using the {\it Hubble Space Telescope} (HST). For this experiment, they used a new technique to create synthesized narrow-band filters from partially overlapping broad-band filters (see figure 3 in \cite{Hayes2016}). The target is a compact, highly star-forming galaxy ($SFR\approx 30-40$\,M$_\odot$~yr$^{-1}$), which is an extreme outlier both in star formation rate and H$\alpha$ equivalent width at these redshifts. 
The HST observations reveal extended \ion{O}{VI} emission from a ring located between $10-20$\,kpc from the central source (Figure~\ref{fig:earlyZimg}, left). At the same time, the innermost region exhibits strong absorption in the \ion{O}{VI} and H$\beta$ lines. 

Combining information from the \ion{O}{VI} emission and absorption under some assumptions (i.e., the gas is collisionally ionized), the authors inferred that the gas is clumpy and has a density of order $0.5$\,cm$^{-3}$. Gas with a temperature of $T\approx 10^{5.5}$\,K  at the peak of the cooling curve and an inferred density of the order of $1$\,cm$^{-3}$ cools very rapidly, on time scales of $\approx 1$\,Myr. At this temperature, it is improbable that the \ion{O}{VI} emitting gas has been lifted by an outflow and transported outward from the nuclear star-forming regions. It is more plausible that we are witnessing a cooling medium that passes through the $T\approx 10^{5.5}$\,K  phase as the hotter X-ray-emitting coronal phase constantly replenishes it.

The study by Rubin et al. \cite{Rubin2011} (see also \cite{Martin2013}) provides an intriguing example of spectroscopic observations that reveal extended metal line emission in the CGM. Through slit spectroscopy, they detected spatially-extended \ion{Mg}{II} emission beyond the stellar continuum of the galaxy, on scales $\approx 7$\,kpc (Figure~\ref{fig:earlyZimg}, right). The target is, again, an intense starburst galaxy at $z \approx 0.694$, with a P-Cygni profile in \ion{Mg}{II} and \ion{Fe}{II$^*$} emission. 
Based on the galaxy's spectral properties and extended emission, the authors conclude that an outflow is responsible for the P-Cygni profile. The near side of the flow produces the blue-shited absorption, and the far side is responsible for the red-shifted emission. 

Moreover, the gas moving above and below the galaxy can originate through the scattering of continuum photons that produce the observed extended emission. This picture is corroborated by the lack of detection of [\ion{O}{II}] and H$\gamma$, which would originate from recombination or collisional excitation of the gas. Intriguingly, the observations of extended UV metal lines on scales reaching $\gtrsim 7$\,kpc provide one of the few direct constraints on the minimum spatial extent an outflow can reach and its morphology.
Therefore, observations of the CGM in emission via metal lines are a valuable complement to the diagnostics that can be obtained in absorption. 

\subsection{Towards statistical samples with IFS observations}\label{sec:uvlstats}

The improved sensitivity for imaging emission lines on large scales, possible thanks to the deployment of IFSs at 8m-class telescopes, has finally enabled the possibility of imaging the extended metal-enriched CGM in a series of environments, including galaxy groups, normal galaxies, and quasars.  Among the first examples proposed in the literature, there is emission arising from galaxy groups detected at moderate redshifts, $z\lesssim 2$, via \ion{Mg}{II}, [\ion{O}{II}], and [\ion{O}{III}] emission lines. Section~\ref{sec:environment} presents an overview of these works; here, we focus on the emergence of the statistical analysis in large samples of galaxies and quasars.

\begin{figure}[b]
\centering
\includegraphics[scale=.3]{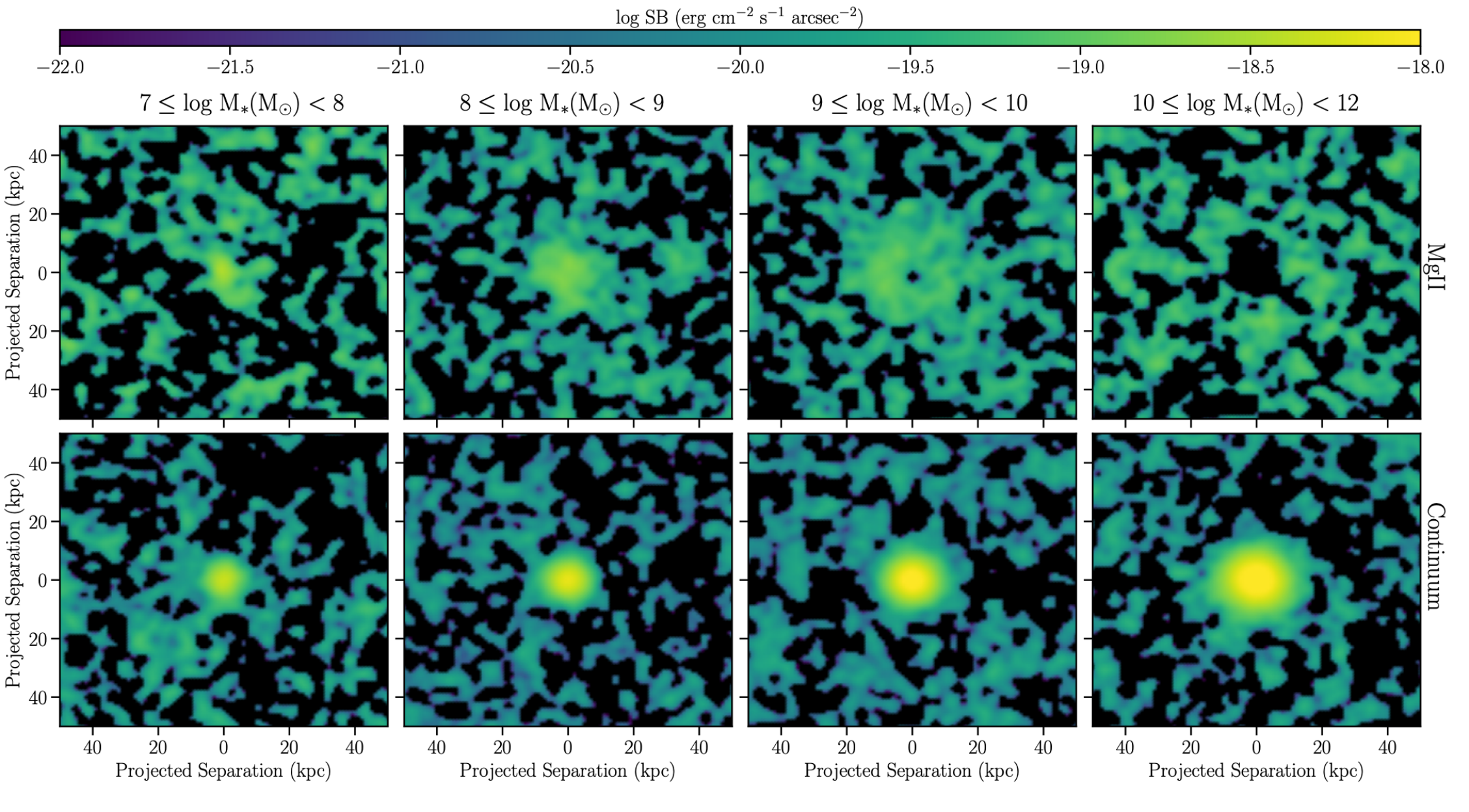}
\caption{Maps of \ion{Mg}{II} (top) and continuum (bottom) emission from stacks of $0.7 \lesssim z \lesssim 1.5$ galaxies in various mass bins. These continuum-subtracted narrow-band images reveal the detection of extended metal emission on scales beyond the stellar continuum, with a profile that progressively increases towards higher stellar masses. \ion{Mg}{II} absorption is also visible in the central regions of the more massive galaxies. {\it Credits: Dutta R. et al., 2023, MNRAS, 522, 535. Reproduced with permission.}}
\label{fig:mgiimagg}       
\end{figure}

Considering metal lines from typical star-forming galaxies, the study by Dutta et al. \cite{Dutta2023} provides the first systematic search of extended emission in a sample of $\approx 600$ galaxies with stellar masses $M_* \approx 10^6-10^{12}$\,M$_\odot$ at $0.7 \lesssim z \lesssim 1.5$ using MUSE data from the MAGG and MUDF surveys \cite{Lofthouse2023,Fossati2019}. 
The continuum-subtracted narrow-band images reconstructed from the data cubes reveal the detection of extended [\ion{O}{II}] and \ion{Mg}{II} emission on scales beyond the stellar continuum, reaching $\approx 50$\,kpc from the central sources. The surface brightness profile progressively increases its extent with increasing stellar mass (Figure~\ref{fig:mgiimagg}). At the highest masses, prominent \ion{Mg}{II} absorption also arises from the inner parts, where the ions can absorb and scatter photons. 

Moreover, the profiles are brighter and more extended at $z\gtrsim 1$ compared to the lower redshift, a trend that can be attributed to the higher SFR at early times. Tentative evidence of a brighter and more extended profile in group galaxies compared to isolated systems is also reported. When comparing the observed profiles with the predictions from the TNG simulations discussed above \cite{Nelson2021}, good agreement is found in terms of the ballpark surface brightness levels in the region of $\approx 10^{-19}-10^{-20}$\,\sbunit. However, the simulated profiles are brighter than the observed ones in the center and decline more steeply, a feature that can be attributed to the lack of radiative transfer processing in the simulations. In fact, in the data, the profile of the resonant \ion{Mg}{II} ion is observed to decline much more gently than [\ion{O}{II}], another sign that radiative transfer effects play a role in shaping the observed emission. It is also interesting to note that the ratio of [\ion{O}{II}] to \ion{Mg}{II} is constant at $\approx 3$ for distances of $\approx 20-40$\,kpc from the center, implying that a significant fraction of the observed \ion{Mg}{II} is produced {\it in-situ} and not just entirely scattered outward from the ISM. 

When examining individual detections without resorting to stacks, Dutta et al. found extended [\ion{O}{II}] emission up to distances of $\approx 50$\,kpc around $\approx 1/3$ of the sample of group galaxies. In contrast, extended \ion{Mg}{II} emission is found only around two quasars that reside in groups at $z\approx 1$. IFSs thus provide the first images of metals in galaxy halos from outflows or environmental processes. However, deeper data are required for mapping individual halos in \ion{Mg}{II} around normal galaxies.

Metal emission from the CGM of halos hosting bright quasars has also been reported using IFSs in large samples at $z\gtrsim 3$ \cite{Fossati2021,Guo2020}. In a stack of $\approx 30$ bright quasars at $z=3-4.5$ from the MAGG survey, Fossati et al. \cite{Fossati2021} searched for metal emission in the CGM, detecting \ion{C}{IV} $\lambda 1549$\AA\ and reporting an upper limit (with a hint of detection) for \ion{He}{II} $\lambda 1640$\AA, out to distances of $40-50$\,kpc when reaching surface brightness levels of  $\approx 5\times 10^{-20}$\,\sbunit. These results agree with the analysis performed by Guo et al. \cite{Guo2020} of a stack of 80 quasars using archival data \cite{Borisova2016,ArrigoniBattaia2019}. In this study, \ion{He}{II} and \ion{C}{III}] is also confidently detected extending to $50-60$\,kpc at a surface brightness level of $\approx 10^{-20}$\,\sbunit. By applying photoionization modeling of gas illuminated by an AGN, Guo et al. infer metallicities 
between $0.5 < Z/Z_\odot < 1$, with a hint of radial gradient moving outward. As expected, they also found a decrease in the ionization parameter at progressively larger radii. However, substantial caveats apply to the inference of metallicity via emission lines, particularly where \lya\ and \ion{C}{IV} are used as tracers because of their resonant nature that complicates the modeling (see \cite{Fossati2021}). 
Nevertheless, these metallicities are compatible with what is inferred with absorption lines of background quasars piercing the CGM of foreground AGNs, which is reported to be $\approx 0.25Z_\odot$ out to distances of 200\,kpc \cite{Lau2016}. Thus, absorption and emission measurements jointly support the view of a quasar CGM enriched by outflow events.

\begin{figure}[b]
\centering
\includegraphics[scale=.28]{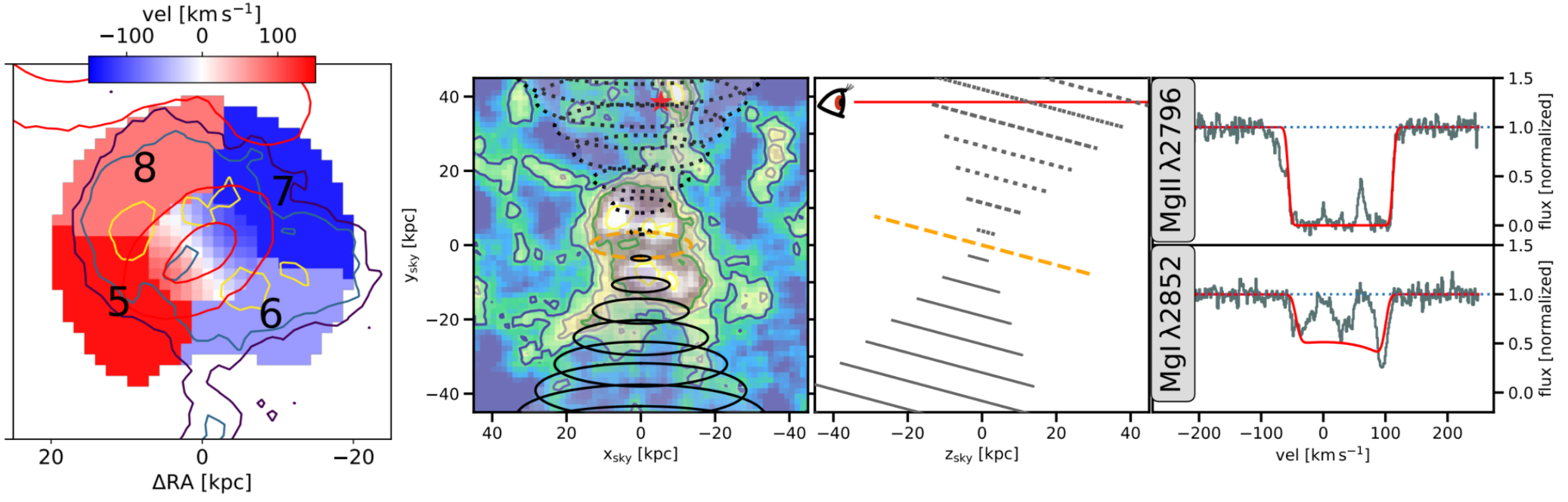}
\caption{Example of the power of combining absorption and emission diagnostics in a $z\approx 0.7$ galaxy for which \ion{Mg}{II} is detected both in MUSE data and against a background sightline. Left: Kinematic map of the ISM and of the CGM, showing how the halo and the disk corotate on scales extending to $\approx 40$\,kpc. Right:  Map of \ion{Mg}{II} emission (first panel) with superimposed a schematic representation of a biconical outflow. The position of the quasar sighline is marked by a red star. A single outflow model repoduces simultaneously the emission and absorption kinematics (third panel). {\it Credits: Zabl J. et al., 2021, MNRAS, 507, 4294. Reproduced with permission.}}
\label{fig:absemmgii}       
\end{figure}

\subsection{Combining emission and absorption diagnostics}\label{sec:uvlemabs}

The combination of observations of the CGM in absorption and emission expands the constraints that can be obtained on the diffuse gas. 
Absorption spectroscopy is a sensitive technique that can probe extremely low column densities, approaching $10^{12}$\,cm$^{-2}$ with the current instruments. It is thus ideal to trace the most diffuse CGM and IGM out to large distances from galaxies. By design, this experiment is quite limited in recovering information on the gas spatial distribution. However, the study of extended sources such as lenses has improved the prospects of spatial mapping (e.g., \cite{Lopez2018}). In contrast, emission measurements provide a powerful technique for mapping the spatial extent and morphology of the gas. However, with few exceptions, this technique is currently limited to the densest portion of the CGM and IGM. Given this complementarity,  absorption and emission can jointly offer improved inference on the underlying physical conditions of the gas, such as temperature, density, and metallicity. 

Beyond a statistical approach in which absorption and emission diagnostics are combined for a class of objects, the literature already contains examples of the power of combining the two techniques in individual systems. The case of a starburst galaxy analyzed by \cite{Hayes2016} and discussed above offers such an example. 
The observed surface brightness of \ion{O}{VI} can be written as
\begin{equation}
    \Sigma_{\ion{O}{VI}} = \tilde{\Lambda} n_e^2 \left(\frac{Z}{Z_\odot}\right)D_{\ion{O}{VI}}
\end{equation}
where $\tilde{\Lambda}$ is the normalised cooling rate at solar metallicity for \ion{O}{VI}, $n_e$ the electron density, and 
$D_{\ion{O}{VI}}$ the thickness of the emitting gas along the line of sight (in units of length).
When the same gas is probed in absorption, the observed column density depends again on the gas density and the thickness of the material via
\begin{equation}
    N_{\ion{O}{IV}} =  n_e D_{\ion{O}{VI}} \left(\frac{n_O}{n_H}\right)\,
\end{equation}
where the last term is the oxygen per proton ratio that is in equilibrium with the \ion{O}{VI}. 
Combined, these two equations can be solved for $n_e$ and $D_{\ion{O}{VI}}$, given observed values for $\Sigma_{\ion{O}{IV}}$ in emission and $N_{\ion{O}{IV}}$ in absorption. Assumptions must be made on the cooling rate (known for a collisionally excited gas at a given temperature) and on the gas metallicity (taken from the galaxy metallicity). For the specific case of the starburst galaxy under exam, Hayes et al. infer $n_e \approx 0.5$\,cm$^{-3}$ and $D_{\ion{O}{VI}}\approx 10$\,pc, but the technique is generally applicable to any systems where absorption and emission can be combined. 

In the era of IFSs, where higher sensitivity in emission is possible, this technique's applicability increases. The study by Zabl et al. \cite{Zabl2021} provides an example of the potential of combined emission and absorption diagnostics. These authors reported the detection of extended \ion{Mg}{II} emission from the halo of a $z\approx 0.7$ galaxy probed by a sightline at a projected impact parameter of 40\,kpc. The \ion{Mg}{II} emission is detected to a surface brightness limit of $\approx 10^{-18}$\,\sbunit\ out to 25\,kpc from the central galaxy and it is particularly intense along the minor axis, which is suggestive of an outflow origin. The quasar sightline, also in the direction of the minor axis, intercepts a strong \ion{Mg}{II} absorber with $\approx 2~$\AA\ equivalent width. The MUSE data in emission trace the halo kinematics, revealing a good correlation of the gas kinematics in the direction of the major axis with the ISM rotation field. Along the minor axis, instead, the \ion{Mg}{II} kinematics in emission are redshifted and blueshifted above and below the galaxy, as expected for a biconical outflow.
The absorption system offers another data point at larger distances, and a single kinematic model of a biconical wind can simultaneously match the absorption and emission signal (Figure~\ref{fig:absemmgii}). 

In addition to the kinematic analysis, the luminosity and morphology of the emitting \ion{Mg}{II} region, together with the absorption signal, provide additional clues on the density distribution of the gas and the emission mechanism. In this particular system, a biconical outflow with emission powered by scattering of the continuum photons struggles to reproduce the density expected from the \ion{Mg}{II} absorption, underestimating it by approximately two orders of magnitude. Moreover, a pure scattering model produces under-luminous halos compared to the observed surface brightness level. This implies that other {\it in situ} mechanisms are necessary to power the CGM emission. Shock models with low velocities ($\approx 100-200$\,km~s$^{-1}$) appear able to provide the missing \ion{Mg}{II} photons. 

Although the joint application of absorption and emission studies in the CGM is in its early days, these examples demonstrate the power that future surveys combining absorption and emission signals will have to analyze the underlying physical properties of the multiphase CGM.

\section{Connecting galaxies, the CGM, and the IGM}
\label{sec:galaxies}

The techniques reviewed in the previous sections (absorption spectroscopy and imaging or IFS in emission) offer complementary views of the multiphase CGM, including constraints on the gas ionization state, temperature, metallicity, density structure, and morphology. Ultimately, the physical information on the gas must be linked to the physical properties of the surrounding galaxies to fulfill the overarching goal of this field: constraining how the baryon cycle occurring within the CGM and IGM influences, shapes, and is shaped by the evolution of galaxies. 

To this end, absorption and emission diagnostics of the halo gas must be combined with diagnostics derived from spectroscopic surveys of galaxies in the same fields. Although the link between the CGM emission and the associated galaxies is often unambiguous, the correspondence between the CGM and the galaxies themselves might not always be trivial. Indeed, the link between galaxies and the absorbing gas can be achieved in several ways. For nearby and particularly extended galaxies, one can search and map the CGM with imaging or spectroscopy, also using multiple background sightlines. In this case, the association is relatively straightforward. However, one should never forget that the distance along the line of sight is not known for gas seen in absorption. Hence, the possibility of intercepting unrelated structures projected in velocity space with the galaxy of interest is always possible. As galaxy clusters on a wide range of masses, including small satellites, one should always bear in mind the presence of additional overlapping systems.  

At progressively increasing distances, the number of available background sightlines per galaxy decreases, and it is - for mere geometric reasons - progressively more challenging to find sightlines at very close separations. Thus, it becomes more complicated to associate individual galaxies with individual absorbers, given the presence of clustered satellites or groups. Assumptions can be made on which galaxy hosts the absorber(s), or one can start associating galaxies and absorbing gas statistically through, e.g., correlation functions or measurements of the covering factor. 

In this Section, we will review some significant examples of studies that link the properties of the CGM and IGM to galaxies, starting with our own Galaxy and moving to progressively higher distances, reaching the end of reionization at $z\approx 5.5$. 

\subsection{The CGM of the Milky Way}

Every extragalactic sightline detected by our instruments has traveled through the CGM of the Milky Way and bears the signatures of the Galaxy's halo. Likewise, halo stars are more local yet powerful background sources to probe the Milky Way CGM. UV spectroscopy, starting with the Far Ultraviolet Spectroscopic Explorer (FUSE) first and then with the STIS and the Cosmic Origins Spectrograph (COS) spectrographs on board HST, has provided a rich dataset to study the CGM of the Milky Way in great detail. Here, we focus on some recent highlights of the dynamic literature on the subject, referring the reader to other comprehensive reviews of the issue \cite{Putman2012}.

\begin{figure}[b]
\centering
\includegraphics[scale=.28]{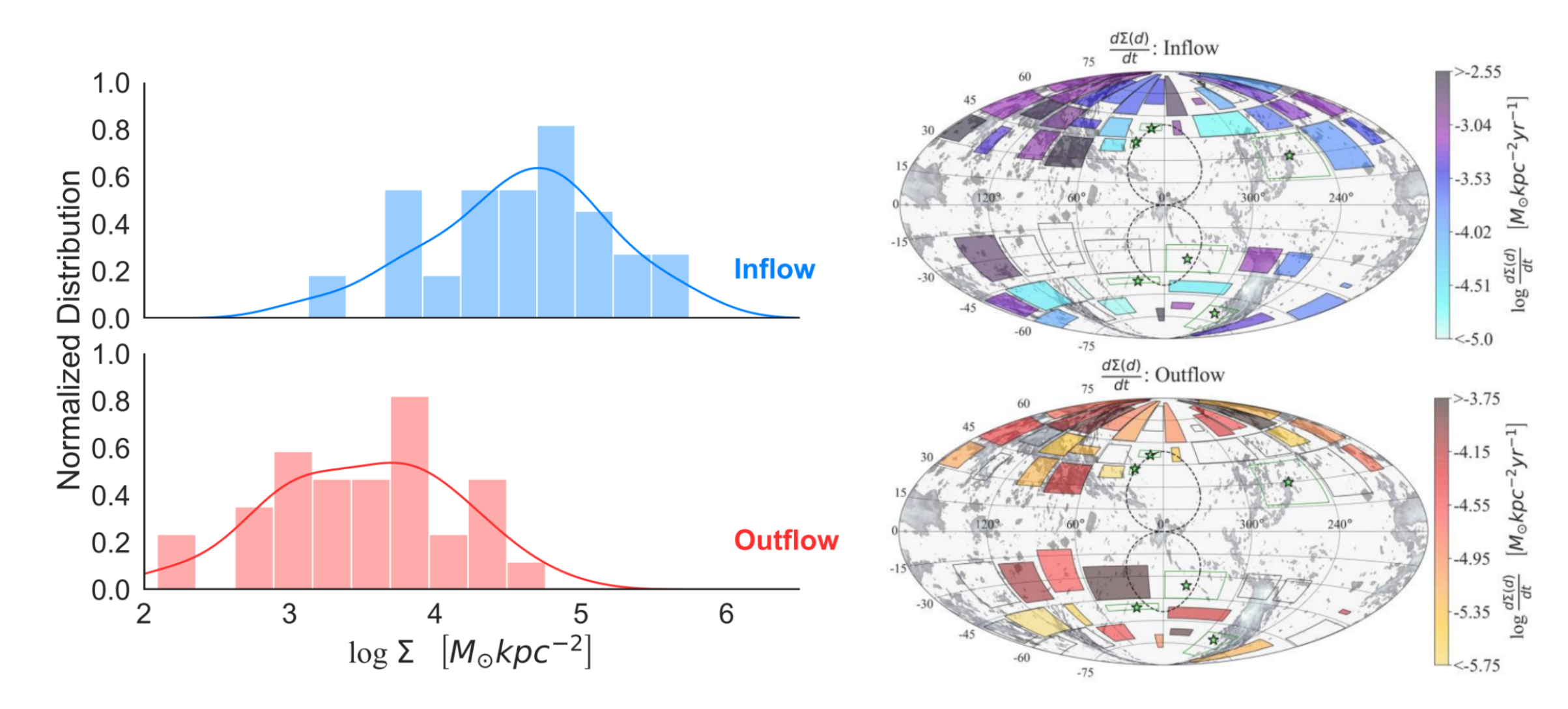}
\caption{Left: Distribution of inflowing (top) and outflowing (bottom) gas surface mass densities inferred in absorption from stacks of spectra. Right: Maps of the local surface mass rate density for inflowing (top) and outflowing (bottom) gas. The Milky Way exhibits a net inflow rate over the outflow rate, with substantial regional variability. {\it Credits: Clark S. et al., 2022, MNRAS, 512, 811. Reproduced with permission.}}
\label{fig:mwcgm}       
\end{figure}

The work by Clark et al. \cite{Clark2022} provides a recent snapshot into the flows of cool gas in and out of the Milky Way, leveraging all-sky observations from the COS archive. By coadding sightlines at short projected separations, these data become sensitive to low-column-density
halo gas, also suppressing the contamination of unrelated extragalactic absorption. By identifying blue or redshifted components centered on the local standard of rest, statistics on the incidence of inflows and outflows can be compiled (Figure~\ref{fig:mwcgm}). However, observations probe only the column density, and some assumptions must be made to transform this observable into a mass surface density.

First, using ions as tracers, one must infer the total column density along the line of sight. This can be achieved by adding the column density of ions in the dominant ionization stages, such as \ion{Si}{II}, \ion{Si}{III} and \ion{Si}{IV}, thus circumventing the need for ionization corrections (Section~\ref{sec:absorption}). Once the element column density (e.g., silicon) is known, the total hydrogen column density is inferred as
\begin{equation}
    N_H=\frac{N_{Si}}{(Z/Z_\odot)(\textrm{Si/H})\zeta}\,,
\end{equation}
where $\zeta$ accounts for dust depletion in the Milky Way halo, and $(\textrm{Si/H})$ is the silicon solar abundance. 
Based on the results of the literature, Clark et al. assume a metallicity of 20 percent solar for the inflows and 50 percent for the outflows. 
The mass surface density is then
\begin{equation}
    \Sigma_H= 1.4 m_H N_{H}\Omega d^2\,,
\end{equation}
where $1.4m_H$ is the gas-particle mass corrected for helium, $\Omega$ is the solid angle of the region within which sightlines are binned, and $d$ is the distance at which the absorbing gas is located. Differently from any other galaxy, the distance at which the absorbing gas lies in the Milky Way CGM can be constrained by combining information on the signal detected towards extragalactic sources and halo stars. Most of the absorption is thus constrained to be at $\approx 5-15$\,kpc \cite{Lehner2011}, given that the rate of detections in stars and AGNs is comparable. 

The results of this analysis are shown in Figure~\ref{fig:mwcgm}. A net inflow characterizes the Milky Way, since the surface mass density of the inflowing gas exceeds by $\approx 1$\,dex that of the outflowing gas. However, the outflowing mass rate, obtained by multiplying the mass surface density by the mean outflow velocity, is not negligible at $\log\frac{\dot\Sigma}{\textrm{M$_\odot$~kpc$^{-2}$~yr$^{-1}$}}=-4.79\pm0.14$. This value can be compared with the inflowing mass rate of $\log\frac{\dot\Sigma}{\textrm{M$_\odot$~kpc$^{-2}$~yr$^{-1}$}}=-3.55\pm0.08$ \cite{Clark2022}. When studying spatial variation in the sky, inflows and outflows are characterized by a sizeable region-by-region variation of $\approx 2$\,dex for inflows and $\approx 1$\,dex for outflow. 

High-velocity clouds (HVCs) observed with 21cm emission are a prime source of infalling material satisfying the inferred accretion rates. However, the largest group (Complex C) at $\lesssim 10$\,kpc can account for only $\approx 1/10$ of the inflow rate needed to maintain the current star formation activity of the Milky Way of $\approx 1$\,M$_\odot$~yr$^{-1}$. However, a significant fraction of ionized clouds are observed in the spectra of UV stars and AGNs \cite{Lehner2011,Lehner2012}. These ionized clouds trace the outskirts of neutral HVCs and regions where \ion{H}{I} emission is absent. As UV spectroscopy is sensitive to small neutral fractions, these ionized HVC contain a significant mass budget, at least comparable if not above the 21cm HVC population. Ionized HVCs with low radial velocities $\lesssim 170$\,km~s$^{-1}$, which are seen in stars and AGNs and therefore are located at $\approx 5-15$\,kpc, contribute to a factor up to 10 more mass than complex C, and can satisfy the accretion rate needed to sustain the present-day Milky Way activity. A second population of ionized HVCs, with radial velocities $\gtrsim 170$\,km~s$^{-1}$ that lie at more considerable distances, will provide a second wave of inflowing material in the future.

Our Galaxy also allows studying the effects of environmental interactions on the CGM (see Section~\ref{sec:environment}). The Magellanic system \cite{Fox2014}, a coherent structure composed of the Magellanic Stream, the Magellanic Bridge, and the Leading Arm, is particularly noteworthy. This system arises from the dynamic interactions between the two Magellanic Clouds, which displaced the satellites' gas inside the Milky Way halo. The Magellanic system is observed across about one-quarter of the sky via 21cm emission and against a large set of extragalactic sightlines in absorption.  
Fox et al. \cite{Fox2014} conducted a detailed analysis of the ionization structure of the Magellanic system following a method similar to the one described above for the study of HVCs. Their analysis indicates that the ionized phase probed by UV absorption carries a substantial mass fraction, $\approx 3$ times that of the neutral phase. Collectively, the gas in the Magellanic system is $\approx 2\times 10^{9}$\,M$_\odot$ (for an assumed distance of $\approx 50$\,kpc), which is above the \ion{H}{I} content of the Magellanic Clouds. Interactions have thus displaced the majority of their gas content towards the Milky Way. The Magellanic system is expected to reach the galaxy in about $0.5-1$\,Gyr, at which point the Milky Way SFR could increase by $3-5$ times.
However, this is a generous estimate since the system is not expected to collapse monolithically since a fraction of the cool gas will interact with the hot corona during its journey.

In addition to the material removed from the Magellanic Clouds ISM, the Large Magellanic Cloud (LMC) is sufficiently massive, $\approx 10^{11}$\,M$_\odot$, to host gas at $T\approx 10^{5.5}$\,K out to $\approx 100$\,kpc for an isolated galaxy. UV spectroscopy reveals ionized gas traced by \ion{O}{VI}, as well as \ion{C}{IV} and \ion{Si}{IV}, in the direction of the LMC, with a radial profile that declines with increasing projected distance from the LMC, at least to $\approx 35$\,kpc. The presence of these ions, and especially \ion{O}{VI}, provides direct evidence of a warm corona that participates in the larger Magellanic ecosystem \cite{Krishnarao2022}. 

A further additional feature of the Milky Way CGM, which might be typical in many distant galaxies, is the presence of the so-called Fermi bubble, a biconical gamma-ray emitting structure originating from the galactic center. Initially discovered by the Fermi satellite, this extended double-cone emission has been subsequently observed at multiple wavelengths, including in X-rays \cite{Predehl2020}, and studied through UV absorption lines \cite{Fox2015,Bordoloi2017,Ashley2020}. The analysis of absorption line kinematics reveals that the Fermi bubble traces gas with high velocities over what is expected from co-rotating gas in the Milky Way disk or halo. This signal is thus consistent with an outflow origin. The wind age is inferred to be short, $\lesssim 10$\,Myr, making an outflow from the central AGN the most likely candidate for the driving mechanism. A wind powered by star formation would last longer, $\approx 30-100$\,Myr.

On the one hand, these studies show the detailed information that can be inferred from the CGM of our Galaxy, combining multiple diagnostics in emission and a large sample of background sightlines. On the other hand, although peculiar to just one galaxy, the study of the Milky Way provides a cautionary tale for interpreting individual sightlines piercing the halos of single galaxies at more considerable distances. The local signal probed by a sightline is expected to be highly varied due to the many gaseous substructures that halos can host. Thus, large samples are required to derive meaningful and robust conclusions on the average nature of the CGM.   

\begin{figure}[b]
\sidecaption
\includegraphics[scale=.27]{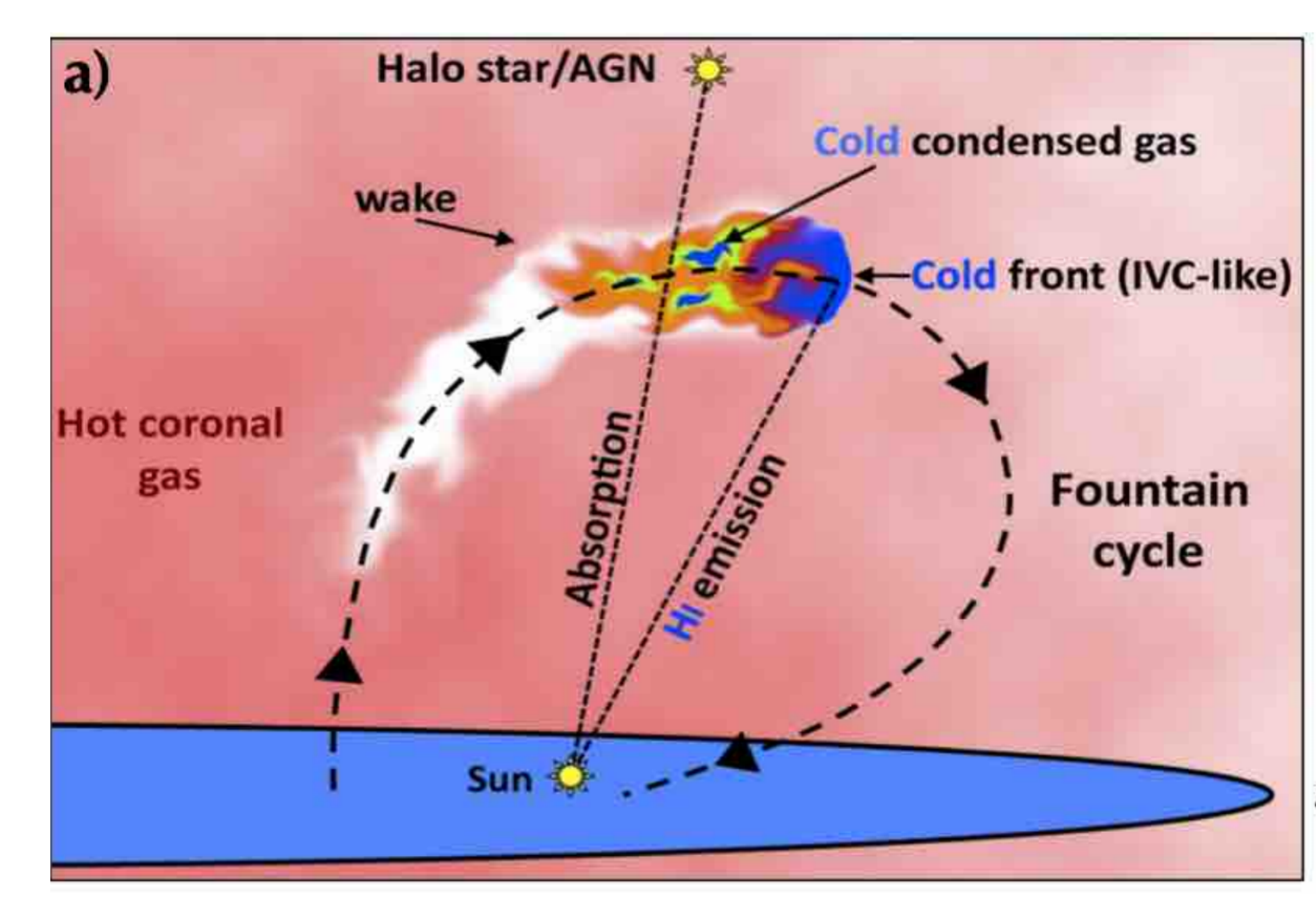}
\caption{Schematic representation of a galactic fountain model. Cold gas is lifted into the halo by supernova feedback and interacts with the hot medium, exchanging angular momentum and slowing down before falling back onto the disk. The mixing of cold enriched and hot metal-poor phases induces a net cooling from the halo gas, triggering a supernova-driven inflow. {\it Credits: Fraternali F. et al., 2013, ApJL, 764, L21. Reproduced with permission.}}
\label{fig:galfountain}       
\end{figure}

\subsection{The disk-halo transition of nearby galaxies}

In external nearby galaxies, when remaining at distances characteristics of the Local Volume ($\approx 11$\,Mpc), it has been possible to explore, through a combination of mainly emission diagnostics, the properties of the inner and denser CGM at the interface with the ISM. This region, also termed the disk-halo transition, has been the subject of extensive and detailed kinematic analysis, mainly using optical emission lines (e.g., H$\alpha$,[\ion{N}{II}],[\ion{Si}{II}]) and 21cm data. 

For example, using IFS spectroscopy with the Wisconsin-Indiana-Yale-NOIRLab (WIYN) Observatory, Heald et al. \cite{Heald2006,Heald2007} have mapped the kinematic of the CGM above and below the disk of edge-on nearby galaxies, reconstructing the vertical velocity structure and its gradient. These observations reveal a lagging halo, whereby the rotational velocity decreases with increasing distance from the disk with a mild slope of $\approx 15-30$\,km~s$^{-1}$~kpc$^{2}$, signaling a progressive transition from a coherent rotation in the disk to a more static halo.    
This kinematics information, coupled with similar information derived from 21cm datacubes and data across the electromagnetic spectrum, provides a pivotal diagnostic to test models for material circulation from the disk to the halo and back.

Several models can be proposed for the disk-halo interface (see a dedicated review in \cite{Fraternali2017}). The first and simplest model assumes gas in hydrostatic equilibrium between a disk and a hot halo, which would yield a temperature too high for the gas observed in a neutral phase. A model in which all the extraplanar gas is infalling is irrealistic, as the implied inflow rates would be at least an order of magnitude higher than the present-day SFRs of nearby galaxies. This leaves as a viable model the galactic fountain scenario (Figure~\ref{fig:galfountain}) in which gas is ejected from the star-forming disk into the halo before returning to the disk. 
The simplest incarnation of this model is a ballistic fountain, in which gas is shot from the disk in a cool phase and travels under the effect of the halo potential before raining back on a parabolic trajectory. 
As this gas conserves angular momentum, the rotational velocity of the extraplanar gas is systematically much higher than that observed in the data, implying that this simple model constitutes an acceptable starting point, but is incomplete.

\begin{figure}[b]
\includegraphics[scale=.36]{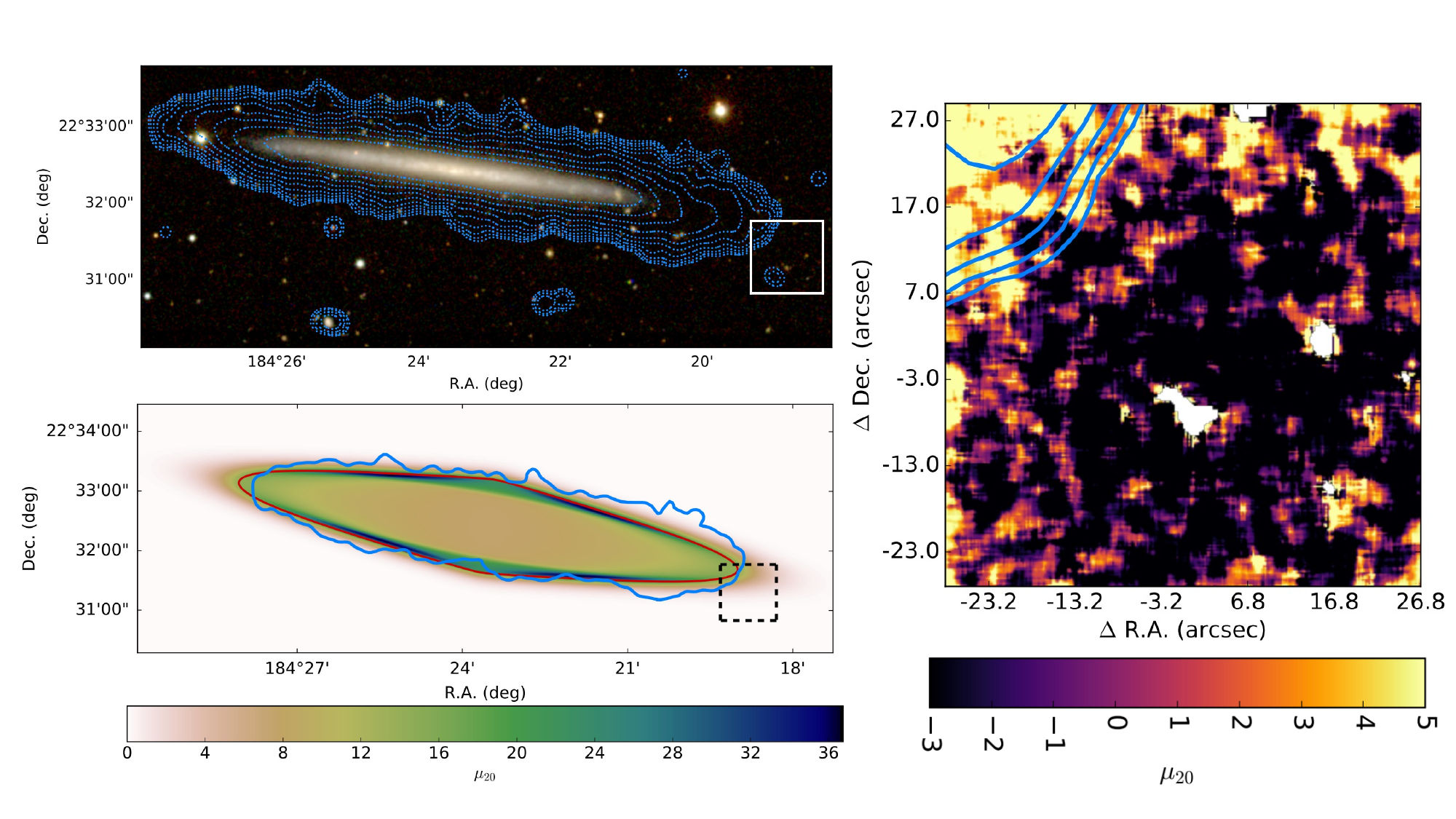}
\caption{Left: Image of the starlight from the edge-on UGC 7321
galaxy (top panel), also showing contours from 21cm measurements. The position of the MUSE pointing is shown in white. Radiative transfer calculations for the \ha\ surface brightness (bottom panel) predict a brightening of the \ha\ flux at the ionization front. Right: The observed \ha\ surface brightness map shows emission overlapping with the radiation front in the top left corner. In these panels, the line surface brightness is expressed in units of $10^{-20}~\rm erg~s^{-1}~cm^{-2}~arcsec^{-2}$. {\it Credits: Fumagalli M. et al., 2017, MNRAS, 467, 4802. Reproduced with permission.}}
\label{fig:uvbmuse}       
\end{figure}

A more detailed treatment of the disk-halo interface is possible with high-resolution numerical simulations that follow the hydrodynamic and thermodynamics of the cool gas ejected by supernovae in the halo. The detailed treatment of the small-scale physics at work in this complex interface is non-trivial also numerically. However, the key results from the extensive literature on the subject are as follows. The gas is lifted as cold from the disk but starts mixing with the hot corona once it reaches the halo. In the process, the angular momentum is transferred from the cold gas to the halo; hence, the extraplanar cool gas slows down. This critical element brings the model and the observed kinematics into agreement, solving the acute shortcoming of the ballistic model. 
As the generally enriched, cold gas mixes with the hotter, more metal-poor halo gas, 
condensation from the corona starts. With time, the corona rains back on the disk, transferring mass from the corona to the disk. In short, the cold gas shares angular momentum with the hot corona, and mass is transferred back from the corona to a cold phase that reaches the disk. 
In some simulations, by mass, cooling from the hot phase dominates over mixing. The analysis of the total gas in the system reveals a steady increase of the cold gas fraction with time, implying that clouds that rain back onto the disk effectively provide a channel for fresh gas accretion at the expense of the coronal gas. The fountain model thus indicates a fountain-driven or supernova-driven accretion \cite{Fraternali2017}.

In the era of sensitive IFSs at 8m-class telescopes, ultra-deep spectroscopic mapping of the disk-halo transition provides a novel avenue for exploring two related topics: where the \ion{H}{I} disk ends and what the amplitude of the UVB is \cite{Adams2011,Fumagalli2017}.
Radiative transfer calculations of atomic hydrogen systems (clouds or disks, \cite{Fumagalli2017,Sykes2019,Dove1994}) embedded in an ionizing radiation field show that the location of the ionization front, where the gas transitions from primarily neutral to primarily ionized, is a function of the photoionization rate. For a fixed underlying density distribution, observations of the location of the edge of the \ion{H}{I} disk would thus constrain the amplitude of the UVB and, conversely, knowing $\Gamma_{H^0}$, one can infer the underlying physical properties of the disk. Both quantities are poorly determined; therefore, ambiguity remains when using only the \ion{H}{I} data. 
Further information can be provided by recombination radiation. The surface brightness of fluorescent radiation peaks at the ionization front, with an intensity that is directly proportional to the amplitude of the UVB (see figures 10 and 11 in \cite{Sykes2019}).

High-sensitivity IFS observations that target the edge of the \ion{H}{I} disk can, jointly with 21cm data, constrain the amplitude of the UVB through detection of, e.g., H$\alpha$. This experiment has been attempted in two highly inclined galaxies \cite{Adams2011} using the VIRUS-P spectrograph on the McDonald 2.7m telescope. However, no detection was made, which resulted in a limit on the UVB of $\Gamma_{HI}\lesssim  2\times 10^{-14}$\,s$^{-1}$. Using MUSE, subsequent follow-up observations \cite{Fumagalli2017} have re-evaluated the previous limit on the photoionization rate with an improved geometric model for the disk, achieving a detection of \ha\ from the disk edge (Figure~\ref{fig:uvbmuse}). Through radiative transfer calculations, this detection was converted into an estimate of the UVB of $6\times 10^{-14}$\,s$^{-1}$. However, the very low surface brightness makes this type of analysis difficult. Hence, possible systematics (e.g., sky subtraction) could be present. More ultradeep observations of recombination radiation from nearby galaxies are required to constrain the UVB via emission measurements firmly.
Contributions to this field are expected from dedicated facilities, such as the Dragonfly Spectral Line Mapper \cite{Chen2022}.

\subsection{The CGM of low-redshift ($z\lesssim 0.5$) galaxies}

In low-redshift galaxies but at more considerable cosmological distances, our ability to image the CGM directly in optical rest-frame lines, such as \ha\ emission, is reduced by the cosmological dimming that affects surface brightness measurements. UV spectroscopy becomes the leading diagnostics for multiphase CGM at these redshifts. Angular scales start to shrink; consequently, the number of sightlines available reduces to an average of one per galaxy. A further difficulty is represented by the fact that hydrogen and the primary metal transitions lie at observed UV wavelengths, making space observations necessary.
A significant technological development in this field has been the installation of COS on board HST in 2009. COS is a medium-resolution spectrograph that operates in the wavelength range $815-3200~$\AA\ and is characterized by exquisite UV sensitivity, thanks to its simple optical design that reduces the number of reflecting surfaces encountered by the light path. 
Therefore, COS has been instrumental in shaping our view of the low-redshift cool/warm CGM. 

Several surveys have taken advantage over the years of the outstanding capabilities of COS, targeting a wide range of galaxy types, including AGNs \cite{Berg2018}, dwarfs \cite{Bordoloi2014}, and starbursts \cite{Heckman2017}. Among the first and better-know programs is the COS-Halo survey \cite{Tumlinson2013} that targeted 39 $z\approx 0.5$ quasars piercing within 150\,kpc the halo of 44 $z\approx 0.15-0.35$ galaxies with stellar masses $\log (M_*/M_\odot)\approx 9.5-11.5$, including both early-type and late-type systems. When considering \ion{H}{I}, only minor differences are found between star-forming or passive galaxies, as both populations are characterized by a comparable strength and distribution of the absorbing gas, with covering factors of $100$ percent for late-type galaxies and $75$ percent for early-type galaxies. 
When comparing the gas line-of-sight velocity relative to systemic, the majority of the absorption is found at $\lesssim 200$\,km~s$^{-1}$ implying that most of the gas probed by \ion{H}{I} is bound and fills the halo of these $L_*$ galaxies.

Conversely, when looking at the warm-ionized gas traced by \ion{O}{VI}, 
a dichotomy becomes apparent \cite{Tumlinson2011}, with star-forming galaxies exhibiting a near-ubiquitous presence of highly-ionized oxygen in their halos ($27/30$ galaxies show a detection) that is not found in passive systems (only $4/12$ galaxies are detected). These observations have spurred an intense debate in the literature on multiple uncertain aspects. First, while it is likely that this is collisionally-ionized and dense gas around a temperature of $\approx 10^5-10^6$\,K, a cooler ($<10^5$\,K) diffuse and photoionized phase could still host \ion{O}{VI} \cite{Stern2016,Tumlinson2011}. The exact origin of \ion{O}{VI} in star-forming galaxies is also debated. Although the evidence that \ion{O}{VI} correlates with the specific SFR would imply a causal connection between the presence of a halo rich in ionized oxygen, for instance, through a galactic outflow origin \cite{Stinson2012}, the virial temperature of the late-type galaxies is more favorable for \ion{O}{VI} ionization than of early-type galaxies, suggesting that the correlation may be more indirect and of cosmological reason \cite{Oppenheimer2016}. Finally, the fate of this ionized phase once galaxies quench remains unclear, with possible mechanisms acting to deplete the halo of \ion{O}{VI} (tidal or ram-pressure stripping in groups, accretion onto the central galaxy, or change in temperature that makes the fifth ionization stage of oxygen unfavorable \cite{Tumlinson2011}). 
X-ray observations can add further insight into the different CGM of passive and active galaxies \cite{Chadayammuri2022}.

\begin{figure}[b]
\sidecaption
\includegraphics[scale=.22]{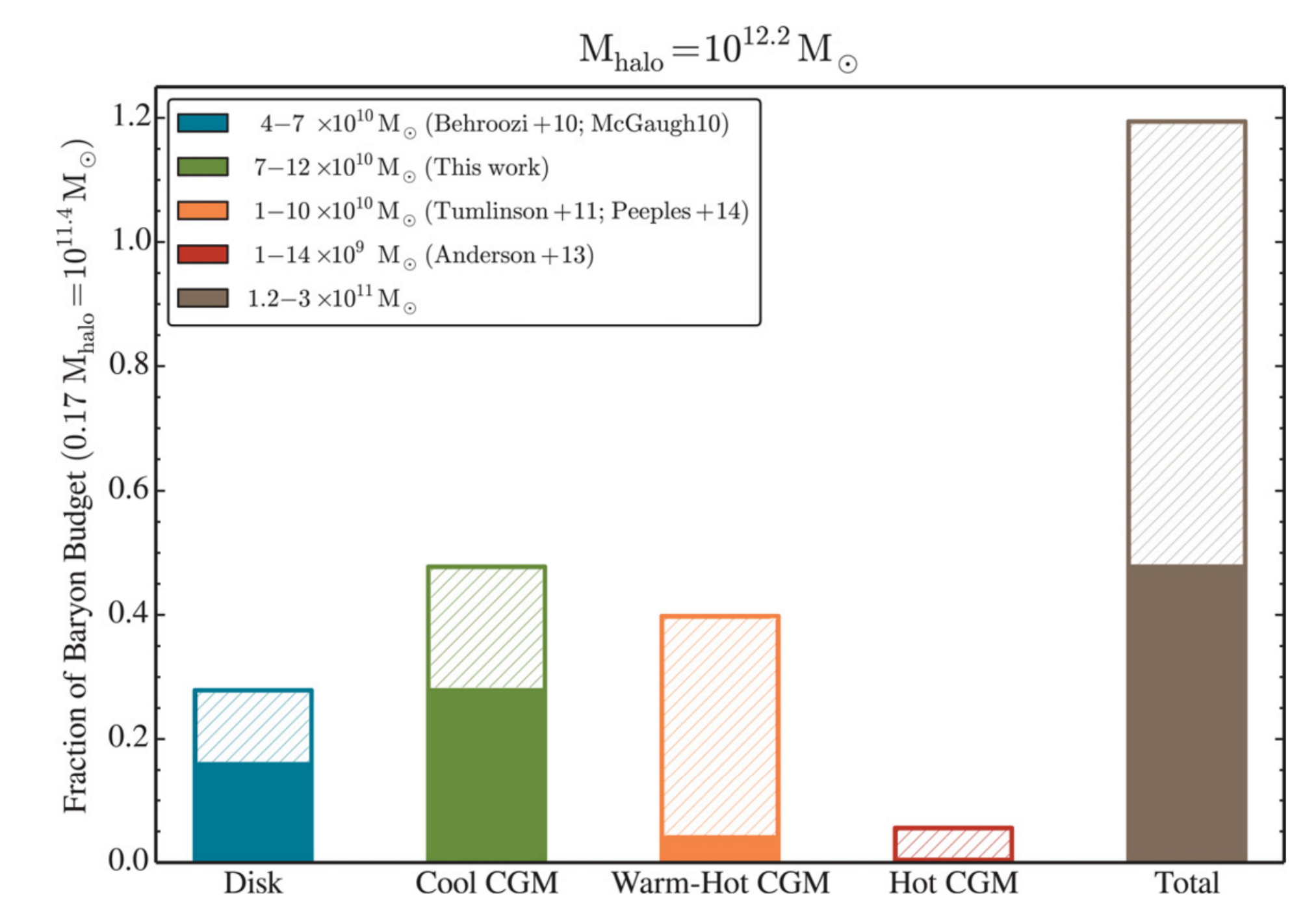}
\caption{The fraction of baryons in different components for a halo with mass $10^{12.2}$\,M$_\odot$ at $z\approx 0.2$. Solid bars express lower limits, while the hatched bars indicate possible additional contributions. The cool and warm CGM could harbor half of the total baryons. Recent X-ray measurements have revised the estimates for the hot CGM upward compared to what is in the figure (see text for detail), making the CGM a significant repository of baryons compared to the disk.  {\it Credits: Werk J.~K. et al., 2014, ApJ, 792, 8. Reproduced with permission.}}
\label{fig:halombudget}       
\end{figure}

In tandem with the highly-ionized \ion{O}{VI} and the neutral \ion{H}{I} phases, follow-up absorption spectroscopy of the COS-Halo sightlines also reveals the presence of a cool, metal-enriched and ionized CGM \cite{Werk2014,Prochaska2017}. Through photoionization corrections, the gas is inferred to have a median metallicity of $\approx 1/3$ solar, and a wide distribution ranging between $\approx 1/50$\,Z$_\odot$ and $\gtrsim 3$\,Z$_\odot$. Photoionization modeling also allows constraining the mass budget of the CGM in $L_*$ galaxies at these redshifts (Figure~\ref{fig:halombudget}, \cite{Werk2014}). For a $10^{12}$\,M$_\odot$ halo with $\approx 20$ percent baryon fraction, the cool CGM could account for $\approx 40$ percent of the baryons, exceeding what is locked inside the disk. 
The warm component traced by \ion{O}{IV} adds a further $1-10\times 10^{10}$\,M$_\odot$ to the mass budget, where the extensive range reflects the fact that \ion{O}{VI} is a sub-dominant ion and, not only substantial ionization corrections are needed, but assumptions about the ionization mechanism, density, and temperature must be made \cite{Tumlinson2011}. 
This mass budget is not free of significant uncertainties, for instance, in the assumptions of the ionizing radiation field \cite{Keeney2017}.

Further constraints on the hot gas arise from observations of absorption in X-ray spectra, such as the high-significance detection in a stack of three sightlines using {\it XMM-Newton} and {\it Chandra} data \cite{Nicastro2023}, or in individual spectra \cite{Mathur2023}. By detecting \ion{O}{VII}, these authors were able not only to confirm the presence of a hot medium extending to $\approx 100$\,kpc from the central galaxy but also to infer that the hot halo harbors a very significant fraction of baryons, with masses of the order of $10^{11}$\,M$_\odot$ for an L$_*$ galaxy, once assumptions on the gas metallicity and size of the halo are made. Although it is difficult to apply this technique to large samples due to the current capabilities of X-ray satellites, these studies demonstrate the significant promise of future high-energy missions in pinning down the properties of the multiphase CGM at high temperatures.

\subsection{The CGM of intermediate-redshift ($z\approx 0.5-2$) galaxies}

\begin{figure}[b]
\includegraphics[scale=.23]{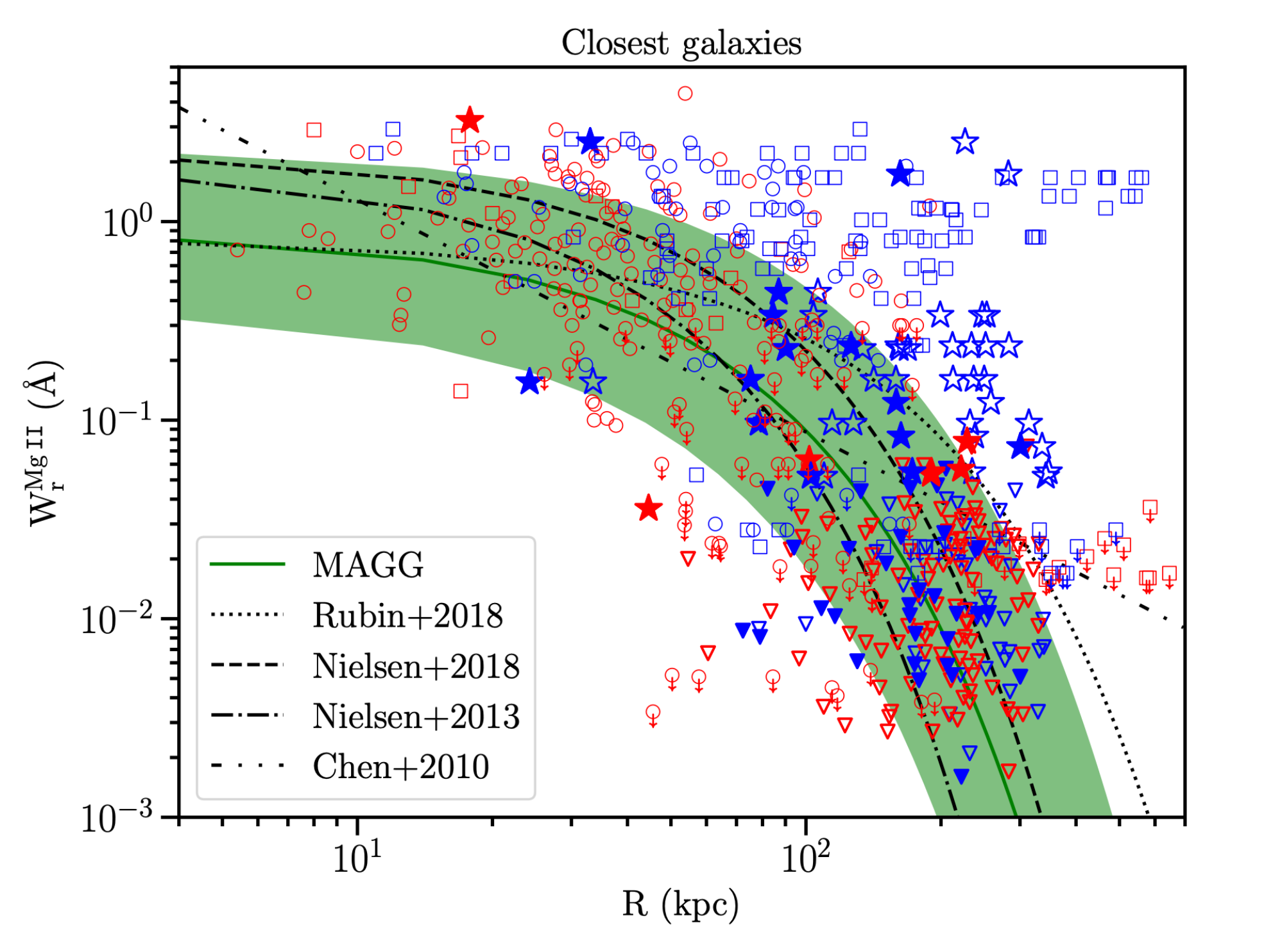}
\sidecaption
\caption{Compilation of \ion{Mg}{II} equivalent width measurements (data points) as a function of distance from galaxies in a large compilation of literature samples (see figure 12 in \cite{Dutta2020} for details). Also shown (lines) are scaling relations derived in various studies. The \ion{Mg}{II} equivalent width declines radially and drops around $\approx 100$\,kpc, but a large scatter is present due to the patchy nature of the CGM and the effects of the environment. {\it Credits: Dutta R. et al., 2020, MNRAS, 499, 5022. Reproduced with permission.}}
\label{fig:mgiitrend}       
\end{figure}

Above redshift $z\approx 0.5$ and up to $\approx 2$, the \ion{Mg}{II} doublet enters the optical region of the spectrum. In this redshift range, a rich literature exists on connecting the cool CGM traced by this ion and galaxies detected in spectroscopy using multi-object spectrographs (MOS) \cite{Chen2010,Bordoloi2011,Weiner2009,Rubin2014,Lan2014} and more recently IFSs \cite{Schroetter2019,Dutta2020,Dutta2021,Qu2023}.
Thanks to these comprehensive datasets, several trends have been studied in the literature, although not without some discrepancies. 

As traced by \ion{Mg}{II}, the cool CGM appears ubiquitous around galaxies in this redshift range \cite{Chen2010} and fills halos up to $\approx 300-400$\,kpc \cite{Dutta2020}. The absorption strength declines as a function of radius (Figure~\ref{fig:mgiitrend}), a trend confirmed by all studies, but with a large scatter in the mean relation due to the patchy nature of the multiphase CGM and the effects of the environment (see Section~\ref{sec:environment}). More luminous (massive) galaxies are characterized by a more extended cool halo \cite{Chen2010,Nielsen2013}, which can be modeled with a power-law function of the form $R\propto L^b$. 
Compared to the cool gas, the more highly-ionized phase traced by \ion{C}{IV} appears to be more extended, with a covering fraction twice as high at a given distance \cite{Dutta2021} (but see \cite{Schroetter2021}).

This trend, combined with the decline in equivalent width with radius, provides a natural way to define the characteristic size of the CGM as a function of mass. Wilde et al. \cite{Wilde2023} present a proof of concept of the possibility of formalizing a CGM size, although at a lower redshift than considered in this section ($z\lesssim 0.5$). With \ion{H}{I} absorption probing the halos of a sample of $\approx 7,000$ galaxies, these authors identify a characteristic radius $R_{CGM}$ that defines the transition from the CGM to the IGM by modeling the signal in the absorber–galaxy cross-correlation function adding a Gaussian profile at small radii (see also \cite{Wilde2021}).  
In the mass interval $10^{8}\lesssim M_*/M_\odot \lesssim 10^{10.5}$, Wilde et al. find $R_{CGM}\approx 2R_{vir}$.
This is comparable to the splashback radius derived from numerical simulations and indicates a cool CGM that extends beyond the virial radius. 

When exploring trends between the galaxy's physical properties and the absorbing gas, correlations with the strength of the \ion{Mg}{II} absorption and the SFR and stellar mass have been found \cite{Chen2010,Nielsen2013}, where stellar mass might be the critical parameter driving the correlation \cite{Bordoloi2011,Dutta2020} of galaxies on the main sequence. Although not seen by all authors \cite{Chen2010,Nielsen2013,Dutta2020}, possibly due to the properties of the individual samples under analysis, large compilations of both active and passive galaxies \cite{Bordoloi2011,Lan2014} show a marked dependence of the \ion{Mg}{II} absorption on the galaxy activity, with blue and star-forming galaxies showing a $2-10$ times higher covering factor within 50 kpc than passive ones \cite{Lan2014}. 
When analyzing these trends as a function of redshift, between $\approx 0.3-1.5$, there appears to be no or only a weak time evolution of the cool metal-enriched CGM \cite{Chen2010,Dutta2020,Lan2014}.

Kinematic analysis can provide further insight into the nature of the cool gas in the context of the baryon cycle, primarily through the down-the-barrel technique, which measures the motion of the gas in front of a galaxy relative to the systematic velocity.
Blue-shifted \ion{Mg}{II} absorption, indicative of outflows, is ubiquitous in star-forming galaxies around $z\approx 1$ across a wide range of stellar mass and SFR \cite{Weiner2009,Rubin2014}.
In a few occasions, totaling $\approx 5$ percent of the entire samples, red-shifted \ion{Mg}{II} absorption is also detected \cite{Rubin2012,Martin2012}, providing a rare direct detection of cool and metal-enriched gas that is accreting onto galaxies.

\begin{figure}[b]
\includegraphics[scale=.3]{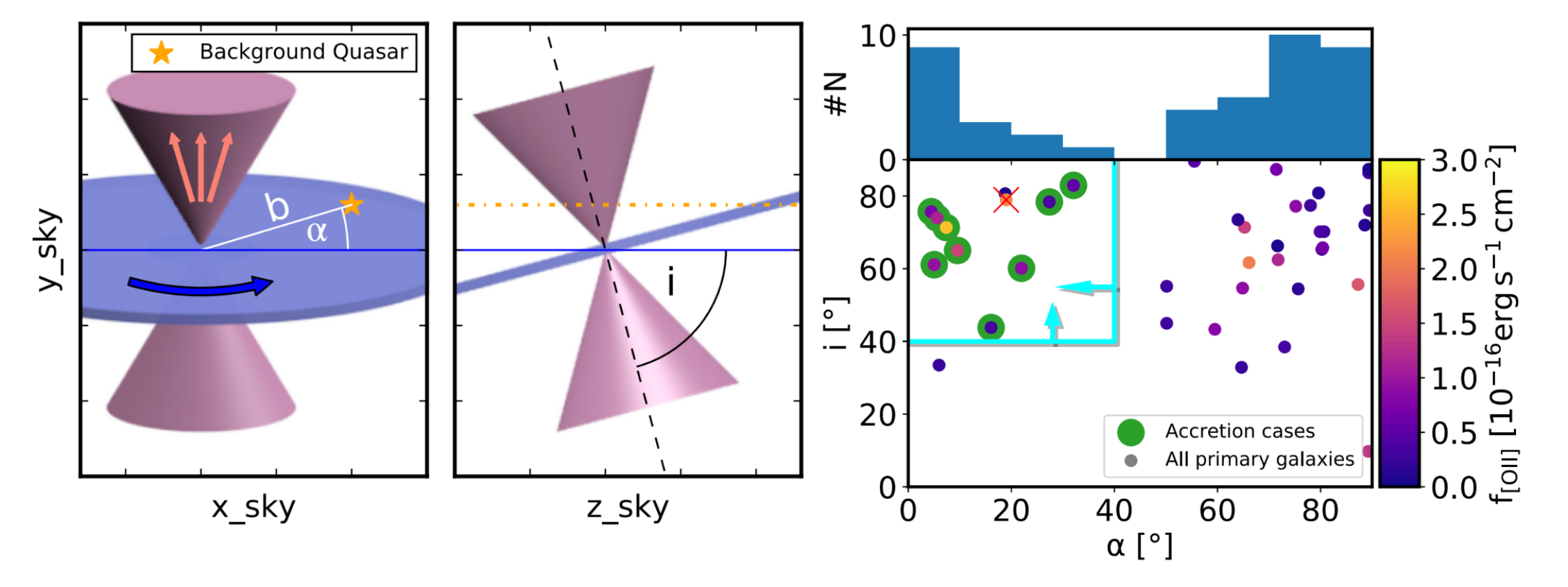}
\caption{Angular dependence of the \ion{Mg}{II} absorption relative to the galaxy disk. Left: A model of an extended disk that corotates with the galaxy in the presence of a bipolar outflow. The angle $\alpha$ specifies if the sightline is close to the minor or major axis of the galaxy, and the angle $i$ describes the direction relative to the rotation axis. Right: The strong \ion{Mg}{II} absorbers identified at close separations from galaxies in the MEGAFLOW survey cluster preferentially along the minor or major axis. {\it Credits: Zabl J. et al., 2019, MNRAS, 485, 1961. Reproduced with permission.}}
\label{fig:mgiiangle}       
\end{figure}

Geometrical considerations further characterize the origin of the absorbing gas. For face-on galaxies, up to $\approx 30\,\deg$ of inclination, the detection rate of outflows in absorption approaches 90 percent. At the same time, it is less than 50 percent in edge-on systems with an inclination of $\gtrsim 50\,\deg$, implying that a significant fraction of the absorbing gas originates from biconical outflows, common in star-forming galaxies at these redshifts \cite{Rubin2014}. A recent detection of \ion{Mg}{II} emission, suggestive of a double-cone geometry, strengthens this argument \cite{Guo2023}.  
Orientation can also be used with background sightlines to add constraints on the nature of the absorption, albeit more indirectly, as the kinematic information of the velocity relative to the galaxy redshift is lost in this experiment. Several authors have reported an azimuthal dependence of the incidence of \ion{Mg}{II} absorption. 

For instance, Bordoloi et al. \cite{Bordoloi2011} show that, within 50\,kpc of star-forming galaxies, \ion{Mg}{II} absorption is stronger on average for background sightlines near the minor axis of a disk than near the plane of the disk, suggesting a biconical outflow that can be more easily intersected in the former than in the latter case. Similar findings have been reported by the MEGAFLOW survey conducted with MUSE when selecting high equivalent width \ion{Mg}{II} absorbers at close separation from the central galaxy \cite{Schroetter2019}: \ion{Mg}{II} is not isotropically distributed, but is characterized by a bimodal distribution in the azimuthal angle defined by the sightline and the galaxy's major axis, with an excess of strong absorbers found in the proximity of the minor axis. At more considerable distances from the central galaxy and in weaker absorbers, this signal vanishes \cite{Dutta2020}.
A second population, located instead in the direction of the galaxy plane, is identified more naturally with cool gas in an extended, possibly corotating disk that contributes to the accretion onto the galaxy (Figure~\ref{fig:mgiiangle}, \cite{Zabl2019}). However, this interpretation is less straightforward than that established by direct kinematic measurements, such as in \cite{Rubin2012} and \cite{Rubin2014}.

The rich data set collected on the cool CGM at $z\approx 1$ in connection with the associated galaxies paints a coherent picture of the distribution, kinematics, and physical origin of the gas probed in absorption. 

\subsection{The CGM of galaxies at cosmic noon}

When approaching cosmic noon, around $z\approx 2-3$, detecting galaxies becomes increasingly challenging. Samples have been generally limited in mass and biased towards blue, star-forming galaxies that are more easily seen at rest-frame optical wavelengths. Before the advent of IFSs, the primary populations selected for the study of the CGM were LBGs, first detected using the dropout technique in imaging surveys, and then studied with follow-up spectroscopy using MOS. 
However, since the deployment of MUSE and KCWI, the need for pre-selection has been removed, and untargeted searches of lower-mass star-forming galaxies are now possible through emission line detections. Next, we review our understanding of the CGM of these two populations. 

\begin{figure}[b]
\centering
\includegraphics[scale=.33]{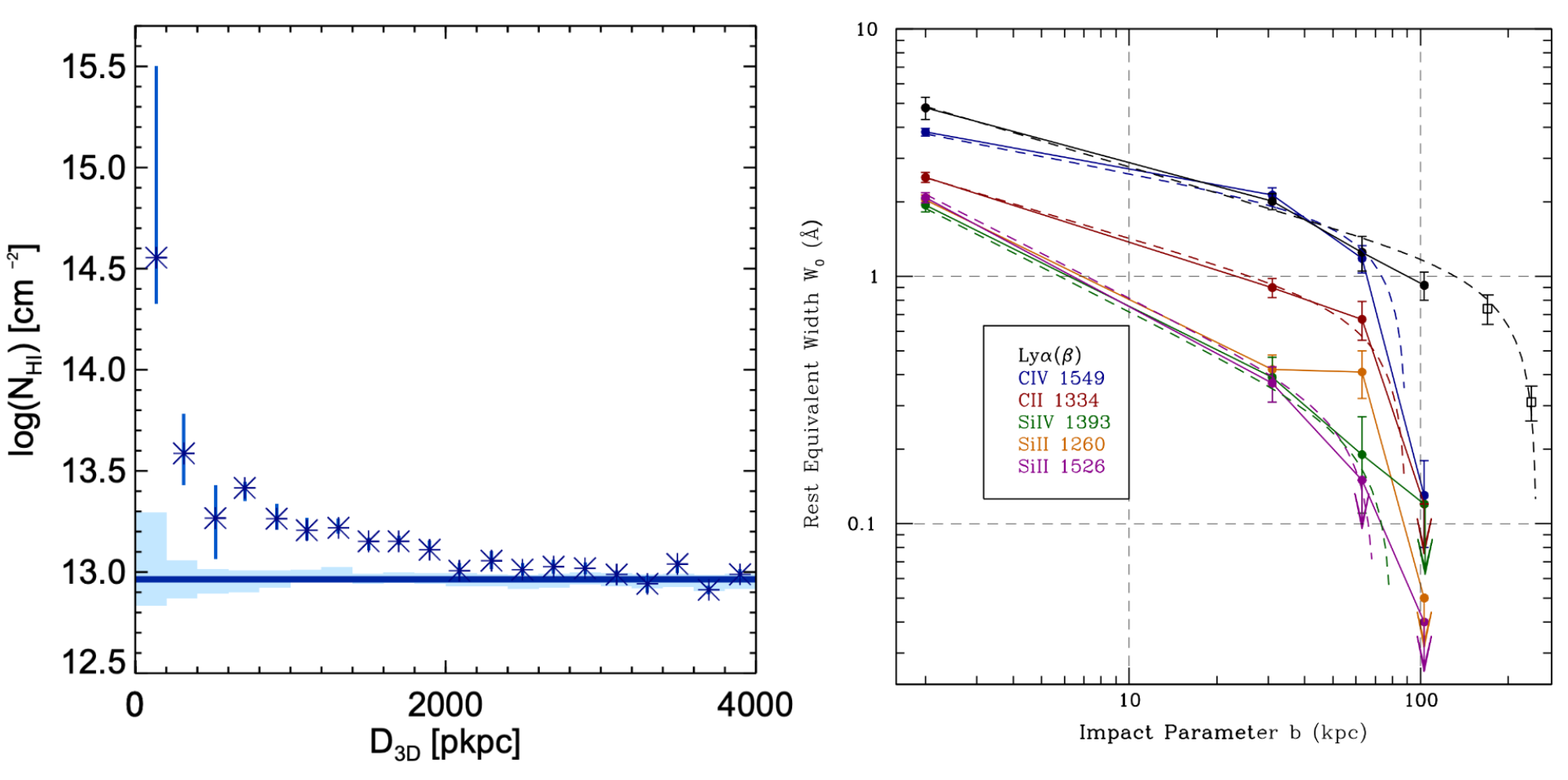}
\caption{The hydrogen and metal distribution in the CGM of LBGs from the KBSS survey. Left: Radial profile of \ion{H}{I} as a function of the 3D distance from the central LBGs. Star-forming galaxies reside in gas-rich regions extending to several megaparsecs. Right: Equivalent width profiles of metal lines from stacks of background galaxies that can be used to probe the CGM of foreground systems at varying projected distances. The halos of LBGs are filled with enriched material in various ionization stages. {\it Credits: Rudie G.~C. et al., 2012, ApJ, 750, 67; Steidel C.~C. et al., 2010, ApJ, 717, 289. Reproduced with permission.}}
\label{fig:cgmLBG}       
\end{figure}

\subsubsection{Lyman break galaxies}

The connection between gas and LBGs, a population of galaxies with $SFR\approx 10-100$\,M$_\odot$ and stellar mass $\approx 10^{10}-10^{11}$\,M$_\odot$ \cite{Erb2006}, has been explored in two critical surveys around $z\approx 2$ and $z\approx 3$, the Keck Baryonic Structure Survey (KBSS) \cite{Steidel2010} and the VLT LBG Redshift Survey (VLRS) \cite{Bielby2011}. 
When studying the cool neutral CGM probed by \ion{H}{I} absorption in large samples of galaxy-absorber pairs, LBGs are found in an overdensity of neutral hydrogen that extends to $\gtrsim 2$\,Mpc with a declining column density profile (\cite{Rudie2012}, Figure~\ref{fig:cgmLBG}). The strongest absorbers, with a typical column density that is three orders of magnitude higher than what is found in the IGM, lie within the inner $\approx 100$\,kpc, i.e. on the scale of the halo, and are tightly correlated with the position of galaxies, implying that they are tracing the CGM rather than the large-scale structures that host these galaxies \cite{Bielby2017}. 
The gas kinematics inside the inner CGM is dominated by peculiar velocities, with the line Doppler widths being $\approx 50$ percent larger than typically found at a comparable column density. This finding is explained by higher temperatures or non-thermal motion, such as turbulence from accretion shocks or galactic winds.

When looking at the metal-enriched gas around LBGs, signatures of outflows become apparent. For instance, by stacking $\approx 100$ galaxies at $z\approx 2.3$ for which systemic redshifts are accurately known from the non-resonant \ha\ line, Steidel et al. \cite{Steidel2010} 
studied the profile of strong interstellar absorption lines down the barrel, i.e., against the galaxy continuum. These authors identify the ubiquitous presence of outflows, similar to what is seen at lower redshift \cite{Weiner2009}, with typical velocities of $\approx 150-200$\,km~s$^{-1}$. The line profiles extend blueward for up to $700-800$\,km~s$^{-1}$, indicating that fast-moving outflows are commonly present in highly star-forming galaxies at $z\approx 2.5$ and can move enriched material through the CGM and inside the IGM. 

The extent of the enriched material around galaxies can be studied in the transverse direction, i.e., examining the gas metal content in galaxy-quasar pairs or in galaxy-galaxy pairs, which, being more common, offer a better sampling of the innermost regions of the CGM. In the transverse direction, Steidel et al. identified the presence of an enriched gas up to scales of $\approx 80-100$\,kpc from galaxies, with a radially decreasing profile in the equivalent widths of the strongest interstellar lines (Figure~\ref{fig:cgmLBG}). This observation reinforces the idea that the enriched gas is circulating inside the CGM. However, since the relative kinematic information is lost in the transverse direction (given the ambiguity in the Doppler shift of gas entering or leaving a galaxy from the front or back side), models must be invoked to discern the underlying mechanisms responsible for the observed gas distribution. 

\begin{figure}[b]
\centering
\includegraphics[scale=.38]{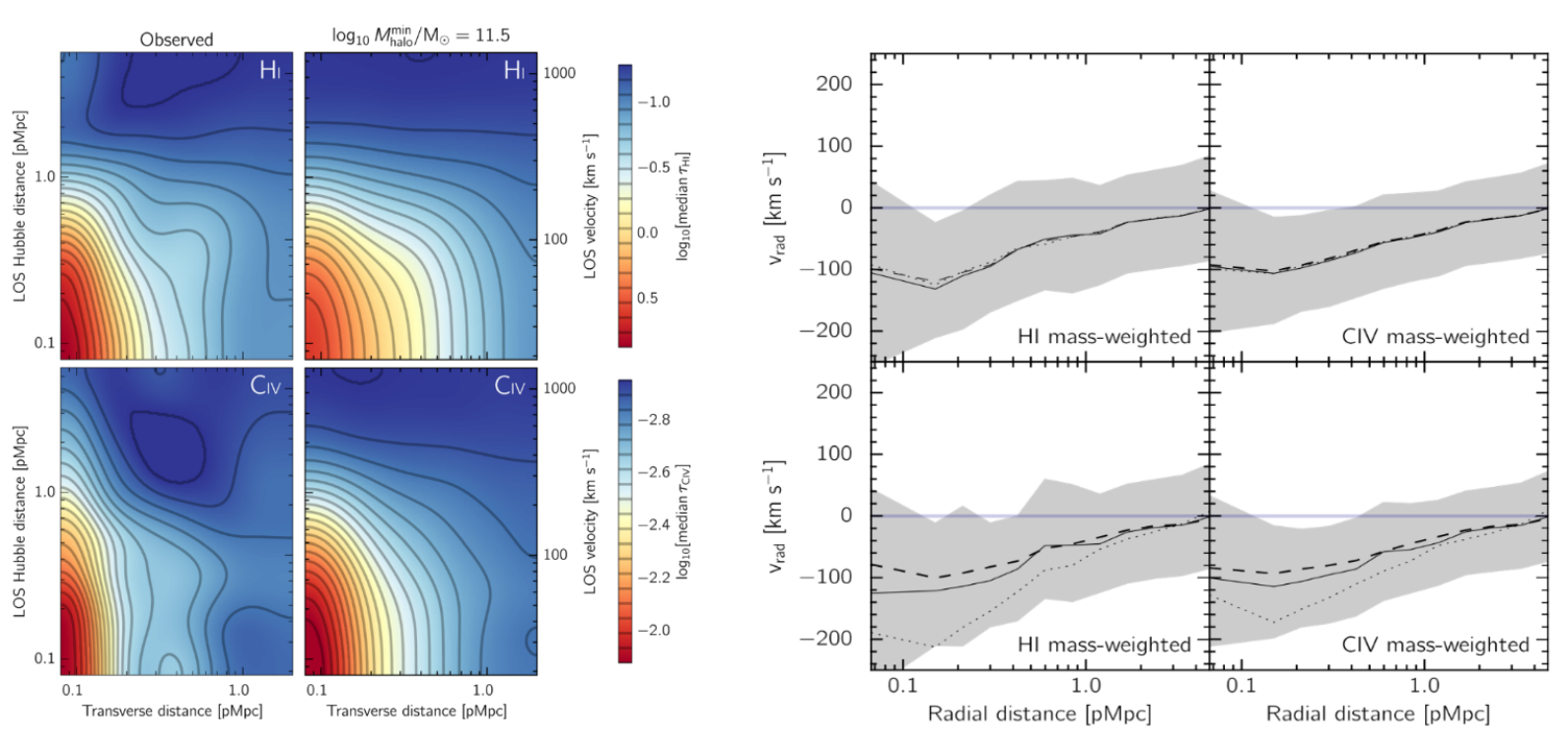}
\caption{Analysis of the CGM of LBGs using the pixel optical depth method. Left: Map of the optical depth of \ion{H}{I} and \ion{C}{IV} along the line of sight and in the transverse direction from observations and simulations of halos with masses $10^{11.5}$\,M$_\odot$. A good agreement is found. Right: Profiles of the radial velocity of the gas (\ion{H}{I} left, \ion{C}{IV} right) in simulated halos. 
The top and bottom panels adopt a minimum halo mass of $10^{11.5}$\,M$_\odot$ and stellar mass $10^{9.5}$\,M$_\odot$. A net inflow in the CGM of LBGs is apparent. {\it Credits: Turner M.~L. et al., 2017, MNRAS, 471, 690. Reproduced with permission.}}
\label{fig:LBGpixels}       
\end{figure}

In a series of papers by Turner et al. \cite{Turner2014,Turner2017}, this problem has been studied using a statistical analysis of the gas distribution via the pixel optical depth technique compared to the results of numerical simulations. This technique consists of examining the velocity-dependent absorption (pixel-by-pixel) of sightlines crossing the halo of a galaxy at a given projected distance. With these data, the 2D optical depth map can be reconstructed as a function of transverse separation and line-of-sight velocity from galaxies (Figure~\ref{fig:LBGpixels}). 
These authors identify a marked absorption signal compared to random locations, reaching $\approx 180$\,kpc in the transverse direction and $\pm 240$\,km~s$^{-1}$ along the line of sight for the most common ions. 
Once detectability effects are considered, the extent of the metal-enriched region is comparable to that of hydrogen in the transverse direction. Along the line of sight, higher ionization species (\ion{C}{IV}, \ion{Si}{IV}, \ion{O}{VI}) are more extended than neutral hydrogen. 
For most ions, the signal along the line of sight is more elongated than in the transverse direction, hinting at additional peculiar velocities due to outflows and inflows or viral motion. Indeed, these local velocities distort the mapping between distance and speed along the line of sight expected for pure Hubble flow, thus breaking the symmetry between the transverse and the line-of-sight direction. 

To interpret the observed signal, these authors resorted to the analysis of numerical simulations from the Evolution and Assembly of GaLaxies and their Environments (EAGLE) suite \cite{Schaye2015}. By generating mock pixel-optical-depth maps extracted from simulations mimicking observational techniques, Turner et al. found that halos with masses $\approx 10^{12}$\,M$_\odot$ in EAGLE best match the observations, including the characteristic anisotropy. 
When studying the mass-weighted radial velocities of the gas around the simulated galaxies, most of the gas appears infalling rather than outflowing. 
Based on this result, the marked distortions in the 2D maps of optical depth can be attributed to a significant infalling component and not outflows. This implies that not all the enriched gas seen at a given distance from LBGs is necessarily moving away from the galaxy, but that a substantial component is (re)accreting in the form of enriched material. 

The analysis of the CGM in large samples of highly star-forming galaxies at cosmic noon confirms that the baryon cycle, in the form of both inflows and outflows, is a characteristic and ubiquitous feature in star-forming systems. 

\subsubsection{\lya\ emitters}

Before the deployment of IFSs, detecting emission lines from galaxies at lower mass than LBGs was particularly challenging, because of the faint continuum level, making pre-selection and follow-up MOS spectroscopy difficult, if not impossible. Significantly few low-mass galaxies were associated with absorption line systems at $z\gtrsim 2$ either by narrow-band imaging \cite{Fynbo2000} or by serendipitous discoveries \cite{Crighton2015}. New large-format IFSs, and especially MUSE at the VLT, have drastically reshaped our view of CGM in lower-mass galaxies with stellar masses $\approx 10^7-10^9$\,M$_\odot$ \cite{Liu2023} by enabling the detection of moderately bright \lya\ emission even in the absence of detectable continuum \cite{Fumagalli2016b}.

\begin{figure}[b]
\includegraphics[scale=.3]{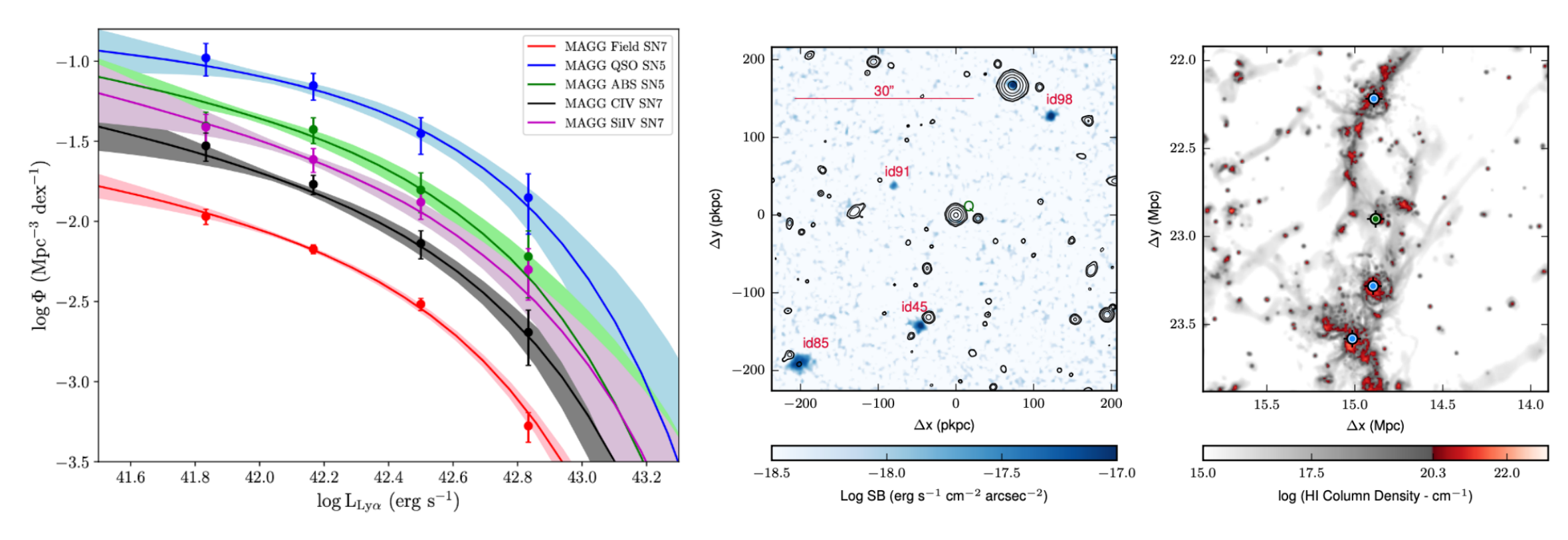}
\caption{The gas environment of LAEs. Left: Luminosity functions of LAEs computed in the field (red), near absorbers (\ion{H}{I} green, \ion{C}{IV} grey, \ion{Si}{IV} purple), and quasars (blue). Compared to the field, an excess of clustering of LAEs around absorbers is evident. Right: Example of MUSE observations (left panel) in a field hosting a strong \ion{H}{I} absorber at the position of the quasar (marked as Q). Multiple LAEs (in blue) aligned along a filament are detected. This type of alignment is consistently predicted by simulations (right panel). In this image, red is for the optically thick gas, and blue symbols are for the LAEs. {\it Credits: Galbiati M. et al., 2023, MNRAS, 524, 3474; Mackenzie R. et al., 2019, MNRAS, 487, 5070. Reproduced with permission.}}
\label{fig:cgmlae}       
\end{figure}

To date, two main MUSE large programs have focused on the study of the CGM in these \lya\ emitting galaxies (often referred to as  LAEs): the MAGG survey \cite{Lofthouse2023,Dutta2020,Lofthouse2020,Galbiati2023} and the MUSEQuBES survey \cite{Muzahid2021,Banerjee2023}. 
Thanks to the analysis of a sample of $\approx 2,000$ LAEs and hundreds of absorbers (\ion{H}{I}, \ion{C}{IV}, and \ion{Si}{IV}), a coherent view of the gas distribution around LAEs at $z\approx 3-4$ has emerged. 

LAEs are not randomly distributed in space, but cluster preferentially around strong absorption lines with $\log(N_{HI}/\textrm{cm$^{-2}$})\gtrsim 17$, and LAEs also lie preferentially near metal line absorbers compared to random regions.
This can be seen, for example, from the study of the LAE luminosity function in different environments (Figure~\ref{fig:cgmlae}), where the normalization in the case of galaxies near absorbers is a factor $\approx 3-5$ higher than the galaxies detected in the field, i.e., without any selection on absorption lines \cite{Galbiati2023}.  
A similar positive excess of LAEs near strong hydrogen and metal absorbers is also visible in the cross-correlation of galaxies and absorbers, with LAEs near \ion{H}{I} exhibiting the strongest signal. 

The detection rate of galaxies near optically thick absorbers is relatively high, with $\approx 60$ percent of the absorption systems having at least a galaxy association within the MUSE field of view. The same is true for the strongest metal absorbers, with equivalent width $\gtrsim 0.3~$\AA, while the detection rate progressively decreases with decreasing equivalent width. 
Frequently, multiple LAEs are found in proximity to the same absorber (see also the next Section). Despite the clear physical association between LAEs and gas clouds, the typical projected distance between the sightline and galaxies is above $50-100$\,kpc, ruling out the inner CGM ($\lesssim 2R_{vir}$) as the origin of the absorbing gas \cite{Lofthouse2023}. 
Despite the intrinsically low covering fraction of the absorbing gas, optically-thick \ion{H}{I} clouds within $\approx 100$\,kpc of LAEs easily account for the entire population of LLSs seen in cosmological surveys of quasars \cite{Lofthouse2023}.  

In this mass scale, strong absorbers are, therefore, primarily tracing the outer CGM and portions of the IGM in proximity to galaxies. Further evidence in support of this scenario emerges from studying the distribution of galaxies and absorbers inside the field of view. Previous results based on the analysis of small samples pointed to filamentary structures of multiple LAEs aligned with the absorbing gas (Figure~\ref{fig:cgmlae}, \cite{Mackenzie2019}), which resemble the cosmic web predicted in cold dark-matter simulations. Expanded statistical analysis in large samples confirms this trend \cite{Lofthouse2023,Galbiati2023}. When studying the relative orientation of LAEs and \ion{H}{I} absorbers, a preference is found for aligned structures (i.e., where emitters and gas are found at close angular separation relative to each other). Concerning \ion{C}{IV}, this preferential alignment is only visible for the strongest absorbers, with the weaker systems being more isotropically distributed.  

The similarity between the results of optically thick \ion{H}{I} absorbers and \ion{C}{IV} systems is not surprising, as the two families of absorption lines trace, for the most part, the same physical systems. A significant fraction of LLSs shows strong \ion{C}{IV} absorption and stacking of Lyman series lines in high equivalent-width \ion{C}{IV} systems reveals saturated lines with noticeable wings typical of LLSs. The weaker \ion{C}{IV} systems are instead connected to the lower column density \ion{H}{I} absorbers. 

The picture emerging from this analysis defines a gas environment near LAEs composed of the CGM of individual galaxies, where strong absorbers arise. Due to the compact nature of these low-mass galaxies, strong absorption line systems primarily trace the outer parts of the CGM and the surrounding filaments that connect multiple galaxies. Weaker metal absorbers are instead probing more isotropic regions surrounding the LAEs, such as lower-density bubbles enriched by previous episodes of star formation.

\subsection{The CGM in the reionization epoch}

Studying the CGM of high-redshift galaxies at $z\gtrsim 5$, at the edge of and within the reionization epoch, is an exciting new frontier in galaxy evolution. Due to the vast cosmic distance, this study requires highly sensitive IR detectors. Moreover, as the young Universe becomes progressively denser and more neutral, even the hydrogen lines of the strongest systems start to suffer from blending with a thick forest of \lya\ lines, requiring high-resolution spectroscopy of intrinsically faint quasars. 

Despite these challenges, pioneering efforts have been conducted using MOS and IFSs in optical and sub-mm wavelengths. ALMA has been used to serendipitously detect a $z\approx 6$ galaxy associated with a DLA \cite{dodorico2018}, allowing a rare opportunity to study the chemistry of the CGM in a galaxy at the edge of cosmic reionization. At optical wavelengths, both MOS and IFS studies \cite{Meyer2020,Diaz2021,Bielby2020} highlighted the role of low-mass galaxies in both enriching and ionizing the surrounding gas, with early signs of positive correlations between ionized hydrogen and metal near star-forming systems. 

These pioneering studies point to young low-mass galaxies growing bubbles of ionized gas in their surroundings while shaping the chemical properties of the IGM through galactic outflows. This field is in its infancy, but JWST, with its superior sensitivity at IR wavelengths, is expected to make a remarkable contribution, as already shown by the analysis of the first available datasets \cite{Bordoloi2023,Wu2023}.

\section{The role of environment in the CGM}
\label{sec:environment}

The role of the environment in affecting the evolution of galaxies has been the subject of a vast body of literature in the past 40 years, with undisputable evidence showing that the properties of galaxies, and in particular their ability to form stars, are deeply connected with the density of galaxies in the surroundings. Indeed, the pioneering analysis by Dressler et al. \cite{Dressler1980} showed an evident decrease in the fraction of star-forming (spiral) galaxies and a corresponding rise in the population of passive (elliptical) galaxies as a function of increasing environmental density. A more modern take on the problem is to analyze a large sample of galaxies from deep redshift surveys \cite{Peng2010}. In these datasets, the excess of passive galaxies over a reference low-density environment compared to the active population -- the so-called environmental quenching efficiency -- increases in progressively denser environments, both in the local Universe and up to $z \approx 1$. Thus, environmental quenching occurs consistently in large-scale structures as they assemble over time.  

\begin{figure}[b]
\centering
\includegraphics[scale=.28]{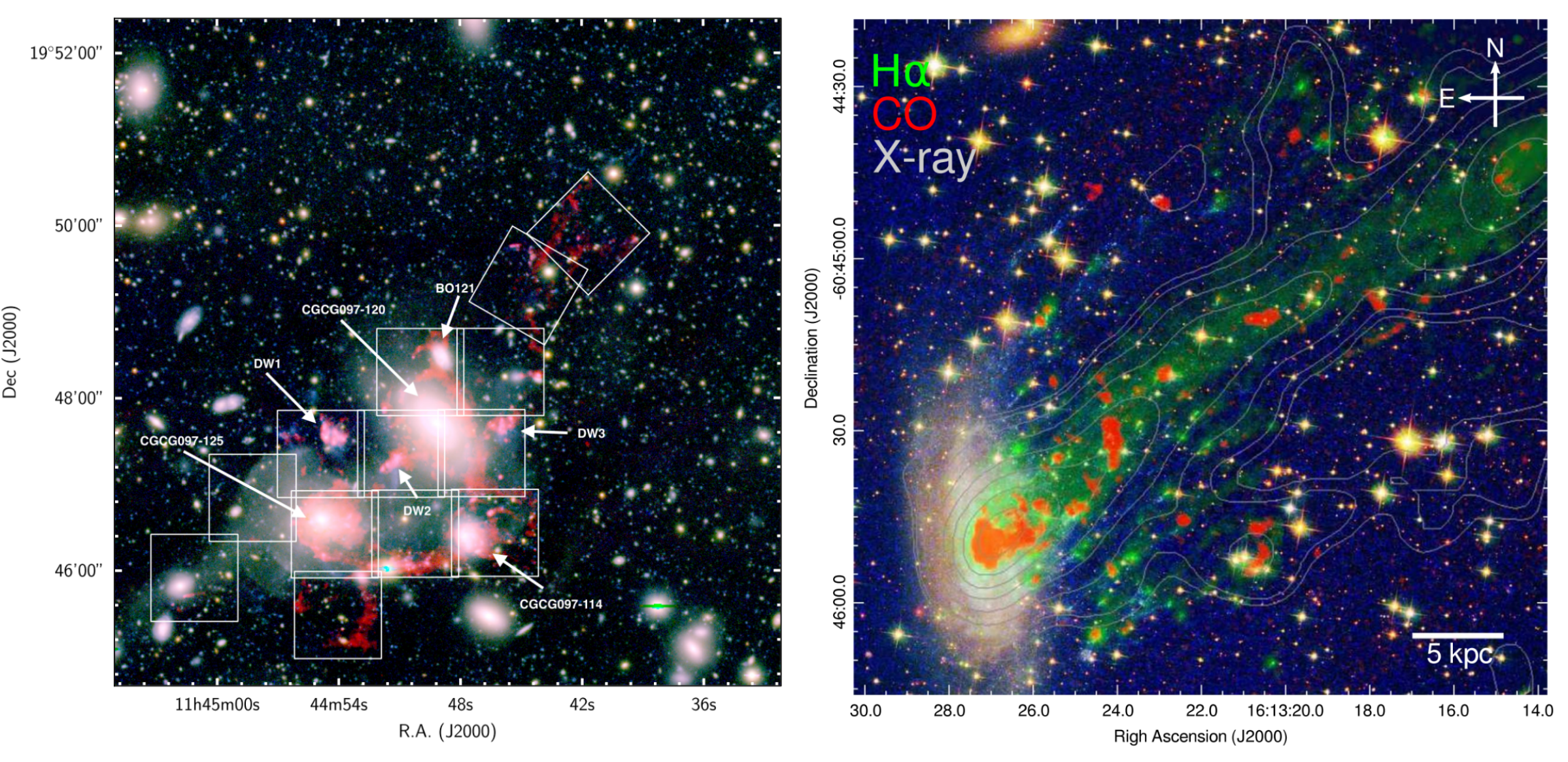}
\caption{Examples of environmental effects on the gaseous envelopes of galaxies as seen by MUSE observations. Left: A compact group, dubbed BIG, infalling into the A1367 cluster. In red is visible the \ha\ emission of warm gas that is displaced by tidal forces from the ISM and into the CGM of individual galaxies and the intragroup medium. Right: Iconic example of ram pressure stripping acting on the dwarf galaxy ESO137$-$001 that is infalling into the Norma cluster. The multiphase gas traced by \ha, CO, and in the X-rays (as labeled) is dispersed within the intracluster medium through the CGM of the galaxy.  {\it Credits: Fossati M. et al., 2019, MNRAS, 484, 2212; J{\'a}chym P. et al., 2019, ApJ, 883, 145. Reproduced with permission.}}
\label{fig:envexamples}       
\end{figure}

Despite several open questions on the exact mechanisms responsible for this transformation, it is well-established that galaxies evolve differently in different environments. As galaxies alter their star formation when residing in dense environments, it is natural and expected that transformations also occur in the CGM, which mediates the galaxy's ability to acquire and eject gas from intergalactic space. However, CGM studies have mainly focused on the role of the central galaxies and their correlation with the surrounding gaseous envelopes, rarely considering the role of the galaxy environment in shaping the CGM. Rather than omission, this approach has been dictated by the modest availability of dense spectroscopic surveys suitable for characterizing the environment jointly with the CGM properties. Large field-of-view IFSs have changed the perspective on this problem, enabling the study of environmental perturbation on the CGM in emission and absorption. 

This Section reviews, at first, the leading families of mechanisms that transform galaxy properties (Section~\ref{sec:envmec}). Next, it examines emerging examples in the literature of recent studies that focus on absorption and emission to diagnose how environmental perturbations shape the CGM (Sections~\ref{sec:envemi} and ~\ref{sec:envabs}) and concludes with a brief overview of current results from numerical simulations that interpret recent observational evidence in the context of galaxy evolution theories (Section~\ref{sec:envsim}). 

\subsection{Main agents of environmental transformation}\label{sec:envmec}

Environmental mechanisms capable of transforming galaxies have been studied in great detail over the past decades, and excellent reviews on the subject can be found in the literature \cite{Boselli2006,Boselli2022}. These processes are usually classified into two broad families. On the one hand, gravitational perturbations involve interactions between two galaxies or satellite galaxies and parent halos, as seen in close encounters within groups (Figure~\ref{fig:envexamples}, left). On the other hand, hydrodynamic perturbations concern the interaction between gas phases inside and outside galaxies (Figure~\ref{fig:envexamples}, right). Both families can displace significant amounts of gas and, in some cases, stars from the galaxies, quenching the systems' activity and altering their morphology. 
The main processes in both classes are discussed next, closely following the review by Boselli \& Gavazzi \cite{Boselli2006}, which provides a comprehensive introduction to the subject. 

\subsubsection{Gravitational interactions} 

Three main perturbing mechanisms can be found within the family of gravitational interactions: tidal interactions between galaxies, cluster-galaxy interactions, and multiple galaxy encounters.

Tidal interactions occur when two galaxies are at sufficiently close separation so that one galaxy (the perturbed system) feels the tidal force arising from the potential of the companion galaxy, which is the perturber. 
For two galaxies with mass $M_1$ and $M_2$, size $R_1$ and $R_2$, at relative distance $D_{12}$, the perturbation $P_T$ induced by the tidal force on galaxy 2 is 
\begin{equation}
    P_T \propto\frac{M_1}{M_2}\left(\frac{R_2}{D_{12}}\right)^3\,.
\end{equation}
This equation highlights the main characteristic of this perturbation: being a tidal force, the efficiency of the disturbance scales with the third power of distance. Hence, close encounters on scales comparable to the one of the galaxy radius produce the most efficient perturbations, being able to displace and remove stars, gas, and dark matter, as seen in many iconic examples of galaxy mergers at first passage (see, e.g., NGC 4676 also known as the Mice Galaxies).
Tidal interactions start affecting the less-bound components in the galaxy outskirts, such as the peripheral or extraplanar
gas and the outer dark matter halo. Thus, this mechanism is expected to significantly modify the CGM by displacing part of the less-bound ISM from the disk components into the halo and ejecting directly halo gas at more considerable distances. 
Tidal forces are also effective in driving gas to the center of a galaxy, enhancing AGN activity, but for short periods. 

The mass dependence in the above equation gives further insight into the process. Low-mass galaxies are more easily perturbed, especially by companions with larger mass ratios. A limiting case is a galaxy interacting with a galaxy cluster, making the galaxy-cluster interaction a particular case of the general tidal perturbation. Due to the largely uneven mass ratio, the tidal effects are significant in clusters. Moreover, within clusters, the high density of galaxies drastically increases the probability of close encounters. One would, therefore, expect that tidal perturbation is the dominant environmental process in rich galaxy clusters. However, this is not the case for at least two reasons. Firstly, these encounters are more complex to spot due to an observational bias: the tidally displaced material is short-lived in a galaxy cluster, as it is quickly lost in the intracluster medium. Secondly, within clusters, the galaxy encounters occur at high velocity, reducing the time for tidal forces to operate and effectively suppressing the efficiency of this mechanism. 

A more cumulative effect can occur for galaxies moving inside a cluster. These systems would repeatedly feel the impact of galaxy-galaxy interactions, which, albeit not dramatically effective individually, can produce a cumulative substantial perturbation. At the same time, the galaxy would continue to interact with the cluster potential. Low-mass systems subject to this continuous interaction are expected to lose a substantial fraction of dark matter, stars, and gas, especially from their outskirts.  More massive galaxies, on the contrary, are subject to less damage in their internal structural composition. The more loosely bound CGM is expected to be significantly affected.

\subsubsection{Hydrodynamic interactions} 

A second family of environmental perturbations concerns the hydrodynamic interactions between gas inside a galaxy and the surrounding intracluster/intragroup medium or the IGM. Three leading mechanisms are ram pressure stripping, viscous stripping, and thermal evaporation. 

Ram pressure stripping is perhaps the most well-known hydrodynamic perturbation. It describes the global pressure exerted on the galaxy's gas content by the presence of an ambient medium, similar to the drag force experienced by a swimmer immersed in water. 
The strength of ram pressure varies with the ambient gas density and the square of the velocity of the galaxy relative to the surrounding fluid, i.e., $P\propto \rho_{\rm amb} v^2$. The pressure generated by this interaction can lift gas from inside the galaxy, giving rise to long tails stretching behind the galaxies (Figure~\ref{fig:envexamples}) if the gravitational pull of the stellar and dark-matter components can be overcome.

In a stellar disk, the equilibrium condition between ram pressure and gravitational restoring force is 
\begin{equation}
    \rho_{\rm amb} v^2 = 2\pi G \Sigma_{\rm gas} \Sigma_{\rm star}\,,
\end{equation}
where $\Sigma_{\rm gas}$ and $\Sigma_{\rm star}$ are the disk stellar and gaseous surface density.
From this equation, it becomes apparent that ram pressure is particularly effective in rich clusters, where the high-velocity dispersions imply fast movements of the galaxies inside the intracluster medium, and hence the gravitational equilibrium can be easily overcome. 
Galaxies with lower values of stellar and gas surface density are more prone to the effects of ram pressure. This perturbation is also highly dependent on the geometry of the system. For instance, a galaxy falling face-on inside a cluster will suffer from a more severe gas depletion than one falling in at high inclination, as in this latter case, one side will shield the other from the drag. Compared to gravitational interactions, for galaxies moving at high speed, ram pressure can act to deplete the gas reservoir of galaxies on very short time scales, of the order of 100 Myr or less. 
As ram pressure can even affect, in some cases, the most-bound molecular gas \cite{Fumagalli2009}, this mechanism can be very effective in altering the CGM of galaxies. 

Viscous stripping is a second hydrodynamic process occurring at the interface of the gas phases of galaxies moving inside a viscous medium and arises due to fluid instabilities at the interface of the two layers. Thus, viscous stripping can act alongside ram pressure. It is also an effective way to deplete fast-moving galaxies inside clusters, with timescales comparable to ram pressure. Two different regimes operate depending on whether the flow is turbulent or laminar. 
In both cases, the mass loss rate depends on the drag force over the relative velocity of the two fluids, 
\begin{equation}
    \frac{\dd M}{\dd t} = F_{\rm drag} v^{-1}\,.
\end{equation}
For the laminar case, $F_{\rm drag} \propto v$, so the mass-loss rate depends only on static properties of the system, such as the galaxy radius $r_{\rm gal}$, the ambient sound-speed $c_{\rm amb}$ and density $\rho_{\rm amb}$, and the mean free path of particles inside the ambient medium $\lambda_{\rm amb}$
\begin{equation}
    \frac{\dd M}{\dd t} \propto \pi r_{\rm gal}\rho_{\rm amb}\lambda_{\rm amb}c_{\rm amb}\,.
\end{equation}
If the flow is instead turbulent, $F_{\rm drag} \propto v^2$ as for ram pressure, and the mass loss depends on velocity as 
\begin{equation}
    \frac{\dd M}{\dd t} \propto r^2_{\rm gal}\rho_{\rm amb}v\,.
\end{equation}
Viscous stripping can affect the galaxy's gas content, and it is plausible to hypothesize that it also alters the evolution and survival of clouds within the CGM. 

Finally, thermal evaporation occurs when hot gas from an ambient medium (primarily intracluster gas) exchanges heat with the cold gas phase inside a galaxy, increasing the cold gas velocity dispersion. The galaxy can lose this gas if the potential well cannot retain it. As a simple order-of-magnitude relation, thermal evaporation becomes significant when the ambient temperature exceeds the galaxy velocity dispersion $kT \gtrsim \sigma^2$.
The mass-loss rate due to this evaporation is similar to the one in the laminar case for viscous stripping, and this process acts on comparable timescales. Magnetic fields, which can alter the heat exchange between phases, and the ambient temperature are the most critical parameters controlling the efficiency of this environmental mechanism.

\subsubsection{Pre-processing and starvation}

The environmental processes discussed above effectively reduce the gas supply that feeds the galaxy activity, ultimately leading to the quenching of the galaxy and its transformation into a passive system. As noted in the previous Section, some of these mechanisms can also temporarily lead to an enhancement in the activity (both in star formation and AGN), a short-lived transient typically followed by quenching. Two additional environmental phenomena are identified and discussed in the literature, pre-processing and starvation, which -- although not individual physical mechanisms as those described above -- constitute a global way for the environment to induce transformations in galaxies. 

The pre-processing mechanism defines the cumulative effects of environmental perturbations inside substructures assembled from $z \approx 0.5-1$ into larger units within our hierarchical Universe. More specifically, pre-processing is believed to be responsible for the leading environmental transformation inside galaxy groups that are infalling towards clusters. Tidal interactions are most commonly found inside these groups, although hydrodynamic processes may still act alongside gravitational ones at earlier times. Pre-processing is not only able to reduce galaxy activity in systems that are already at large distances from cluster cores ($>1R_{\rm vir}$) but also to induce an initial degree of perturbation that accelerates the subsequent effects of the cluster environment onto galaxies. As seen in the iconic example of the so-called ``Blue Infalling Group'' (BIG) in Figure~\ref{fig:envexamples}, pre-processing is highly relevant in transforming the gas content of the CGM of group galaxies by altering the cross section of, e.g., warm ionized gas at close separation from galaxies.

\begin{figure}[b]
\sidecaption
\includegraphics[scale=.28]{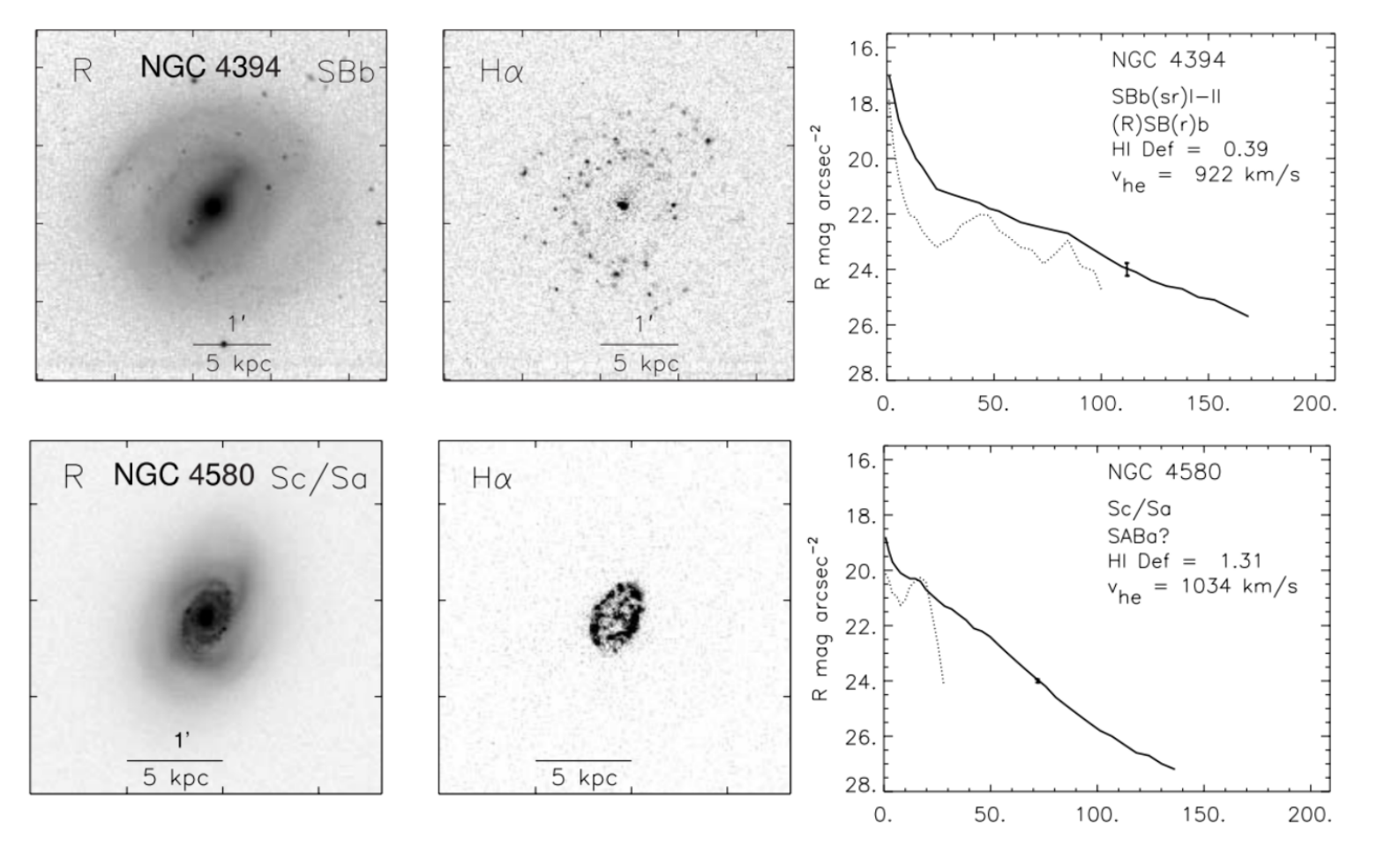}
\caption{Example of starvation (top) versus ram pressure (bottom). For each galaxy, the continuum (left), \ha\ images (center), and surface brightness profiles (right; dotted lines for H$\alpha$ and solid lines for $R-$band) are shown. Ram pressure induces an obvious truncation in the \ha\ map. Conversely, starvation gradually reduces star formation as gas is consumed. {\it Credits: Koopmann R.~A. \& Kenney J.~D.~P., 2004, ApJ, 613, 866. Reproduced with permission.}}
\label{fig:starvation}       
\end{figure}

Particularly relevant for the study of the CGM is the induced quenching of a galaxy activity following the removal of the gas reservoir that can feed it. This mechanism, defined as starvation in the literature, refers to a gas supply cut-off by any environmental process that acts on the outskirts of galaxies without strong and direct perturbation of the central disk. By suppressing the ability of a galaxy to refuel its central region and sustain star formation over time, the system will slowly fade as the available reservoirs are used up and not refilled. Unlike more impulsive processes such as ram pressure, starvation is a slow-acting mechanism that leads to quenching on timescales of a gigayear, which is the typical depletion time-scale of an average galaxy \cite{Bigiel2008}. Moreover, as it does not arise from violent interactions like ram pressure or tidal interaction, starvation rarely leaves marked imprints onto a galaxy gas and stellar content, such as truncations in the gas disk or tidal features in the stellar component. This process gradually offsets the overall gas content and activity level (Figure~\ref{fig:starvation}).  
Acting on the galaxy's outer regions, and in particular on the gas halos, starvation can be an effective mechanism in transforming the activity of galaxies already at large distances from overdensity, such as galaxies orbiting at $\gtrsim 2-3$ core radii from galaxy clusters. Numerical simulations reveal that the loosely bound halo gas can be removed almost entirely on time scales of a gigayear, already at these large distances \cite{Bekki2002}. Of all the processes described so far, this is believed to primarily act on the CGM \cite{Yoon2013}. 

\subsection{Imaging the multiphase gas in dense environments}\label{sec:envemi}

\begin{figure}[b]
\centering
\includegraphics[scale=.14]{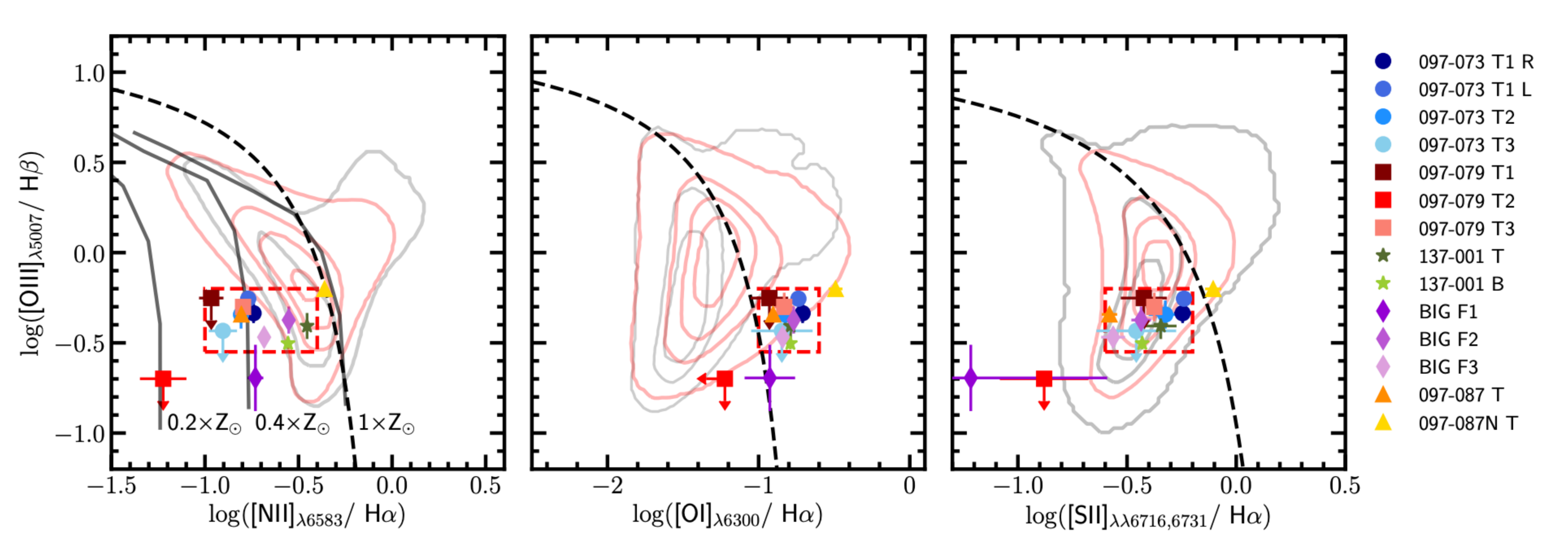}
\caption{Ionization diagnostic diagrams for line ratios derived in composite spectra of a sample of ram-pressure stripped tails. The multiphase gas within the tails shares common traits across different systems. There is evidence of photoionization as data points lie below the dashed line in the left panel, which defines the region typically occupied by photoionized gas in normal star-forming galaxies. The middle panel also reveals an elevated [OI]/\ha\ ratio, a signature of weak shocks. {\it Credits: Pedrini A. et al., 2022, MNRAS, 511, 5180. Reproduced with permission.}}
\label{fig:tailcgm}       
\end{figure}

The study of the multiphase CGM in a dense galaxy environment through imaging is a novel field of research, by large, enabled by IFSs at 8m-class telescopes. In the local Universe, the study of more classical environmental processes has been boosted by IFSs such as MUSE, in tandem with multi-wavelength observations across a wide interval of energies, from the X-rays to the radio. These analyses have clear relevance for our appreciation of how the multiphase CGM is altered and evolves in rich environments, as it offers a high-definition laboratory to map in exquisite detail the hot, warm, and cool components of the CGM near perturbed galaxies. The posterchild of ram pressure stripped galaxy, ESO~137$-$001, provides a clear example of why imaging of local environmentally-perturbed galaxies is particularly relevant to studying the multiphase CGM (Figure~\ref{fig:envexamples}). A panchromatic view of how the ISM is displaced by ram pressure inside the galaxy halo becomes apparent by combining data from {\it Chandra} and {\it XMM} in X-rays \cite{Sun2006}, MUSE in the optical \cite{Fumagalli2014b}, and ALMA at millimeter wavelengths \cite{Jachym2019}.

These data show environmentally-induced gas tails stretching inside the CGM for $\approx 100$\,kpc. Within these tails, traced by a diffuse and turbulent warm-hot medium, seen in both X-rays and \ha, \ion{H}{II} regions can be found external to the disk, both displaced directly from the ISM or even formed {\it in-situ} inside the tail \cite{Fossati2016}. Cold molecular gas is also present in clumpy and filamentary structures that formed {\it in-situ} or were pushed out directly from the galaxy ISM. The features observed in this archetypal galaxy appear general to a large sample of tails studied with MUSE. Extended ionized tails of warm gas are commonly found in gas-rich galaxies entering clusters, with line emission powered by multiple mechanisms, including photoionization, shocks, and heat conduction (\cite{Pedrini2022}, Figure~\ref{fig:tailcgm}). Photoionization is more prominent at close separation from the galaxy disk, where ionized material that was recently stripped is still recombining.

\begin{figure}[b]
\centering
\includegraphics[scale=.23]{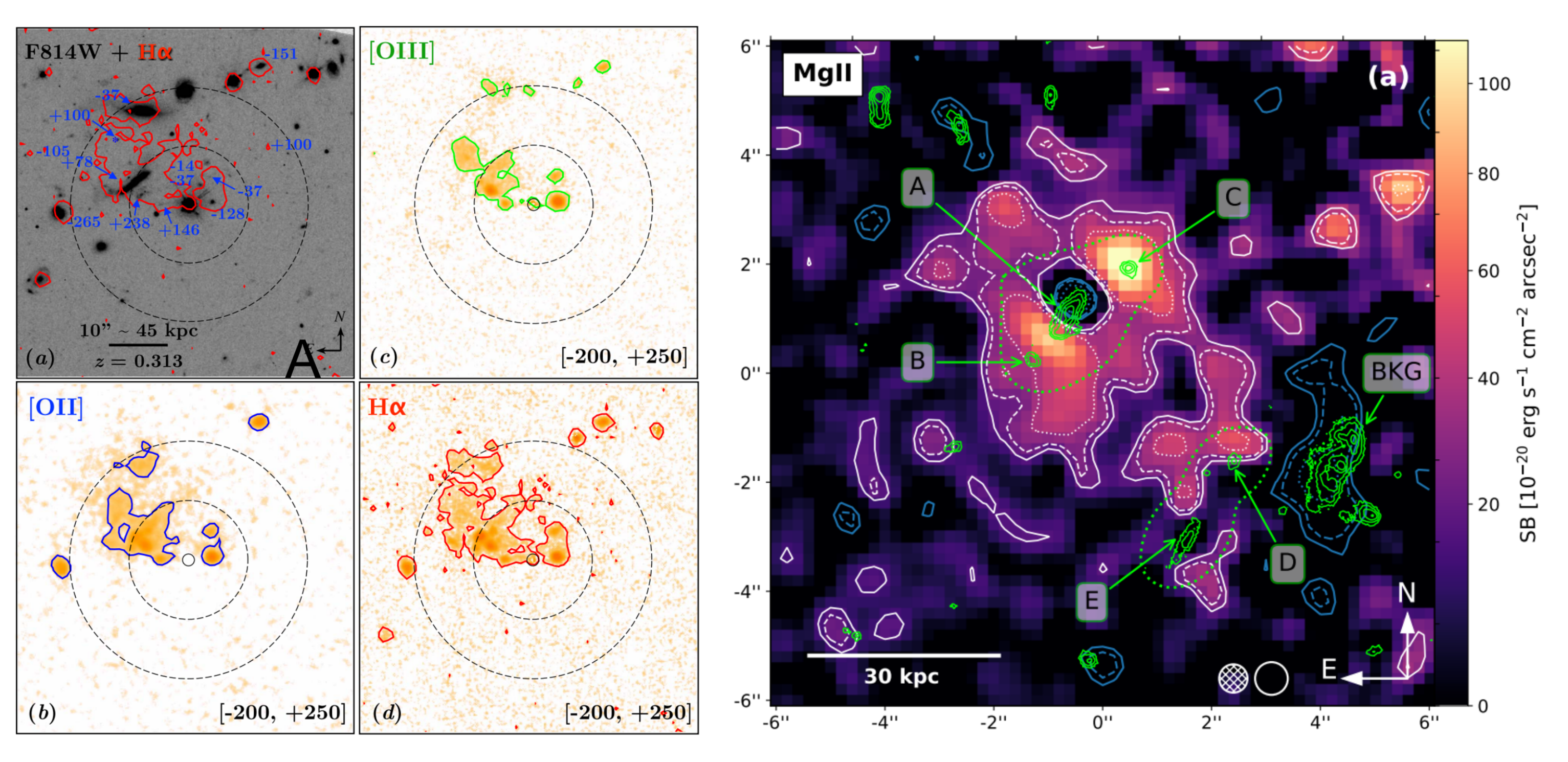}
\caption{Detection of emission lines in gas near high-redshift group galaxies. Left: Hydrogen and metal emission in an extended nebula within a low-mass group at $z\approx 0.3$ unveils material stripped from the galaxies and displaced inside their CGM and within the intragroup medium. The exact morphology of the emitting gas may be affected by the low $S/N$. Right: Extended \ion{Mg}{II} emission from a $z\approx 1.3$ group. Galaxy A also shows a P-Cygni profile in its spectrum, while the background galaxy ``BKG'' reveals \ion{Mg}{II} in absorption at the group location. A combination of tidal interactions and outflows drive enriched gas inside the group on scales that extend beyond what is detected in emission. {\it Credits: Chen H.-W. et al., 2019, ApJL, 878, L33. Leclercq F. et al., 2022, A\&A, 663, A11. Reproduced with permission.}}
\label{fig:groupemission}       
\end{figure}

These studies clearly show that ram pressure can lift ISM material (typically more enriched and dense) inside the CGM and thus trigger additional cooling processes not present in isolated systems. The ability to displace dense and metal-enriched ISM gas inside the halo is not restricted to extreme cases of ram pressure in galaxies infalling inside clusters. It is also evident in images of less harsh environments, such as nearby groups. The IFS study of BIG \cite{Fossati2019b} reveals the presence of an extended ionized gas traced by \ha\ in between the galaxies that compose this compact group (Figure~\ref{fig:envexamples}). Thus, tidal interactions can redistribute the ISM gas inside the halos, enhancing the cross section of warm-ionized and metal-rich gas within the CGM. We can conclude that environmental processes in the nearby Universe are effective agents that perturb the morphology, density, and chemical structure of the CGM.

Similar mechanisms also operate at higher redshift, as new IFS observations in groups reveal. At $z\approx 0.3$ Chen et al. \cite{Chen2019} unveiled a giant nebula detected in multiple hydrogen (\ha, H$\beta$) and metal ([\ion{O}{II}], [\ion{O}{III}], and [\ion{N}{II}]) emission lines within a galaxy group lying at close separation from a background quasar sightline (Figure~\ref{fig:groupemission}, left). Analysis of excitation mechanisms and kinematics suggests that this material is stripped from members of this low-mass group ($\approx 3\times 10^{12}$\,M$_\odot$) and that shocks and turbulent mixing layers in stripped gaseous streams contribute to the ionization state of the gas. Thus, environmental effects effectively release metal-enriched gas from star-forming regions, producing an absorption signal in the background sightline. 

At an even higher redshift, $z\approx 1.3$, Leclercq et al. \cite{Leclercq2022} discovered extended \ion{Mg}{II} emission from a low-mass group of five star-forming galaxies. The nebula extends for $\approx 70$\,kpc side to side (Figure~\ref{fig:groupemission}, right). Still, the presence of absorption in the spectra of background galaxies indicates the presence of enriched gas even beyond the region probed in emission.  An \ion{Mg}{II} P-Cygni profile is also seen in the most massive galaxy, indicating the presence of outflowing gas. Based on the joint analysis of galaxy spectra and nebula emission, these authors infer that outflows and tidal stripping from galaxy interactions add metal-enriched material to the intragroup medium of this structure.

\subsection{Absorption lines as probes of environmental effects on the CGM}\label{sec:envabs}

In the past decade, dense spectroscopic surveys in quasar fields with MOS have begun to uncover signatures of environmental mechanisms on the CGM using absorption spectroscopy \cite{Chen2010,Bordoloi2011,Nielsen2013}. Progress in the study of environmental effects on halo gas has rapidly accelerated in recent years thanks to the capability of IFSs, including slitless spectroscopy from space, to carry out exceptionally complete flux-limited surveys without the need for preselection \cite{Dutta2020,Dutta2021,Qu2023,Peroux2017,Chen2020}.

\begin{figure}[b]
\centering
\includegraphics[scale=.14]{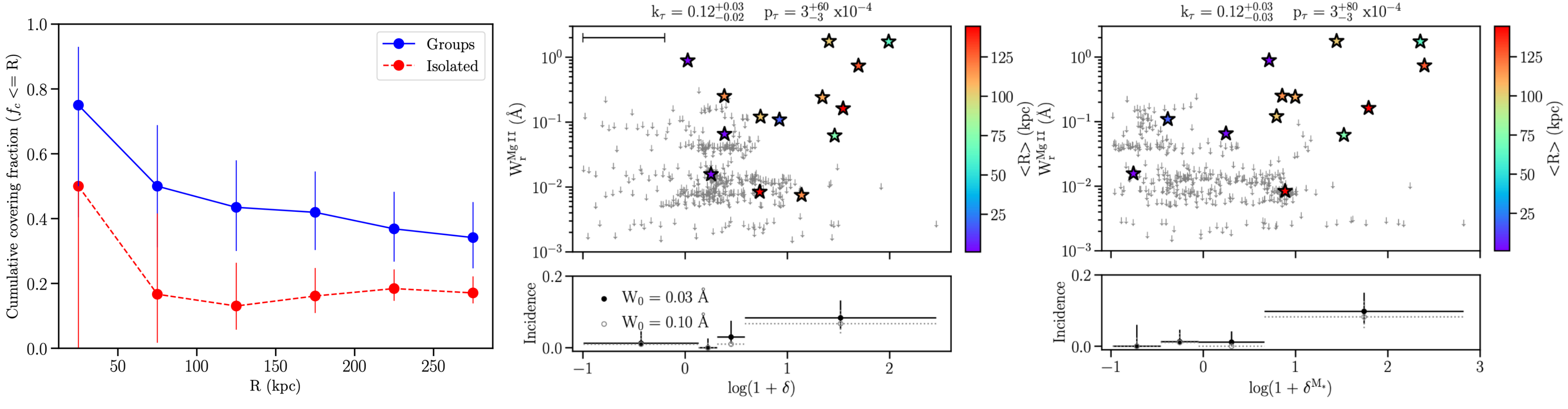}
\caption{Evironmental effects on the CGM of $z\lesssim 2$ galaxies as revealed by absorption spectroscopy. Left: The covering fraction of \ion{Mg}{II} gas in groups is consistently a factor $\gtrsim 2$ higher than in more isolated systems. Middle and Right: \ion{Mg}{II} equivalent width as a function of galaxies' number and stellar mass overdensity. The strongest absorption lines are found in the largest overdensities.  {\it Credits: Dutta R. et al., 2021, MNRAS, 508, 4573. Reproduced with permission.}}
\label{fig:lowzenv}       
\end{figure}

At $z<2$, there is clear evidence that the CGM of galaxies residing in groups (i.e., in the presence of at least a companion) is more extended and exhibits a higher equivalent width of low ions, primarily \ion{Mg}{II}, compared to more isolated systems. For example, the study of the radial profile of the  \ion{Mg}{II} equivalent width around galaxies (Figure~\ref{fig:mgiitrend}) shows a pronounced ($\gtrsim 1$\,dex) scatter at any fixed radius. This scatter is primarily driven by galaxies in groups (blue points in the figure). These systems are found at larger equivalent widths for a fixed radius or large radii for a fixed equivalent width \cite{Chen2010,Dutta2021}. The same result can be recast considering the covering factor, i.e., what fraction of the area around a galaxy is covered by \ion{Mg}{II} gas. The analysis of large samples (Figure~\ref{fig:lowzenv}, left) reveals elevated (by a factor $\approx 2$) covering fractions for group galaxies relative to isolated ones \cite{Dutta2021}. The availability of highly complete surveys has recently allowed the study of the strength of CGM absorption using continuous variables defining the environment \cite{Dutta2021}, thus going beyond a discrete analysis in sub-samples of groups and isolated systems. Based on this approach, the strength of \ion{Mg}{II} absorption appears to correlate with the galaxies' number or stellar mass overdensity (Figure~\ref{fig:lowzenv}, middle and right). 

Although there is clear consensus that the environment directly affects the CGM at these redshifts by boosting the cross-section of cool metal-enriched gas inside the halos, at present, observations do not now constrain the cause for this phenomenology, and different hypotheses have been explored in the literature. Three different scenarios can be considered. First, the enhanced incidence of \ion{Mg}{II} absorption might reflect the fact that, within groups, single sightlines can intercept more halos which, without any particular perturbation in their CGM, give rise to a stronger signal due to a mere superposition along the line of sight. The second scenario invokes the presence of an intragroup medium, considered a component that fills the entire group and is not associated with the CGM of individual galaxy members (satellites). However, some links with the central galaxy must be present (the parent halo).
In the third scenario, galaxy halos are individually perturbed by environmental mechanisms, and this perturbation is reflected in an enhanced \ion{Mg}{II} absorption within groups. 

Early studies based on MOS surveys considered the superposition model as a possible explanation of the excess absorption seen in group galaxies \cite{Bordoloi2011}. Indeed, superposition models constructed by adding the contribution of individually unperturbed halos seem adequate for reproducing the mean absorption profile in group galaxies. Although this effect must be present inside groups because of a simple geometric argument, follow-up investigations have highlighted how the superposition model alone is insufficient to explain the tail of the equivalent width distribution \cite{Dutta2021} and overpredicts the kinematics observed in groups \cite{Nielsen2018}. 
Additional contributions from the intragroup medium or perturbations in the CGM of individual galaxies must be considered.

\begin{figure}[b]
\centering
\includegraphics[scale=.18]{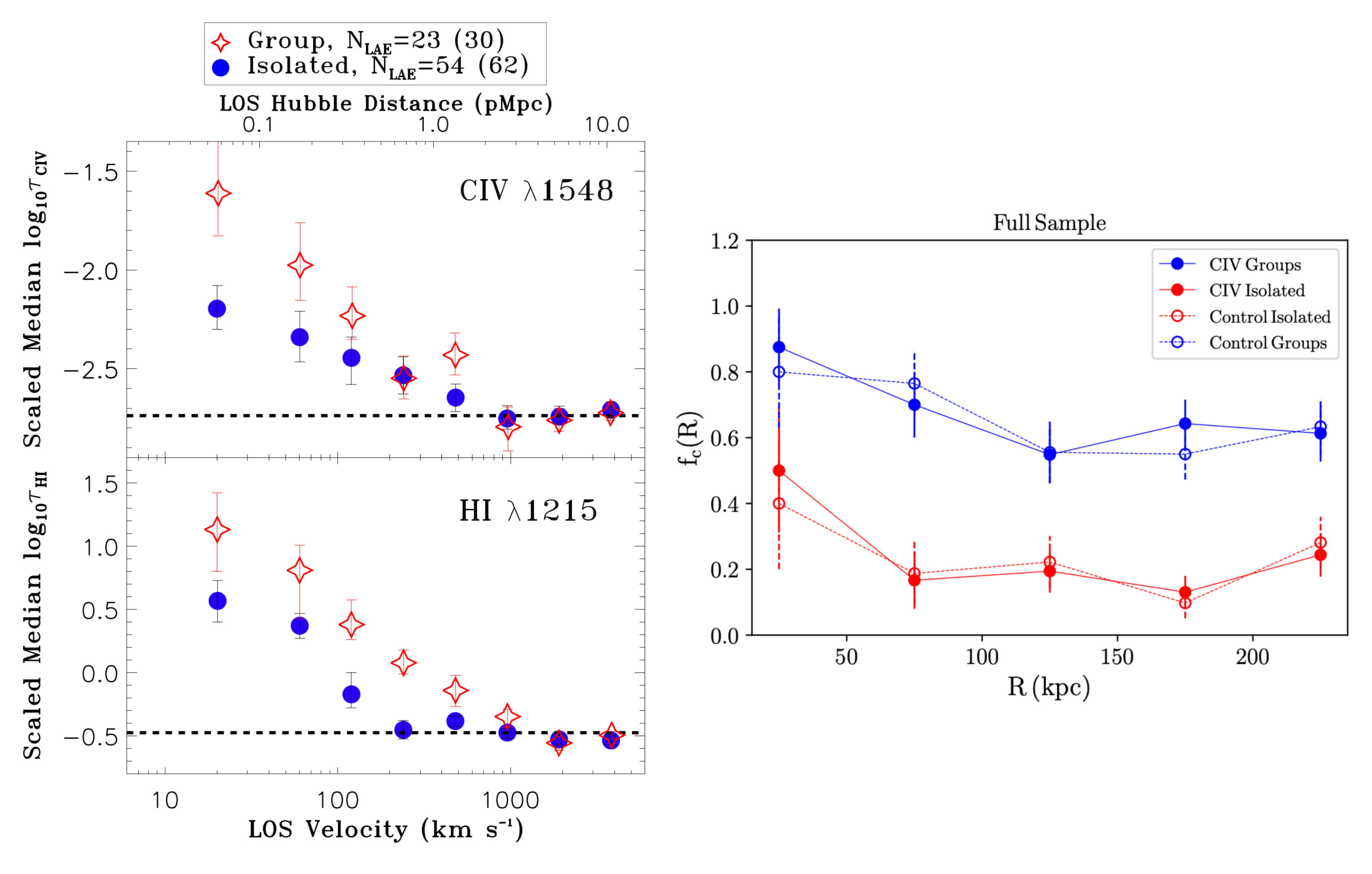}
\caption{Evironmental effects on the CGM of $z\approx 3.5$ galaxies as revealed by absorption spectroscopy. Left: The optical depth as a function of the line of sight velocity in stacks of spectra at the redshift of LAEs in groups (red) and isolation (blue). Right: Covering factors as a function of projected distance from LAEs in groups (now in blue) and isolation (now in red). In both cases, an excess absorption signal is found for group galaxies, hinting at mechanisms responsible for boosting the distribution of cool and ionized material near galaxies in richer environments. {\it Credits: Muzahid S. et al., 2021, MNRAS, 508, 5612; Galbiati M. et al., 2023, MNRAS, 524, 3474. Reproduced with permission.}}
\label{fig:highzenv}       
\end{figure}

The intragroup medium is a clear source of additional absorption for groups composed of a massive parent halo and additional sub-haloes, as implied by several studies \cite{Nielsen2018}. However, several pieces of information suggest that environmental perturbations on the CGM of individual galaxies can be responsible for the exceptionally high equivalent widths found in group galaxies, more than can be attributed to the intragroup medium. Firstly, by analogy of what is revealed in emission (see previous section), it is plausible to expect that tidal and hydrodynamic perturbation displace cool dense and metal-enriched gas at large distances from the gaseous disks, effectively boosting the cross-section of strong absorbers inside the CGM of individual galaxies. Moreover, multiple studies reveal how the properties of the absorbing gas found in the groups are more closely correlated with the presence of nearby galaxies than with the group as a whole \cite{Dutta2020,Dutta2021,Qu2023}. Finally, the tomographic experiment by Fossati et al. \cite{Fossati2019} in which galaxy background sightlines are used to probe the absorption in foreground groups reveals a clear preference for strong absorption to arise from the CGM of group members rather than from the geometric center of the groups. 

Together, these observations confirm that a combination of superposition models, intragroup medium, and environmental mechanisms are required to explain the excess of cool absorption seen in the CGM of group galaxies at $z\lesssim 2$ and that environmental mechanisms acting on individual halos are a particularly effective way to provide high equivalent-width absorbers in excess of what a superposition model would suggest.

Moving to a higher redshift, the role of the environment has been explored between $z\approx 3-4$ with MUSE by studying the distribution of \ion{H}{I} and \ion{C}{IV} near LAEs (Figure~\ref{fig:highzenv}). Similarly to the results at lower redshifts, observations reveal an elevated covering factor of hydrogen and metals in groups compared to isolated galaxies \cite{Lofthouse2023,Galbiati2023,Banerjee2023}. The same result is found when stacking \ion{H}{I} and \ion{C}{IV} absorption at the redshift of isolated or group LAEs \cite{Muzahid2021}, with the latter subsample being associated with higher optical depth (or higher equivalent width).  
Finally, the analysis of cool metal-enriched gas probed by \ion{Mg}{II} at comparable redshifts (Galbiati et al., submitted) once again reveals a marked excess of absorption in the group environment. 

Although independent observations confidently confirm the higher cross section of cool and more ionized gas, the interpretation of the observed signal is less clear. In this early epoch, massive virialized structures have yet to assemble, and the environmental mechanisms at work in massive groups or clusters are unlikely to operate precisely as seen at $z\lesssim 1.5$. 
This question remains open because of the lack of detailed modeling on the subject (see the next Section). Some hypotheses suggest that the observed excess in groups can be attributed to the large-scale structures (e.g., filaments) in which these galaxies are embedded. 

In the nearby Universe, at $z\lesssim 0.5$, absorption spectroscopy can also assess the gas distribution within the cluster environment. When considering the \ion{H}{I} distribution of cluster galaxies, various authors \cite{Yoon2013,Burchett2018} have found a marked reduction in the neutral hydrogen covering factor compared to field galaxies. This indicates that most neutral gas is either removed or ionized in clusters. Hence, the cold and natural components of the CGM can be easily truncated or stripped, likely contributing to the reduction of star formation through mechanisms like starvation.  
When comparing and contrasting the cool enriched gas within clusters, luminous red galaxies, and emission-line galaxies in considerable samples from the Sloan Digital Sky Survey (SDSS), Anand et al. \cite{Anand2022} confirm the lower covering factor of \ion{Mg}{II} within clusters compared to active galaxies. However, the cluster environment appears to have more cool gas than the halos of luminous red galaxies. This suggests that cool gas can be injected inside the cluster halos by infalling galaxies that lose some of their ISM components via stripping.

\subsection{The CGM and the environment in simulations}\label{sec:envsim}

\begin{figure}[b]
\centering
\includegraphics[scale=.20]{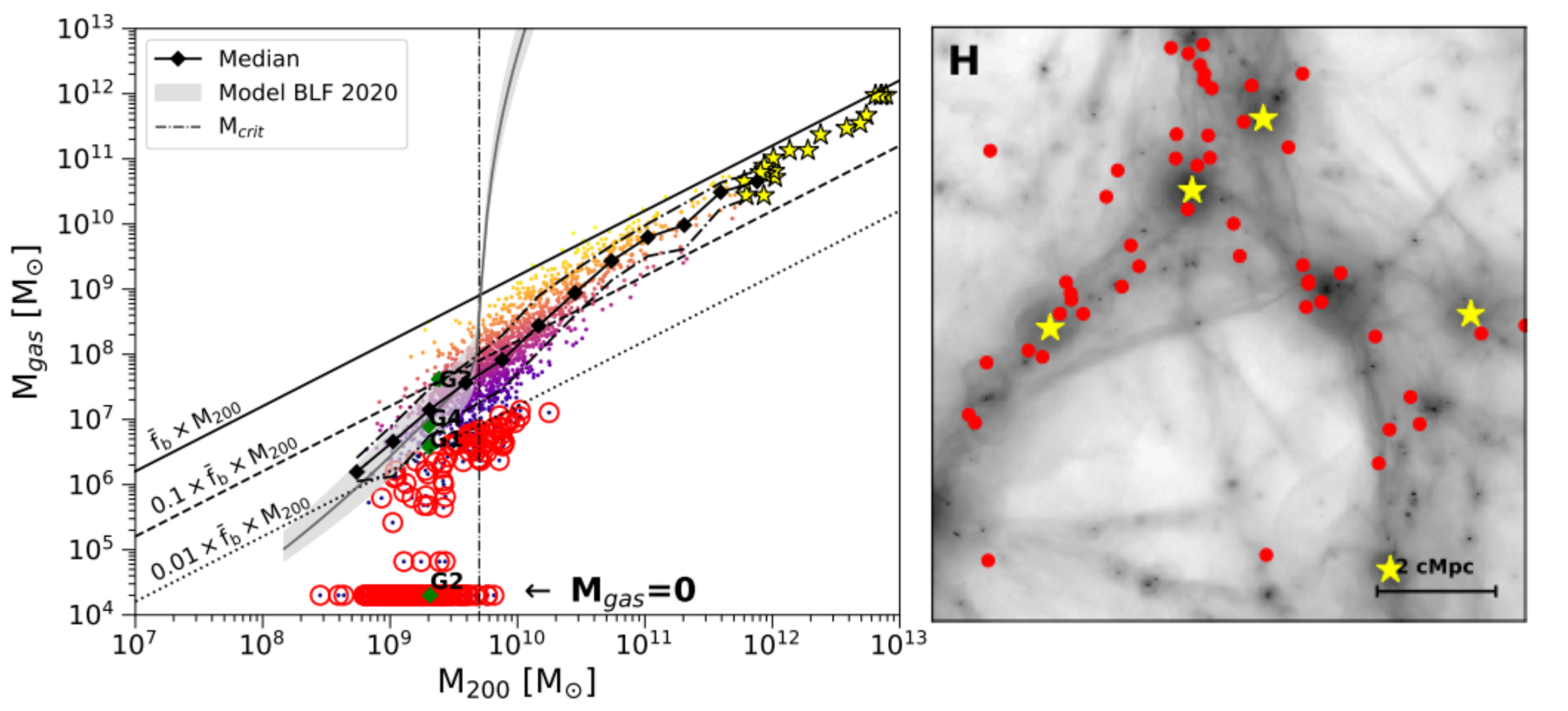}
\caption{Predictions of the environmental effects of the cosmic web on low-mass galaxies in the local Universe. Left: The analysis of the gas fraction as a function of halo mass reveals a $\lesssim 10^{10}$\,M$_\odot$ population of gas-deficient systems (circled in red), including halos completely devoided of gas (marked with $M_{gas}=0$). The color coding reflects the gas fraction, where the lightest colors are used for the universal baryon fraction and the darkest ones for gas-free systems. Right: These gas-poor halos reside in filaments and have lost much of their gas in the cosmic web before approaching massive halos. {\it Credits: Herzog G. et al. 2023, MNRAS, 518, 6305. Reproduced with permission.}}
\label{fig:cgmsimenv}       
\end{figure}

As argued in the previous Section, studying the environmental effects on the CGM is a novel investigation direction made possible by the recent availability of highly complete, flux-limited surveys in quasar fields. For this reason, theoretical investigations of the subject are not widely examined in the literature, but there is a valuable opportunity for further exploration of the topic.

Various studies have focused in the past decade on appreciating how the environment transforms the evolution of galaxies, especially in idealized and bespoke simulations \cite{Roediger2008,Tonnesen2009,Roediger2011,Tonnesen2012}. Although hydrodynamic and gravitational interaction effects are naturally tracked in these simulations, the emphasis has often been on galaxy properties, such as the ISM depletion or the quenching of star formation. Furthermore, initial conditions tailored to CGM studies should be considered beyond what is currently explored in these idealized setups.
Notable exceptions focusing specifically on the CGM of the Milky Way exist \cite{Salem2015,Setton2023,Roy2023}.

In more cosmological settings, the effects of the environment on the CGM can be studied in tandem with the galaxy transformation since no fine-tuning of the initial condition is required. However, one should be mindful of the limitations imposed by resolution in such calculations. 
Recent results from higher-resolution TNG simulations offer new insight into this problem. When considering the evolution of infalling satellites inside the halos of characteristic mass $\approx 10^{12}$\,M$_\odot$, Engler et al. \cite{Engler2023} identified ram pressure and tidal stripping as the primary mechanisms able to reduce the gas reservoirs of the satellites and consequently their star formation. Focusing explicitly on the so-called jellyfish galaxies undergoing ram pressure stripping using tracer particles, Rohr et al. \cite{Rohr2023} confirmed that ram pressure is the leading mechanism that depletes the gas reservoir in low-mass satellites accreting into massive ($10^{13}-10^{14}$\,M$_\odot$) structures and that galaxies infalling at early times (about half of the satellite population) are completely devoid of gas by $z\approx 0$.

Remaining in cosmological settings, Nelson et al. \cite{Nelson2021} considered the evolution of the \ion{Mg}{II} emission as a function of the environment. Interactions within galaxies affect the \ion{Mg}{II} distribution, inducing a scatter in the halo properties, particularly in the halo size, at fixed stellar mass.  By studying the halo's total luminosity, size, and shape as a function of galaxy overdensity, these authors found that galaxies in overdense environments host preferentially more extended \ion{Mg}{II} halos in emission, a trend also tentatively suggested by recent observations \cite{Dutta2023}. No significant dependence on luminosity and shape is found. 

Considering the wider large-scale structure environment, the importance of cosmic-web stripping has also been studied in numerical simulations. This process refers to suppressing the star formation in dwarf galaxies moving inside the cosmic web because of gas removal from their halo. This mechanism might render low-mass galaxies in the local volume invisible to \ion{H}{I} surveys \cite{BenitezLlambay2013}.  At the same time, low-velocity encounters between dwarfs and the low-density cosmic web might lead to a re-ignition of the star formation, as ram pressure is weak under this condition, and gas can be accreted and compressed inside the halo, triggering new episodes of star formation \cite{Wright2019}.

The problem of cosmic web stripping has been further investigated in recent work by Herzog et al. \cite{Herzog2023} using a high-resolution box based on the EAGLE model \cite{Schaye2015} with mass resolution $4.5\times 10^4$\,M$_\odot$ for gas and $2.4\times 10^5$\,M$_\odot$ for dark matter ($\approx 100$ times higher than the original EAGLE simulations). These authors identified a population of dwarf galaxies, termed COSWEBs, which lose gas from hydrodynamic interactions with the gaseous filaments and sheets of the cosmic web (Figure~\ref{fig:cgmsimenv}). Cosmic web stripping can reduce or even inhibit star formation entirely in $\gtrsim 70$ percent of the systems suffering from this environmental interaction.

The theoretical investigation of the role of the environment in shaping the halo gas is still in its early days. As the number of studies on how the environment affects the multiphase CGM as a function of cosmic time grows, theoretical work becomes essential to offer a framework for interpreting the empirical findings. Numerical simulations focused explicitly on how the CGM is affected by environmental mechanisms become a natural direction for the field.

\begin{acknowledgement}
MF thanks the 52nd Saas-Fee Advanced Course organizers, lecturers, and participants for a formative, engaging, and fun experience. MF acknowledges the fundamental contribution of Dr. Alejandro Benitez-Llambay, who has prepared and delivered the hands-on sessions at \url{https://alejandrobll.github.io/resources/saas-fee/hands-on.html}. Many thanks also to Prof. Rajeshwari Dutta, who has contributed to preparing the material presented in these lectures and the associated hands-on sessions. MF would like to thank Prof. Hsiao-Wen Chen, Prof. Rajeshwari Dutta, and Mr. Davide Tornotti for their useful comments on this manuscript. Some research presented in this contribution was made possible by funding from the European Research Council (ERC) under the European Union's Horizon 2020 research and innovation program (grant agreement No 757535) and by Fondazione Cariplo (grant No 2018-2329).

\end{acknowledgement}

\bibliographystyle{styles/spphys} 
\bibliography{refs} 

\end{document}